%% file: main.tex
\newif\ifOOPSLA
\OOPSLAfalse 

\documentclass[acmsmall,screen,nonacm]{acmart}

\input{packages}
\input{macros}

\title{Validating Concurrent Correctness Witnesses with Ghosts by Thread-Modular Abstract Interpretation\\(Extended Version)}

\author{Michael Schwarz}
\affiliation{%
  \institution{National University of Singapore}
  \city{Singapore}
  \country{Singapore}
}
\email{m.schwarz@nus.edu.sg}

\author{Zhendong Ang}
\affiliation{%
  \institution{National University of Singapore}
  \city{Singapore}
  \country{Singapore}
}
\email{zhendong.ang@u.nus.edu.sg}

\author{Umang Mathur}
\affiliation{%
  \institution{National University of Singapore}
  \city{Singapore}
  \country{Singapore}
}
\email{umathur@nus.edu.sg}

\input{abstract}

\begin{document}
\maketitle

\input{1-intro}
\input{2-defs.tex}
\input{3-tmai.tex}
\input{4-approach}
\input{5-mutex.tex}
\input{6-phases.tex}

\input{6a-further-refine.tex}
\input{7-unassume.tex}
\input{8-experiments}
\input{9-related}
\input{10-conclusion}
\paragraph{Acknowledgments}
This work is partially supported by the National Research Foundation, Singapore, and Cyber Security Agency of Singapore
under its National Cybersecurity R\&D Programme (Fuzz Testing <NRF-NCR25-Fuzz-0001>). Any opinions, findings and
conclusions, or recommendations expressed in this material are those of the author(s) and do not reflect the views
of National Research Foundation, Singapore, and Cyber Security Agency of Singapore.

\paragraph{Disclosure of Use of AI in Research}
The authors used \textsc{ChatGPT-Codex} interactively to assist with implementing the approach in
\textsc{Goblint}. No AI-generated code was incorporated verbatim.

\paragraph{Data Availability Statement}
Our implementation in \Goblint is OpenSource and a corresponding artifact is available at
\url{https://doi.org/10.6084/m9.figshare.34411389}.

\bibliographystyle{ACM-Reference-Format}
\bibliography{bibliography}
\appendix
\input{appendix/5b-proof.tex}
\input{appendix/crafted.tex}
\input{appendix/cyclic-phases.tex}
\end{document}

%% file: packages.tex
\usepackage{algorithm}                 
\usepackage[noend]{algpseudocode}      
\usepackage{balance}                   
\usepackage{xspace}
\usepackage{comment}
\usepackage{enumitem}
\usepackage{etoolbox}
\usepackage{halloweenmath}
\usepackage{pgfplots}
\usepackage{fontawesome}
\pgfplotsset{compat=1.18}
\usepgfplotslibrary{groupplots}
\usepgfplotslibrary{statistics}

\usepackage{subcaption}

\usepackage[capitalize]{cleveref}
\usepackage{booktabs}
\usepackage{array}
\usepackage{colortbl}

\usepackage{circledsteps}

\usepackage{listings}
\usepackage{xcolor}
\usepackage{mdframed}

\usepackage[frozencache]{minted}
\setminted{escapeinside=@@,fontsize=\footnotesize,mathescape,linenos,texcomments}

\usepackage{todonotes} 

\usepackage{tikz}
\usetikzlibrary{
    arrows.meta,      
    positioning,      
    shapes.geometric, 
    shadows,          
    shadows.blur,
    decorations.pathmorphing, 
    calc,             
    fit,              
    backgrounds,       
    tikzmark,
    positioning,
    chains,
    patterns,
    patterns.meta
}

\usepackage{tcolorbox}
\tcbuselibrary{skins} 
\tcbset{
    mygraybox/.style={
        colback=gray!15, 
        colframe=gray!60, 
        rounded corners, 
        boxsep=5pt, 
        arc=5pt, 
        top=3pt, bottom=3pt, left=3pt, right=3pt, 
        boxrule=0.5pt 
    }
}

%% file: macros.tex
\definecolor{codegreen}{rgb}{0,0.6,0}
\definecolor{codegray}{rgb}{0.5,0.5,0.5}
\definecolor{codepurple}{rgb}{0.58,0,0.82}
\definecolor{backcolour}{rgb}{0.95,0.95,0.92}

\lstdefinestyle{mystyle}{
    backgroundcolor=\color{backcolour},
    commentstyle=\color{codegreen},
    keywordstyle=\color{magenta},
    numberstyle=\tiny\color{codegray},
    stringstyle=\color{codepurple},
    basicstyle=\ttfamily\footnotesize,
    breakatwhitespace=false,
    breaklines=true,
    captionpos=b,
    keepspaces=true,
    numbers=left,
    numbersep=5pt,
    showspaces=false,
    showstringspaces=false,
    showtabs=false,
    tabsize=2
}
\newcommand{\code}[1]{\texttt{\detokenize{#1}}}

\newtheorem{remark}{Remark}
\newtheorem{proofsketch}{Proof Sketch}
\AtEndEnvironment{remark}{\qed}
\AtEndEnvironment{example}{\qed}

\newcommand{\sem}[1]{\llbracket #1 \rrbracket}

\newcommand{\RQ}[1]{\textbf{RQ#1}}

\newcommand{\LocalUnknowns}{{\mathcal{L}}}
\newcommand{\GlobalUnknowns}{{\mathcal{G}}}

\makeatletter
\newcommand{\anonorfinal}[2]{%
  \if@ACM@anonymous
    #1
  \else
    #2
  \fi
}
\makeatother

\newcommand{\GoblintAnon}{\anonorfinal{\textsc{WizWoz}}{\textsc{Goblint}}\xspace}
\newcommand{\SVComp}{\textsc{SV-COMP}\xspace}
\newcommand{\Goblint}{\textsc{Goblint}\xspace}
\newcommand{\Ultimate}{\textsc{UGemcutter}\xspace}

\newcommand{\Domain}{\ensuremath{\mathcal{D}^\sharp}}

\newcommand\Mu{\mathrm{M}}

\newcommand*{\Nodes}{\ensuremath{\mathcal{N}}}

\newcommand*{\Globals}{\ensuremath{\mathcal{G}}}

\newcommand*{\lock}[1]{\code{lock(#1)}}
\newcommand*{\unassume}[1]{\code{unassume(#1)}}
\newcommand*{\unlock}[1]{\code{unlock(#1)}}

\newcommand*{\ghost}{\mathghost}

\newcommand*{\GhostGlobalDeclarations}{\ensuremath{\mathcal{D}}_{\ghost}}
\newcommand*{\GhostUpdate}{\ensuremath{\mathbb{U}}}
\newcommand*{\GhostInvariant}{\ensuremath{\mathbb{I}}}
\newcommand*{\Witness}{\ensuremath{(\GhostGlobalDeclarations, \GhostUpdate, \GhostInvariant)}}

\newcommand{\aSem}[1]{\left\llbracket #1 \right\rrbracket^\sharp}

\newcommand{\aSemB}[1]{\left\llbracket #1 \right\rrbracket^\sharp_{B}}
\newcommand{\aSemEB}[1]{\left\llbracket #1 \right\rrbracket^\sharp_{\mathcal{B}}}
\newcommand{\aSemInt}[1]{\left\llbracket #1 \right\rrbracket^\sharp_{\mathbb{Z}}}
\newcommand{\semInt}[1]{\left\llbracket #1 \right\rrbracket_{\mathbb{Z}}}
\newcommand{\aSemDigest}[1]{\left\llbracket #1 \right\rrbracket^\sharp_{\Digests}}

\newcommand{\Digests}{{\mathcal{A}}}
\newcommand{\Phases}{\mathcal{P}}
\newcommand{\PhaseMaps}{{\bar\Phases}}
\newcommand{\ExtendedPhaseMaps}{{\hat\Phases}}
\newcommand{\DomainPhaseMap}{\ensuremath{\mathcal{D}^\sharp_{\PhaseMaps}}}
\newcommand{\ConcCtx}{{\mathcal{C}^\sharp}}
\newcommand{\succs}{\textsf{succs}}
\newcommand{\psuccs}{\textsf{psuccs}}
\newcommand{\lift}{\textsf{lift}}
\newcommand{\unlift}{\textsf{unlift}}
\newcommand{\phaseBound}{\texttt{phase\_bound}}
\newcommand{\absPointsToIn}[2]{#1 \mapsto^\sharp_{#2} }
\newcommand{\TIDs}{\mathcal{I}^\sharp}

\newcommand{\unique}{\textsf{unique}}
\newcommand{\projTIDs}{\pi_{\TIDs}}
\newcommand{\projConcCtx}{\pi_{\ConcCtx}}
\newcommand{\phaseGhostTIDs}{\TIDs_{\mathghost}}

\newcommand{\DomainAux}{\ensuremath{\mathcal{D}^\sharp_{\mathit{aux}}}}
\newcommand{\projDomainAux}{\pi_{\DomainAux}}

\newcommand{\projToPhaseMap}{\pi}

\newcommand{\aSemP}[1]{\left\llbracket #1 \right\rrbracket^\sharp_{\PhaseMaps}}
\newcommand{\stmt}{\textsf{stmt}}

\newcommand{\aTrue}{\textsf{true}^\sharp}
\newcommand{\aFalse}{\textsf{false}^\sharp}
\newcommand{\Traces}{\mathcal{T}}

\newcommand{\Let}{\textbf{let}}

\newcommand{\In}{\textbf{in}}

\newcommand{\atomicBegin}{\texttt{atomic\_begin}}
\newcommand{\atomicEnd}{\texttt{atomic\_end}}

%% file: abstract.tex
\begin{abstract}
Concurrent programs are notoriously difficult to write and to reason about, and
even tools based on sound theoretical foundations can produce incorrect results
due to implementation bugs.
Correctness witnesses have been proposed as a way to increase the confidence in the verdicts
of software verifiers by enabling independent validation.
Recently, correctness witnesses have been extended to the concurrent setting,
by introducing ghost variables to allow encoding more expressive invariants.
While different conceptual approaches to producing witnesses with ghosts exist, validation techniques
are still lacking; instead of leveraging the information encoded
in the ghosts to guide the validation process, existing validators merely treat the
ghost instrumentation and invariants over ghost variables as additional proof obligations.
We propose a novel framework for the validation of concurrent correctness witnesses
that exploits the semantic information encoded in ghost variables.
It combines a checking phase, which establishes the semantic content of the ghost instrumentation,
with an exploitation phase making use of this information.
Both phases are formulated as thread-modular abstract interpretation.
In particular, we consider two different classes of ghost variables and
techniques for exploiting them, one based on \emph{resource invariants} and one based on
\emph{phases}.
Phase ghosts guide the degree to which the abstract interpretation considers interleavings
instead of always abstracting them coarsely.
In this way, the abstract interpreter can validate witnesses containing
invariants beyond the reach of unguided thread-modular analysis,
while still retaining its scalability benefits.
We have implemented our framework in a static analyzer and
evaluate it on witnesses encoding resource invariants obtained
from a state-of-the-art software verifier, as well
as on hand-crafted witnesses encoding phase information.
\end{abstract}

%% file: 1-intro.tex
\section{Introduction}
Concurrent programs are prone to subtle bugs. This has prompted the development of a wide range of automated
techniques, from those that \emph{catch bugs} during execution such as fuzzing or dynamic analysis, to \emph{sound} approaches
that can establish the \emph{absence} of bugs by considering \emph{all} possible executions.
The latter class encompasses approaches as diverse as those based on commutativity reasoning, automata-based model checking,
Petri nets, sequentialization, or abstract interpretation.
While the underlying theories are elegant and simple, their implementations are often in
large and complex~\cite{DBLP:journals/fmsd/CousotCFMMR09,DBLP:journals/sttt/MonatOM24,DBLP:journals/cacm/BaudinBBCKKMPPS21,CAV26b,DBLP:conf/tacas/SaanESBHTVS24a}
software systems.
Thus, even tools firmly based on a sound theory have been shown to occasionally produce incorrect results
(see, e.g., \cite{DBLP:conf/kbse/KaindlstorferIW24,ErhardBHKSSSTV2025,DBLP:conf/issta/KlingerCW19,DBLP:conf/cav/VishwanathanSNN23}
for some recent work that has identified implementation bugs in mature static analyzers).

To increase trust in the verdicts of such \emph{software verifiers}, the community
has adopted the idea of \emph{witnesses}, which originated in certifying algorithms~\cite{DBLP:journals/csr/McConnellMNS11},
where tools produce not only an answer, but also a proof object (\emph{witness}) supporting their claim.
This witness can be independently \emph{validated} by other tools --- ideally employing a different approach than the tool
producing the witness.

In witness-based verification workflows, a \emph{verifier} emits both a verdict and a witness,
both of which are passed to the \emph{validator}.
Here, we focus on the more difficult case of correctness witnesses, i.e., witnesses that support
the claim that a program is free of errors, as opposed to violation witnesses that support the
claim that a program contains an error.
The International Competition on Software Verification (\SVComp)~\cite{svcomp25,svcomp24,svcomp26}
is a prominent and community-maintained setting in which this workflow is deployed for {C} verification.

Witness format design has evolved from an initial control-flow-automata-based representation~\cite{DBLP:conf/sigsoft/0001DDH16}
to a simpler format that annotates program points with invariants expressed as side-effect-free {C} expressions~\cite{DBLP:conf/spin/AyaziovaBRSS24}.
This shift reflects a move toward better semantic portability across tools, directly tying witnesses to the semantics
of the source language without relying on an intermediate representation, but also limits the expressiveness of the witness format.

Recently~\cite{ErhardBHKSSSTV2025}, the format was extended with \emph{ghost variables}. This allows providing correctness
witnesses for concurrent programs, where invariants about in-scope program variables at each program point
do not suffice to encode meaningful information about the interactions between threads.
This work also establishes that an exchange of invariants
between tools based on different verification approaches and underlying concrete semantics of concurrent programs is possible.

While the semantic gap between different approaches to concurrent program verification is bridged
by ghost variables
(see \cref{fig:positioning}), the question of how to \emph{efficiently} validate such witnesses remains open.
The only approach that has been proposed~\cite{ErhardBHKSSSTV2025}
works by transforming the witness and the original program into a program with additional assertions corresponding
to the provided invariants, which is then analyzed using a standard verification technique.
Thus, the validator is only burdened by the additional invariants provided by the witness, rather than using the semantic
information encoded in the ghost variables and the invariants to guide the validation process.

\begin{figure}
\centering
\small
\begin{tikzpicture}[>=latex, font=\footnotesize, every node/.style={inner sep=2pt}]
    \node[
        draw=none,
        align=center,
        inner sep=0pt
    ] (witness) at (0,0) {%
        \tikz[baseline=(ghost.base)]{
            \node[
                inner sep=3pt,
                preaction={fill=orange!20},
                preaction={
                    pattern={Lines[
                        angle=-45,
                        distance={16pt/sqrt(2)},
                        xshift=8pt,
                        line width=4.4pt
                    ]},
                    pattern color=blue!20
                }
            ] (ghost) {{\Large $\text{\faFile}_{\mathghost}$}};
        }\\
        Ghost Witness\\
        \tiny{for exchange across tools}\\
        \tiny{and semantic divide~\cite{ErhardBHKSSSTV2025}}
    };

    \node[font=\bfseries] at (-4.65, 2.9) {Verifiers};
    \node[draw,dotted,rectangle, minimum width=3.1cm, minimum height=0.8cm, text width=3.0cm, align=center, anchor=east, fill=orange!10] (comm)     at (-3.1,  2.0) {Commutativity\\Reasoning};
    \node[draw,dotted,rectangle, minimum width=3.1cm, minimum height=0.8cm, text width=3.0cm, align=center, anchor=east, fill=orange!10] (automata) at (-3.1,  1.0) {Automata-Based\\Model Checking};
    \node[draw,      rectangle, minimum width=3.1cm, minimum height=0.8cm, text width=3.0cm, align=center, anchor=east, fill=orange!10] (petri)    at (-3.1,  0.0) {Petri-Net Based\\Approaches};
    \node[draw,      rectangle, minimum width=3.1cm, minimum height=0.8cm, text width=3.0cm, align=center, anchor=east, fill=blue!10] (thread)   at (-3.1, -1.0) {Thread-Modular\\Abstract Interpretation};
    \node[draw,dotted,rectangle, minimum width=3.1cm, minimum height=0.8cm, text width=3.0cm, align=center, anchor=east, fill=orange!10] (seq)      at (-3.1, -2.0) {Sequentialization-Based\\Techniques};

    \node[font=\bfseries] at (4.65, 2.9) {Validators};

    \node[draw, rectangle, minimum width=3.8cm, minimum height=0.8cm, text width=3.0cm, align=center, anchor=west, fill=orange!10] (commR)     at (3.1,  2.0) {Commutativity\\Reasoning};
    \node[draw, rectangle, minimum width=3.8cm, minimum height=0.8cm, text width=3.0cm, align=center, anchor=west, fill=orange!10] (automataR) at (3.1,  1.0) {Automata-Based\\Model Checking};
    \node[draw, rectangle, minimum width=3.8cm, minimum height=0.8cm, text width=3.0cm, align=center, anchor=west, fill=orange!10] (petriR)    at (3.1,  0.0) {Petri-Net Based\\Approaches};
    \node[draw, rectangle, very thick, minimum width=3.8cm, minimum height=0.8cm, text width=3.7cm, align=center, anchor=west, font=\bfseries, fill=blue!10] (threadR)   at (3.1, -1.0) {Thread-Modular\\Abstract Interpretation};
    \node[draw, dotted, rectangle, minimum width=3.8cm, minimum height=0.8cm, text width=3.7cm, align=center, anchor=west, fill=orange!10] (seqR)      at (3.1, -2.0) {Sequentialization-Based\\Techniques};

    \coordinate (w1) at (-1.2,  0.60);
    \coordinate (w2) at (-1.2,  0.30);
    \coordinate (w3) at (-1.2,  0.00);
    \coordinate (w4) at (-1.2, -0.30);
    \coordinate (w5) at (-1.2, -0.60);

    \coordinate (w1r) at (1.2,  0.60);
    \coordinate (w2r) at (1.2,  0.30);
    \coordinate (w3r) at (1.2,  0.00);
    \coordinate (w4r) at (1.2, -0.30);
    \coordinate (w5r) at (1.2, -0.60);

    \draw[dotted, -latex] (comm.east)     -- (w1);
    \draw[dotted, -latex] (automata.east) -- (w2);
    \coordinate (petriMid) at ($(petri.east)!0.5!(w3)$);
    \draw                 (petri.east) -- (petriMid);
    \draw[dotted, -latex] (petriMid) -- (w3);
    \node[above] at (petriMid) {\cite{SchusseleZLK26}};
    \draw[-latex]         (thread.east)   -- (w4)
        node[midway, above, xshift=-3pt] {\cite{ErhardBHKSSSTV2025, SaanPhD}};
    \draw[dotted, -latex] (seq.east)      -- (w5);

    \draw[-latex] (w1r) -- (commR.west)
        node[midway, yshift=15pt] {$||$, \cite{ErhardBHKSSSTV2025}};
    \draw[-latex] (w2r) -- (automataR.west)
        node[midway, yshift=7pt] {$||$};
    \draw[-latex] (w3r) -- (petriR.west)
        node[midway, yshift=7pt] {$||$};

    \draw[very thick,-latex] (w4r) -- (threadR.west);
    \draw[dotted, -latex] (w5r) -- (seqR.west);

    \node[anchor=center, font=\scriptsize] at (0,-2.85) {%
        {\setlength{\fboxsep}{0pt}%
        \begin{tabular}{@{}c@{\hspace{0.4em}}l@{\hspace{1.2em}}c@{\hspace{0.4em}}l@{\hspace{1.2em}}l@{}}
        \fcolorbox{black}{orange!10}{\rule{0pt}{0.28cm}\rule{0.45cm}{0pt}} & Interleaving Semantics &
        \fcolorbox{black}{blue!10}{\rule{0pt}{0.28cm}\rule{0.45cm}{0pt}} & Local Trace Semantics &
        $||$ = Product Construction, followed by standard verification
        \end{tabular}}%
    };

\end{tikzpicture}
\caption{Ghost witnesses can be used to exchange information between different approaches to concurrent program verification.
Our work corresponds to the part highlighted in bold.}
\Description{Diagram positioning this work.}
\label{fig:positioning}
\end{figure}

Existing validators for ghost witnesses are all based on model checking and consider the possible
interleavings of different threads. These techniques can be quite precise but usually scale rather poorly.
Approaches based on thread-modular abstract interpretation, on the other hand, are fast and scale well by coarsely
abstracting the interactions between threads. However, precise invariants that require more
fine-grained interleaving exploration are usually out of reach for such approaches.

We address this challenge by developing a new approach to validate correctness witnesses with ghosts
based on thread-modular abstract interpretation that uses the information encoded in the ghost variables
and invariants to guide the validation process.
In a sequential setting, 
such guidance is mostly helpful in avoiding overshooting
a fixpoint during widening where the invariants provided by the witness act similarly to
widening thresholds.
For concurrent programs, the guidance provided by the witness can be more substantial, as
ghost variables may, e.g., encode the degree to which interleavings are to be distinguished, which
can be crucial for making an analysis more precise.
\input{ex-phase.tex}
\begin{example}\label{ex:running-disc}
    Consider the program in \cref{fig:running}, ignoring the ghost variables for now. A thread-modular
    abstract interpretation that uses the interval domain and flow-insensitively collects
    all possible values of $g$ will establish the invariant $g \in [1,\infty]$ at the end of the program,
    which is not sufficient to establish that the assertion holds.
    The ghost variables (seen here in gray) provided by a witness, e.g., produced by a model checker,
    can then serve as hints about which \emph{phases} of the program to distinguish,
    based on whether the respective updates to $g$ have taken place or not (indicated
    by $\Circled{\mathrm{A}}$/$\Circled{\mathrm{B}}$ for the main thread and
    $\Circled{\mathrm{X}}$/$\Circled{\mathrm{Y}}$ for $t_1$).
    Using this information to compute one abstraction of $g$ for each possible combination of phases,
    the analysis can establish the invariant $g \in [3,\infty]$ when the program is in the phases
    $\left(\Circled{\mathrm{B}},\Circled{\mathrm{Y}}\right)$ which is when the assert happens.
    Moreover, the abstract interpreter can now confirm invariants from the witness such as
    $\phi_B \equiv (\ghost_{\textsf{main}} = 1) \land ((\ghost_{\textsf{t1}} = 0 \land g \geq 1) \lor (\ghost_{\textsf{t1}} = 1 \land g \geq 2))$
    that were out of reach before.
\end{example}
Unlike an analysis that considers all interleavings, this analysis distinguishes only the phases
that other threads may occupy, as indicated by the witness.

In particular, we want our approach 
to (a) provide a speedy validation of witnesses that do not require reasoning
about interleavings at all, as well as to (b) validate witnesses for programs whose correctness is out-of-reach for
thread-modular abstract interpretation without guidance from a witness by using the information encoded
in the ghost variables.

One of the challenges is that validators must remain \emph{sound} even when given an invalid witness,
i.e., they must not \emph{rely} on information provided by the witness, such as candidate invariants
to actually be invariant.
Furthermore, they must not \emph{rely} on the ghost variables and
updates encoding semantically meaningful information.
We thus propose a two-phase approach that first \emph{checks} whether a ghost
variable is used consistently with one of the supported hypotheses and, after establishing that it is,
\emph{exploits} this information.
In summary, we make the following contributions:
\begin{itemize}
    \item We provide a novel approach to the validation of concurrent correctness witnesses with ghosts
        based on thread-modular abstract interpretation. This
        is the first approach to not just consider the information provided in the witness as additional
        proof burdens.
    \item We propose a two-phase approach: A \emph{check} phase to confirm hypotheses about
        what ghost variables encode, followed by a phase \emph{exploiting} these hypotheses
        during validation.
    \item We instantiate this for two types of ghost variables:
        mutex ghosts to encode resource invariants,
        and phase ghosts to encode information about which interleavings to distinguish.
    \item We show that lifting a sound abstract interpreter with our approach yields a sound validator
        for concurrent correctness witnesses with ghosts.
    \item We implement our approach in \Goblint
        and evaluate it on ghost witnesses
        for \SVComp tasks from prior work~\cite{ErhardBHKSSSTV2025},
        complemented by a study of witnesses using phase ghosts.
\end{itemize}

The rest of this paper is structured as follows: \cref{sec:background} introduces ghost witnesses and their validity,
background on thread-modular abstract interpretation, and, putting both together, describes
how to validate witnesses for single-threaded programs by abstract interpretation.
\cref{sec:approach} outlines our general two-phase approach. \cref{sec:mutex-ghosts} explains how to leverage
witnesses encoding resource invariants,
whereas \cref{sec:phase-ghosts} does so for phases.
\cref{sec:unassume} discusses how to integrate unassume operators into ghost witness validation.
\cref{sec:experiments} describes our implementation and experimental evaluation,
whereas \cref{sec:related} outlines related work and \cref{sec:conclusion} gives future work and concludes.

%% file: ex-phase.tex
\begin{figure}[t]
    \begin{minipage}[t]{0.5\textwidth}
\begin{minted}[escapeinside=@@]{C}
main:                         @{\tikzmark{ExPhaseAtop}}@
    assume(g > 0);
    create(t1);
    // $\phi_A$
    @\colorbox{gray!20}{\strut\texttt{atomic \{}}@ g++; @\colorbox{gray!20}{\strut$\ghost_{\textsf{main}}$ \texttt{= 1; \}}}@  @{\tikzmark{ExPhaseAmid}}@
    // $\phi_B$
    join(t1);
    assert(g > 2);            @{\tikzmark{ExPhaseAend}}@
\end{minted}
\tikz[remember picture,overlay]{
    \draw[line width=0.8pt]
      let \p1=(pic cs:ExPhaseAtop), \p2=(pic cs:ExPhaseAmid) in
      (\x1+2.8em,\y1+0.8ex) --
      (\x1+3em,\y1+0.8ex) --
      (\x1+3em,\y2+0.5ex) --
      (\x1+2.8em,\y2+0.5ex);

    \draw[line width=0.8pt]
      let \p1=(pic cs:ExPhaseAtop), \p2=(pic cs:ExPhaseAmid),
          \p3=(pic cs:ExPhaseAend) in
      (\x1+2.8em,\y2+0.2ex) --
      (\x1+3em,\y2+0.2ex) --
      (\x1+3em,\y3-0.8ex) --
      (\x1+2.8em,\y3-0.8ex);

    \path let \p1=(pic cs:ExPhaseAtop), \p2=(pic cs:ExPhaseAmid) in
      node[draw,circle,inner sep=1pt,fill=white]
      at (\x1+4.2em,{0.5*(\y1+\y2)}) {A};

    \path let \p1=(pic cs:ExPhaseAtop), \p2=(pic cs:ExPhaseAmid),
          \p3=(pic cs:ExPhaseAend) in
      node[draw,circle,inner sep=1pt,fill=white]
      at (\x1+4.2em,{0.5*(\y2+\y3)}) {B};
}
    \end{minipage}\begin{minipage}[t]{0.5\textwidth}
\begin{minted}[firstnumber=9,escapeinside=@@]{C}
t1:                           @{\tikzmark{ExPhaseXtop}}@
    a = 5;                    @{\tikzmark{ExPhaseXmid}}@
    // ...
    // $\phi_X$
    @\colorbox{gray!20}{\strut\texttt{atomic \{}}@ g++; @\colorbox{gray!20}{\strut$\ghost_{\textsf{t1}}$ \texttt{= 1; \}}}@  @{\tikzmark{ExPhaseYmid}}@
    // $\phi_Y$
    // ...
    return;                   @{\tikzmark{ExPhaseYend}}@
\end{minted}
\tikz[remember picture,overlay]{
    \draw[line width=0.8pt]
      let \p1=(pic cs:ExPhaseXtop), \p2=(pic cs:ExPhaseYmid) in
      (\x1+2.8em,\y1+0.8ex) --
      (\x1+3em,\y1+0.8ex) --
      (\x1+3em,\y2+0.5ex) --
      (\x1+2.8em,\y2+0.5ex);

    \draw[line width=0.8pt]
      let \p1=(pic cs:ExPhaseXtop), \p2=(pic cs:ExPhaseYmid),
          \p3=(pic cs:ExPhaseYend) in
      (\x1+2.8em,\y2+0.2ex) --
      (\x1+3em,\y2+0.2ex) --
      (\x1+3em,\y3-0.8ex) --
      (\x1+2.8em,\y3-0.8ex);

    \path let \p1=(pic cs:ExPhaseXtop), \p2=(pic cs:ExPhaseXmid) in
      node[draw,circle,inner sep=1pt,fill=white]
      at (\x1+4.2em,{0.5*(\y1+\y2)}) {X};

    \path let \p1=(pic cs:ExPhaseXtop), \p2=(pic cs:ExPhaseYmid),
          \p3=(pic cs:ExPhaseYend) in
      node[draw,circle,inner sep=1pt,fill=white]
      at (\x1+4.2em,{0.5*(\y2+\y3)}) {Y};
}
\end{minipage}
\caption{Example program (adapted from \cite{SchusseleZLK26}) over mathematical integers with two threads
and two phase ghosts
which we assume are both initialized to $0$. Here, we assume increment operations to be atomic.
The dots represent arbitrary code
that does not access the shared variable $g$ or the ghost variables.}
\label{fig:running}
\end{figure}

%% file: 2-defs.tex
\section{Background and Definitions}\label{sec:background}
We first introduce the notion of ghost witnesses and their validity, and give some background on thread-modular
abstract interpretation, which is the basis of our approach to witness validation.
\subsection{Ghost Witnesses and Their Validity}\label{sec:defs}
The general setup for witness validation is illustrated in \cref{fig:verifier}.
\begin{figure}
\centering
\begin{tikzpicture}[font=\small]
    \node[align=center] (input) at (0,0) {Program $P$\\Property $\varphi$};
    \node[draw, rectangle, minimum width=2cm, minimum height=0.65cm] (verifier) at (3.2,0.5) {Verifier};
    \node[align=center] (witness) at (6.0,0.5) {{\large $\text{\faFile}_{\mathghost}$}\\Ghost witness};
    \node[draw, rectangle, minimum width=2cm, minimum height=0.65cm] (validator) at (9.0,0.5) {Validator};

    \node (validated) at (12.0,1.0) {{\color{green!60!black}\faCheck}\,Validated};
    \node (unknown)   at (12.0,0.5) {{\faQuestionCircle}\,Unknown};
    \node (rejected)  at (12.0,0.0) {{\color{red!60!black}\faTimes}\,Rejected};

    \coordinate (inputsplit) at ([xshift=1.0cm]input.east);

    \draw (input.east) -- (inputsplit);
    \draw[->] (inputsplit) |- (verifier.west);
    \draw[->] (inputsplit) -| (validator.south);

    \draw[->] (verifier.east) -- (witness.west);
    \draw[->] (witness.east) -- (validator.west);

    \draw[->] (validator.east) -- (validated.west);
    \draw[->] (validator.east) -- (unknown.west);
    \draw[->] (validator.east) -- (rejected.west);
\end{tikzpicture}
\caption{Witness Validation: A \emph{verifier} takes a program $P$ and a property $\varphi$ as input
and produces a (ghost) witness. It is passed to a validator, along with the program and the property.
The validator then either validates the witness, rejects it, or reports unknown.}
\Description{Block diagram illustrating the witness validation setup as used in SV-COMP. A verifier takes a program and a property as input and produces a ghost witness.}
\label{fig:verifier}
\end{figure}
%
%
We define concurrent correctness witnesses (\emph{ghost witnesses}) and their \emph{validity}
as follows.
\begin{definition}[Ghost Witness]
    A ghost witness for program $P$ and property $\varphi$ is a tuple $\Witness$:
    \begin{itemize}
        \item $\GhostGlobalDeclarations$: A set of global ghost variables, including types and initial values;
        \item $\GhostUpdate$: Ghost updates, i.e., a function from a statement to a (potentially empty) sequence of updates to ghost variables
            to be performed after executing the statement. The ghost updates are assignments where the right-hand
            side may refer to both in-scope program variables and ghost variables.
            To an observer, the statement and the ghost updates are atomic;
        \item $\GhostInvariant$: Ghost invariants, i.e., a function from program points to 
            invariants over in-scope program variables and ghost variables expressed as side-effect free {C} expressions.
    \end{itemize}
\end{definition}
\begin{definition}[Ghost Witness Validity]
    A witness $\Witness$ for a program $P$ and property $\varphi$ is \emph{valid} if
    \textbf{a)} $P$ satisfies $\varphi$, \emph{and} \textbf{b)}
    in the program $P_{\ghost}$ obtained by inserting ghost variables and updates into $P$,
    when program execution reaches a program point $u$ in state $\sigma$,
    $\GhostInvariant\,u$ evaluates to 
    \texttt{true} in $\sigma$.
\end{definition}
These definitions and our approach are generic in the property $\varphi$, except that it has to be
\emph{robust} w.r.t.\ the insertion of ghost variables and updates in the following sense:
$P$ satisfies $\varphi$ if and only if $P_{\ghost}$ satisfies $\varphi$.\footnote{
    For technical reasons we further require $\varphi$ to be robust w.r.t.\
    introducing fresh local variables for each global variable and, in an atomic block,
    assigning the value of the global variable to the local variable and then
    writing it back to the global.
    All properties considered in \textsc{SV-COMP} are robust in both senses,
    as they refer to the high-level behavior of programs, such as whether a specific function is called
    or whether a signed integer overflow can happen.
}
In our examples, we focus on the property that all assertions in the program hold.

\begin{example}
    Consider again the program in \cref{fig:running} (ignoring the gray parts). From this program $P$
    and a witness $\Witness$ that introduces the ghost variables $\ghost_{\textsf{main}}$ and $\ghost_{\textsf{t1}}$
    (initialized to $0$) and updates them to $1$ when the corresponding assignment to $g$ is executed,
    one obtains a program $P_{\ghost}$, which corresponds to the program in \cref{fig:running} when also considering
    the parts in gray.
\end{example}

\begin{remark}
    Perhaps surprisingly, the empty witness $(\emptyset, \textrm{fun } \_ \to \epsilon,
    \textrm{fun } \_ \to \texttt{true})$ is a valid witness for any property $\varphi$ and program $P$
    that satisfies $\varphi$.
    This choice is deliberate: Not all invariants that verifiers
    infer are expressible as {C} expressions. For instance, there is no way to encode quantifiers,
    and not all invariants about the heap can be expressed as {C} expressions about in-scope variables either.
    Thus, one cannot demand that the invariants on their own imply the property $\varphi$, or even that they are inductive.
    This is part of what makes witness validation challenging.
\end{remark}
\paragraph{Tension between Expressiveness and Portability.}
In general, the witness format design reflects a trade-off between expressiveness and
portability across verification approaches and tools:
Providing more of the producer's internal proof can simplify validation, but requires the validator to understand that proof representation.
For instance, \GoblintAnon incorporates an independent checker that verifies whether the abstract states
computed during analysis satisfy its constraint system (see
\cite{DBLP:conf/iclp/AlbertAPH06,DBLP:journals/tcs/BessonJP06,DBLP:conf/lpar/AlbertPH04,DBLP:conf/esop/BessonJT07}
for other uses of fixpoint certificates in abstract interpretation).
Such certificates depend on the underlying abstract domains and transfer functions, whereas
automata-based approaches may produce interpolant automata. Exchanging these internal certificates
thus requires support for the producer's proof representation.
A further step is to translate internal proof information into independently checkable deductive
certificates, as proposed for the $\mu$-calculus~\cite{DBLP:conf/cav/Namjoshi01} and
{LTL}~\cite{DBLP:conf/fsttcs/PeledPZ01}. Validation then requires support for the shared proof system.
The invariant-based witness format used in \SVComp communicates
properties of the program, together with ghost instrumentation, without prescribing a proof system.
This supports validation across tools and approaches, at the expense of requiring the validator to
reconstruct more of the reasoning. The evolution of the \SVComp witness format
reflects this emphasis on interoperability~\cite{DBLP:conf/sigsoft/0001DDH16,DBLP:conf/spin/AyaziovaBRSS24}.

%% file: 3-tmai.tex
\subsection{Thread-Modular Abstract Interpretation}\label{sec:ai}

We first fix some assumptions and notation for the concrete semantics and the abstract domain:
Assume that a program $P$ is given as a set of labeled control-flow graphs (CFGs), with nodes from some set $\Nodes$ of program
points, which are connected with edges labeled with statements. We assume that there are
statements corresponding to creating new threads, and locking and unlocking of mutexes (which take as an argument
an expression evaluating to a set of addresses of mutexes). Furthermore, there are global variables, which we assume
to be shared between threads, as well as statements to copy values between local variables and globals.
In addition, there are local statements affecting only local variables. We make no further assumptions about
these local statements, but they may, e.g., include assignments to local variables or function calls and returns.

A thread-modular abstraction uses some abstract domain $\Domain$ which we assume to be a complete lattice\footnote{
    We assume complete lattices to simplify the presentation; the requirements are in fact weaker. For instance, domains need
    not provide a \emph{least} upper bound operation. The existence of a \emph{join} operation is sufficient.},
with a monotonic concretization function $\gamma : \Domain \to 2^{\Traces}$, where $\Traces$ is the set of traces of the program.
Let $\sqcup$ denote the least upper bound, and $\sqcap$ the greatest lower bound.
We do \emph{not} require $\Domain$ to be Noetherian, i.e., it may admit infinite ascending chains.

Assume that there is an abstract evaluation function for boolean expressions
$\aSemEB{\cdot} : \Domain \to \{\aTrue, \aFalse, \top\}$.
Here, $\aTrue$ and $\aFalse$ are incomparable and smaller than $\top$.
There also is an evaluation function for pointer-valued expressions to
compute a set of addresses they may point to. We denote the result by a predicate.
For instance,
$\absPointsToIn{ptr}{\sigma^\sharp} \{\&mutex1,\&mutex2\}$ means that the expression
$ptr$, when evaluated in the abstract state $\sigma^\sharp$, must point either to $mutex1$ or to $mutex2$.
Lastly, assume that there is a function $\aSemInt{\cdot}: \Domain \to 2^\mathbb{Z}$ where $\aSemInt{e}\,\sigma^\sharp$
evaluates an expression $e$
to the set of possible integer values it may take in the abstract state $\sigma^\sharp$.
Assume all three functions to be sound w.r.t.\ some concrete counterpart.

We further assume that the abstract interpretation computes abstract thread \emph{id}s from some set $\TIDs$.
These thread \emph{id}s provide a predicate $\unique: \TIDs \to \textbf{bool}$ to determine whether an abstract thread
\emph{id} is \emph{unique}, i.e., whether there is at most one concrete thread at runtime with this abstract thread \emph{id}.
Abstract thread \emph{id}s may, e.g., correspond to task names in embedded settings such as \textsc{OSEK}~\cite{osek2005osek}
or \textsc{ARINC653}~\cite{arinc653} where threads are statically defined (marking all of them unique), or may be derived from the
creation history of threads~\cite{schwarz2023clustered,hammerandnail}
in settings with dynamic thread creation such as \textsc{Pthreads}.
We further assume that there is a function $\projTIDs: \Domain \to 2^{(\TIDs)}$ that maps abstract states to the set of
possible abstract thread IDs of threads the abstract state corresponds to.

Our approach is not only generic in the employed abstract domain, but also in the thread-modular abstract interpretation.
To retain this generality as much as possible, while still providing proofs, we build on the concept of
\emph{mixed flow-sensitivity}~\cite{Stemmler25,DBLP:conf/fm/SeidlVES26} as it can be used as a unifying formalism
for a wide range~\cite{DBLP:journals/jss/StievenartNMR19,Vojdani2010,DBLP:conf/sas/SchwarzSSAEV21,Mine2014,schwarz2023clustered,VojdaniARSVV16}
of approaches to thread-modular abstract interpretation.

In this formalism, the abstract interpretation is formulated as a system of constraints over so-called
\emph{unknowns} (constraint system variables) that range over the abstract domain.
There are two different kinds of unknowns: \emph{local} unknowns from some set $\LocalUnknowns$
and \emph{global} unknowns from some set $\GlobalUnknowns$. The local unknowns are associated with program points
and their abstract values are meant to describe the abstract state of all threads reaching that program point.
In this way, they represent flow-sensitive information.
The global unknowns, on the other hand, accumulate flow-insensitive information about the program, i.e., information
that is not associated with a specific program point but accumulated from information at various (or all) program points.
\begin{example}
    The basic analysis from~\cref{ex:running-disc} which computes an interval abstraction of
    the values that global variables may have at any point in the program (e.g.,~\cite{Vojdani2010,DBLP:journals/jss/StievenartNMR19}),
    would collect, at local unknowns, abstractions of the
    values of \emph{local} variables at that program point and introduce a global unknown for each \emph{global} variable
    to collect an abstraction of all possible values of that global variable.
    For the program in \cref{fig:running}, the analysis introduces a global unknown $[g]$
    for the program global $g$ and computes the interval $[1,\infty]$ for it.
\end{example}
While this is a straightforward use of global unknowns, they need not always correspond to global program variables:
For example, an analysis may introduce a global unknown for each mutex and collect abstractions of the values
global variables had when that mutex was unlocked~\cite{schwarz2023clustered}.

Let $\Lambda = \LocalUnknowns\to\Domain$ be the set of mappings from local unknowns to elements of the abstract domain and
$\Mu = \GlobalUnknowns\to\Domain$ be the set of mappings from global unknowns to such elements.
An edge $(u,\stmt,v)$ in the control-flow graph of a thread then gives rise to a constraint
\[
    (\lambda\,\relax[v], \mu) \sqsupseteq \aSem{\stmt}\,(\lambda\,\relax[u])\,\mu
\]
It describes how the abstract state at program point $v$ is related to the abstract state at program point $u$
and the values accumulated at global unknowns and which contributions to the global unknowns are made
by executing the statement $\stmt$.
$\aSem{\stmt} : \Domain \times \Mu \to (\Domain \times \Mu)$ is called the \emph{transfer function} for the
statement $\stmt$. It returns the contribution to $[v]$ in its first
component, whereas the second component is a map from global unknowns to contributions.
We list only non-$\bot$ bindings for these maps. For instance, $\emptyset$
is shorthand for the map that maps all global unknowns to $\bot$.
In addition to constraints arising from control-flow edges, there is also an initial constraint
\[
    (\lambda\,\relax[u_0], \mu) \sqsupseteq (\sigma_0^\sharp, \mu_0)
\]
providing an initial abstract state $\sigma_0^\sharp$ (typically $\top$)
for the program entry point $u_0$\footnote{For technical reasons, we require that $u_0$ has no incoming edges.}
and an initial contribution $\mu_0$ for the global unknowns
(e.g., ensuring all unknowns corresponding to global variables are initialized with appropriate
abstractions of their initial values).

A tuple $(\lambda,\mu) \in \Lambda \times \Mu$ then is a \emph{solution} of the constraint system if it satisfies all constraints
induced by the control-flow graphs of the program and the initial constraint.

To compute solutions on non-Noetherian domains,
the solver may need to use \emph{widening} because otherwise the iteration may not terminate. We thus
assume that such abstract domains provide a widening~\cite{CousotC77} operator $\nabla : \Domain \to \Domain \to \Domain$
to ensure convergence of fixpoint iteration by deliberately giving up precision. Then, a solver
carefully replaces some least upper bounds $\sqcup$ in the fixpoint iteration with widening $\nabla$
to guarantee termination. Sometimes, some lost precision may be recovered by using a \emph{narrowing}~\cite{CousotC77}
operator $\Delta: \Domain \to \Domain \to \Domain$. 

With this setup, we also use $\absPointsToIn{e}{u}$ as shorthand for evaluating $e$ in the abstract state associated
with program point $u$:
$\absPointsToIn{e}{u} \equiv \absPointsToIn{e}{(\lambda\,\relax[u])}$
i.e., the set of addresses that $e$ may point to at program point $u$ is determined by evaluating
the expression in the abstract state $\lambda\,\relax[u]$ associated with the local unknown at program point $u$ for
a solution $(\lambda,\mu)$ of the constraint system.

\medskip
Throughout the rest of the paper, we assume this base analysis to be sound.

%% file: 4-approach.tex
\subsection{Witness Validation by Abstract Interpretation}\label{sec:validation}
A first, straightforward, approach to validating a correctness witness without ghosts is to run the abstract interpreter
on the program, obtaining a solution $(\lambda,\mu)$ and then checking that
\begin{itemize}
    \item the abstract interpretation can establish that $\varphi$ is satisfied, and
    \item for each program point $u$ the invariant $\GhostInvariant\,u$ from the witness is known to hold:
        \[
            \aSemB{\GhostInvariant\,u}\,(\lambda\,\relax[u]) = \aTrue.
        \]
\end{itemize}
If those conditions hold, the witness is valid. Otherwise, the witness is either invalid or the abstract interpreter
is not precise enough to validate it.

\begin{remark}
An abstract interpreter can never reject witnesses as invalid. 
Abstract interpretation is sound, but incomplete. This carries over
to witness validation: 
There are valid witnesses that cannot be validated.
Interestingly, even if the abstract interpreter can show that the invariant
$\GhostInvariant\,u$ for some program point $u$ does \emph{definitely not} hold
for any state reaching $u$, this witness cannot be rejected.
This is a consequence of over-approximating reachability: $u$ may in fact never be reached, making any invariant
vacuously true. Thus, without additionally \emph{underapproximating} the set of reachable program points,
the validator needs to return \emph{unknown}.
\end{remark}

However, this approach treats the invariants in the witness only as additional proof obligations. This is different
for the approach of~\citet{SaanSESTV24} for the single-threaded setting, where invariants are used to
guide the abstract interpreter towards a fixpoint that validates the witness.
The key idea is to use \emph{unassume} operators to relax abstract states to the information provided by the invariants.
This can speed up fixpoint iteration,
and also prevent widening from losing the information needed to establish the validity of the witness,
acting similarly to widening thresholds~\cite{DBLP:journals/fmsd/CousotCFMMR09,DBLP:journals/fmsd/HalbwachsPR97}.

\begin{example}\label{ex:unassume}
    Consider the single-threaded program on the right. An
    \par\noindent
    \begin{minipage}[t]{0.74\linewidth}
        abstract interpreter that uses the interval domain will start with the abstract state $i \in [0,0]$
        at the loop head and will then widen to $i \in [0,\infty]$. As the guard uses disequality, narrowing
        cannot recover the information that $i$ is bounded from above, and thus the abstract interpreter will not be able to
        establish that the assertion holds.
    \end{minipage}
    \begin{minipage}[t]{0.08\linewidth}
        \begin{minted}[linenos=false]{C}
  int i = 0;
  while(i != 4711) {
    i++;
    assert(i <= 4711);
  }
        \end{minted}
    \end{minipage}~\\

    However, if the witness provides the invariant \code{0 <= i && i <= 4711} at the beginning of the loop,
    and the abstract interpreter uses \emph{unassume} to relax the state to this invariant,
    it finds the precise fixpoint $i \in [0,4711]$ in one iteration
    and establishes the assertion. This illustrates that witnesses can not only speed up the analysis,
    but can also make it more precise.
\end{example}

Technically, the unassume operator is inserted as a statement into the control-flow graph at every program location
with an associated invariant. In the concrete, unassumes are no-ops, whereas in the abstract they relax
the abstract state to the information provided by the invariant.
\begin{definition}\label{def:unassume-sound}
    An unassume operator $\aSem{\unassume{e}}$ on an abstract domain $\Domain$ is sound if,
    for all abstract states $\sigma^\sharp \in \Domain$ and all expressions $e$,
    $\gamma(\sigma^\sharp) \subseteq \gamma(\aSem{\unassume{e}}\,\sigma^\sharp)$, i.e.,
    it only relaxes abstract states, and thus enlarges the concretization.
\end{definition}
\begin{proposition}
    Let $\aSem{\unassume{e}}$ be sound in the sense above. Inserting unassume statements into the CFG
    at locations with witness invariants does not compromise the soundness of the abstract interpretation,
    and hence does not compromise the soundness of witness validation.\qed
\end{proposition}

\section{A Two-Phase Approach to Validating Ghost Witnesses}\label{sec:approach}

For the validation of correctness witnesses with ghost variables, the general setup from \cref{sec:validation} carries over,
except that the program to be considered is the ghost-instrumented program $P_{\ghost}$ instead of the original program $P$.

On top of the challenges outlined above, a new key challenge arises: making use of the information provided
by the ghost instrumentation, rather than treating it only as an additional burden.
Ghosts are not inserted at random, but rather are meant to encode key insights into the reasoning used
by the verifier to establish the correctness of the program. Thus, it is reasonable to expect that they typically
encode information corresponding to some high-level pattern.
In fact, when internal proofs produced by tools are translated into witnesses, the tool designers make some choice
of how to encode the intended high-level pattern using ghosts.
We maintain a set of patterns corresponding to high-level semantic insights ghost variables may encode.
However, a witness does not specify what exactly a given ghost variable is encoding. Even if this information were to be provided,
there still is the possibility that there is a bug in the encoding of the desired high-level pattern into the ghost instrumentation,
meaning it may, in fact, \emph{not} actually encode the intended pattern. Trusting that it does would make the witness
validator unsound.
Thus, we propose a two-phase approach to exploiting the information provided by ghost variables:
\begin{description}
    \item[Check]
        Check that a given ghost variable is used in a way consistent with
        one of the patterns.
    \item[Exploit]
        After having established that the ghost variable encodes semantically meaningful information, exploit this
        information to guide the abstract interpreter. 
        This consists of both shaping the constraint system in a way that uses the information encoded in the ghost variable,
        and using \emph{unassume} during fixpoint computation for this new constraint system.
        We discuss the role of unassume operators in more detail in \cref{sec:unassume}.
\end{description}
This approach of first \emph{checking} whether some hypothesis about
the usage of ghosts is satisfied and then \emph{exploiting} it is in fact not specific to thread-modular abstract interpretation, but
applies to any technique for the validation of correctness witnesses with ghosts.

In this paper, we consider two patterns of the usage of ghost variables:
\begin{enumerate}
    \item[(a)] \textbf{Ghosts for Resource Invariants}:
        Ghost variables that are used to encode resource invariants, e.g., for locks. We discuss this type of ghost variable
        in \cref{sec:mutex-ghosts}.
    \item[(b)] \textbf{Phase-encoding Ghosts}:
        Ghost variables that are used to encode the phase of a program, i.e., whether a certain action has been performed
        by a given thread or not as introduced in \cref{ex:running-disc}.
        We focus on such ghost variables in \cref{sec:phase-ghosts}.
\end{enumerate}
While both patterns can occur in a single witness, and our approach can exploit both types of
ghost variables at the same time, we discuss them separately 
for clarity.
We first discuss ghosts for resource invariants as their validation requires fewer changes to the constraint system,
before turning to phase-encoding ghost variables, which require a more sophisticated approach.

%% file: 5-mutex.tex
\section{Ghosts for Resource Invariants}\label{sec:mutex-ghosts}

\emph{Resource invariants} as proposed by~\citet{DBLP:journals/cacm/OwickiG76} describe a
common pattern in concurrent programs:
Certain properties are guaranteed to hold whenever no thread holds
a given resource, but may be violated when a thread does hold the resource.
An example would be a data structure protected by a mutex: When a thread
gains exclusive access to the data structure by locking the mutex, it may modify the
data structure in a way that some invariant is locally violated. However, when the
thread releases the mutex again, it is required to restore the invariant, so that
other threads once again have a consistent view of the data structure.
Resource invariants are closely related to the concept of \emph{ownership}, 
and correspond to one of the rules in concurrent separation logic~\cite{DBLP:journals/tcs/OHearn07} used for
proving properties of concurrent programs either manually or with proof assistants.

\begin{figure}[b]
\centering
\begin{minipage}[t]{0.45\textwidth}
\begin{minted}[escapeinside=@@,linenos=false]{C}
main:                         @{\tikzmark{ExPhaseAtops}}@
    int g, h = 0;
    while(*) {
      create(t1);
    }
    // $\phi_1$
    @\colorbox{gray!20}{\strut\texttt{atomic \{}}@ lock(&m); @\colorbox{gray!20}{\strut$\ghost_{\textsf{m}}$ \texttt{= 1; \}}}@  @{\tikzmark{ExPhaseAmidz}}@
    assert(g == h);
    g++; h++;
    @\colorbox{gray!20}{\strut\texttt{atomic \{}}@ unlock(&m); @\colorbox{gray!20}{\strut$\ghost_{\textsf{m}}$ \texttt{= 0; \}}}@  @{\tikzmark{ExPhaseAmidz}}@
\end{minted}
\end{minipage}\begin{minipage}[t]{0.45\textwidth}
\begin{minted}[firstnumber=10,escapeinside=@@,linenos=false]{C}
t1:                           @{\tikzmark{ExPhaseXtops}}@
    // There is an unbounded number of t1's
    @\colorbox{gray!20}{\strut\texttt{atomic \{}}@ lock(&m); @\colorbox{gray!20}{\strut$\ghost_{\textsf{m}}$ \texttt{= 1; \}}}@
    int x = 0;
    while(x != 42) x++;
    g = x; h = 42;
    @\colorbox{gray!20}{\strut\texttt{atomic \{}}@ unlock(&m); @\colorbox{gray!20}{\strut$\ghost_{\textsf{m}}$ \texttt{= 0; \}}}@
    // $\phi_2$
    // ...
    return;                   @{\tikzmark{ExPhaseYends}}@
\end{minted}
\end{minipage}
\caption{Program illustrating a mutex-protected resource invariant and corresponding ghost updates.}
\label{fig:mutex-ghost-example}
\end{figure}

\begin{example}
Consider the multi-threaded program in~\cref{fig:mutex-ghost-example}, ignoring the annotations in gray for the moment.
The mutex \code{m} protects the globals \code{g} and \code{h} here: They are only modified when \code{m} is held.
Whenever \code{m} is not held by any thread, \code{g == h} holds, which thus is the \emph{resource invariant} associated
with the mutex \code{m}. This is crucial for establishing the assert in the main thread, as the resource invariant
can be assumed when the mutex \code{m} is locked.
\end{example}

There are various tools that internally compute resource invariants. Early instances include
thread-modular shape analysis~\cite{DBLP:conf/pldi/GotsmanBCS07},
the \emph{privatization}-approach~\cite{VojdaniARSVV16,Vojdani2010,DBLP:conf/sas/SchwarzSSAEV21},
or the work of~\citet{Mine2014}.
Here, the approach implemented in \Goblint~\cite{schwarz2023clustered,schwarzthesis,SaanPhD} is of particular
interest, as the tool already exports resource invariants in the form of witnesses~\cite{ErhardBHKSSSTV2025}.

Unfortunately, the fact that no thread holds a given resource is not expressible as a {C} expression.
To nevertheless express resource invariants in a witness, a ghost variable is introduced for each resource, e.g.,
a mutex, to track whether the resource is currently held by some thread or not.
If $\ghost_m$ is $1$ whenever the mutex is locked, and $0$ whenever it is not held, an invariant of the shape
$\ghost_m = 0 \implies \phi_g$ where $\phi_g$ is a formula involving those global variables of the program that are
only accessed when $m$ is held~\cite{ErhardBHKSSSTV2025} holds throughout the program.
\begin{definition}[Mutex Ghost]
    A ghost variable $\ghost_0$ is called a \emph{mutex ghost} for some mutex $m$ if it has value $0$ whenever $m$
    is not held by any thread, and value $1$ whenever $m$ is held by some thread.
\end{definition}
\begin{example}
Consider the program in~\cref{fig:mutex-ghost-example}. With the ghost code in gray, $\ghost_m$
is a mutex ghost for \code{m}, and the invariant \texttt{$\ghost_m$ == 1 || g == h} is
valid at any program point, e.g., at $\phi_1$ and $\phi_2$.
\end{example}

\subsection{Checking Whether a Ghost Variable Is a Mutex Ghost}
To establish whether a given ghost variable $\ghost_0$ from some ghost witness $\Witness$
is a mutex ghost for some mutex $m$, we build upon the results of a pointer analysis, which, we recall, provides
a predicate $\absPointsToIn{}{u}$, where $\absPointsToIn{x}{u}\,Y$ indicates that the expression $x$
must evaluate to one of the addresses in $Y$ whenever the program point $u$ is reached.
We then check the following conditions:
\begin{enumerate}
    \item The initial value for $\ghost_0$ (as recorded in $\GhostGlobalDeclarations\,(\ghost_0)$) is $0$.
    \item In $\GhostUpdate$, wherever $\ghost_0$ is updated, it is set either to $0$ or to $1$.
    \item Whenever $\ghost_0$ is set to $1$ in $\GhostUpdate$ along some edge originating at program point $u$,
        the edge is labeled with $\lock{e}$ for some expression $e$ and the
        abstract interpretation can establish that whenever the corresponding program point is reached
        $\absPointsToIn{e}{u}\, \{ \&m \}$ holds, i.e., the expression $e$ must evaluate to the address of $m$.
    \item Conversely, whenever $\ghost_0$ is set to $0$ in $\GhostUpdate$ along some edge originating at program point $u$,
        the edge is labeled with $\unlock{e}$ for some expression $e$ and the
        abstract interpretation can establish that whenever the corresponding program point is reached
        $\absPointsToIn{e}{u}\, \{ \&m \}$ holds, i.e., the expression $e$  must evaluate to the address of $m$.
    \item For any lock or unlock operation of the form $\lock{e}$ or $\unlock{e}$ in the program that is not associated
        with an update of $\ghost_0$ in $\GhostUpdate$, originating at some program point $u$, we have
        $\absPointsToIn{e}{u}\,X$ with
        $\&m \notin X$, i.e., it can be established that such operations do not operate on the mutex $m$.
\end{enumerate}
\begin{proposition}
    Provided the predicate $\mapsto^\sharp$ soundly overapproximates points-to sets, a ghost variable $\ghost_0$
    fulfilling the above conditions for some mutex $m$ is a mutex ghost for $m$.
\end{proposition}
\begin{proof}
    The proof is by induction on traces and relies on the soundness of the helper analyses exposing information via the
    $\mapsto^\sharp$ predicate. We omit the details here.
\end{proof}

\begin{remark}
    Our implementation infers for each ghost variable for which mutex (if any) it is a mutex ghost.
    It also handles mutexes inside structures and other complications omitted for brevity.
    Moreover, it is permissive and checks conditions (3) and (4) only for potentially reachable program points.
\end{remark}

\subsection{Exploiting Mutex Ghosts}
Invariants using a mutex ghost have the following shape: $\ghost_m = 0 \implies \phi$ where $\phi$
is a formula involving global and local variables. When given as a {C} expression, this corresponds to
expressions of the shape \texttt{$\ghost_m$ == 1 || e$_\phi$} where \texttt{e$_\phi$} is a {C} expression corresponding to $\phi$.

W.l.o.g., we assume that the invariant associated with some program point $u$ is of this shape, with the ghost variable
appearing in the outermost disjunction.\footnote{
    If this is not the case, expressions can be split up. 
    Such piecewise checking of invariants is sound, but may be less precise than a joint evaluation.
    In our implementation, we did not encounter any issues related to this further abstraction.
}

The validation of an invariant $\GhostInvariant\,u \equiv \texttt{$\ghost_m$ == 1 || e$_\phi$}$
at some program point $u$ where $\ghost_m$ is a mutex ghost for some mutex $m$ then relies on the observation that either:
\begin{itemize}
    \item Some thread holds $m$, in which case $\ghost_m$ evaluates to $1$ and the invariant is trivially true; or
    \item No thread holds $m$, in which case $\ghost_m$ evaluates to $0$ and the invariant reduces to $\texttt{e$_\phi$}$.
\end{itemize}
Let $(\lambda,\mu)$ be a solution of the constraint system arising from the ghost-instrumented program $P_{\ghost}$ and
let $\sigma^\sharp = \lambda\,[u]$ be the abstract state for program point $u$. 
Then, we can check whether the invariant $\GhostInvariant\,u$ is known to hold at program point $u$ by checking whether
\begin{equation}\label{eq:mutex-ghost-check}
    \aSemEB{\ghost_m == 1} \sigma^\sharp = \aTrue \;\lor\;
    \aSemEB{\texttt{e$_\phi$}} \left( \aSem{\lock{&m}} \sigma^\sharp\,\mu \right)_1 = \aTrue
\end{equation}
holds, where $(\cdot)_1$ denotes accessing the first component of a tuple.
The idea is that if no thread holds the mutex $m$, the thread currently at program point $u$ can acquire
$m$.
It may be surprising that acquiring $m$ in the case where no other thread may hold $m$ can make a difference to whether the invariant can
be established or not. However, recall that unlike in the concrete, where acquiring a lock does not change the values of any variable,
this is not true in the abstract: Some analysis may, e.g., upon locking a mutex, pull the resource invariant into the local state of the
thread, as is the case for some thread-modular analyses in \Goblint~\cite{schwarz2023clustered,schwarzthesis,SaanPhD}.

More formally, for this witness-validation technique, we have:
\begin{theorem}\label{thm:mutex-ghost-check}
    Consider some program point $u$ in the ghost-instrumented program $P_{\ghost}$,
    and a solution $(\lambda,\mu)$ of the constraint system.
    Consider further an invariant $\GhostInvariant\,u \equiv \texttt{($\ghost_m$ == 1 || e$_\phi$)}$ associated with
    program point $u$, where $\ghost_m$ is established to be a mutex ghost for some mutex $m$.
    If the formula \cref{eq:mutex-ghost-check} evaluates to true
    for the abstract state $\sigma^\sharp = \lambda\,[u]$ at program point $u$,
    then the invariant $\GhostInvariant\,u$ is known to hold at program point $u$ in the concrete semantics of $P_{\ghost}$.
\end{theorem}
\begin{proof}
The proof proceeds by case distinction on the result of evaluating $\aSemEB{\ghost_m == 1} \sigma^\sharp$:
\begin{itemize}
    \item If $\aSemEB{\ghost_m == 1} \sigma^\sharp = \aTrue$,
        by the soundness of the abstract interpretation and since $(\lambda,\mu)$ is a solution,
        \texttt{($\ghost_m$ == 1)} holds whenever $u$ is reached. Therefore, so does the invariant
        $\texttt{($\ghost_m$ == 1 || e$_\phi$)}$.
    \item Consider the case where $\aSemEB{\ghost_m == 1} \sigma^\sharp \neq \aTrue$ and some
        $\sigma\in \gamma\,\sigma^\sharp$. In $\sigma$, \texttt{$\ghost_m$ == 1} is either true or false.
        \begin{itemize}
            \item
                If $\ghost_m == 1$, \texttt{($\ghost_m$ == 1 || e$_\phi$)} is trivially satisfied for $\sigma$.
            \item
                As $\ghost_m$ is a mutex ghost for $m$, it remains to consider the case where $\ghost_m == 0$ in $\sigma$.
                As $\ghost_m$ is a mutex ghost for $m$, no thread holds the mutex $m$ in $\sigma$.
                Thus, in the concrete, executing the operation $\lock{&m}$ does not block.
                We observe that if $\texttt{e$_\phi$}$ is satisfied in $\sigma$, it is also satisfied in $\sigma'$
                obtained after locking $m$ as this does not block, and the values of variables do not change when acquiring a lock.
                By soundness of the abstract interpretation, and since $(\lambda,\mu)$ is a solution, we have
                \[
                    \sigma\in \gamma\,\sigma^\sharp \implies \sigma' \in \gamma\,(\aSem{\lock{&m}} \sigma^\sharp\,\mu)_1
                \]
                Thus, if $\aSemEB{\texttt{e$_\phi$}} \left( \aSem{\lock{&m}} \sigma^\sharp\,\mu \right)_1 = \aTrue$,
                \texttt{e$_\phi$} holds in $\sigma$. Therefore, so does the invariant \texttt{($\ghost_m$ == 1 || e$_\phi$)}.
        \end{itemize}
        Thus, if $\aSemEB{\texttt{e$_\phi$}} \left( \aSem{\lock{&m}} \sigma^\sharp\,\mu \right)_1 = \aTrue$,
        the invariant \texttt{($\ghost_m$ == 1 || e$_\phi$)} holds in all $\sigma \in \gamma \sigma^\sharp$, i.e.,
        $\GhostInvariant\,u$ holds whenever program execution reaches program point $u$ in $P_{\ghost}$.
        \qedhere
\end{itemize}
\end{proof}

\paragraph{Analysis Cost}
This analysis exploiting the information provided by mutex ghosts does not incur much overhead over attempting to validate
the correctness witness without exploiting mutex ghosts. The analysis checking whether a ghost variable is a
mutex ghost uses a rather simple abstract domain, and the additional overhead when
validating invariants containing mutex ghosts corresponds to executing the transfer function for locking
a mutex once per invariant.

%% file: 6-phases.tex
\section{Ghosts Encoding Phases of the Program}\label{sec:phase-ghosts}
Ghost variables in thread-modular proofs often encode phases of a program, i.e., whether a certain action
has been performed by a given thread or not, or specific regions of code that a given thread may be in.
Generalizing this, phases can alternatively be understood as the states of a \emph{per-thread} finite-state automaton.
%
%
%
Formally, we define a \emph{phase ghost} as follows:
\begin{definition}
    A phase-encoding ghost variable (\emph{phase ghost}) is a ghost variable of integer type
    that is \emph{updated} by exactly one runtime thread, while other threads can only read it.
    Furthermore, its value is bounded from below and above by constants.
\end{definition}
%
%
%
\begin{example}\label{ex:phase-encoding-ghosts}
    Consider again \cref{fig:running} and the discussion in \cref{ex:running-disc}.
    Both ghost variables are phase ghosts, as
    their values are bounded, and each is updated by a single thread.
\end{example}
For validation, we again take a two-phase approach, first checking
whether a ghost variable is a phase ghost and then exploiting this.
Before detailing this, we illustrate how, by exploiting phase ghosts, an abstract interpreter
can confirm invariants previously out of reach. 

\subsection{Exploiting Phase Ghosts in the Running Example}
Consider again the program in \cref{fig:running} and the discussion in \cref{ex:running-disc} and assume
that the ghost witness provides the following invariants\footnote{We use mathematical notation instead of giving {C} expressions here
as it is more concise.}
for the respective program points:
\[
\begin{array}{lll}
    \phi_A & \equiv & (\ghost_{\textsf{main}} = 0) \land ((\ghost_{\textsf{t1}} = 0 \land g \geq 0) \lor (\ghost_{\textsf{t1}} = 1 \land g \geq 1))\\
    \phi_B & \equiv & (\ghost_{\textsf{main}} = 1) \land ((\ghost_{\textsf{t1}} = 0 \land g \geq 1) \lor (\ghost_{\textsf{t1}} = 1 \land g \geq 2))\\[1ex]
    \phi_X & \equiv & (\ghost_{\textsf{t1}} = 0) \land ((\ghost_{\textsf{main}} = 0 \land g \geq 0) \lor (\ghost_{\textsf{main}} = 1 \land g \geq 1))\\
    \phi_Y & \equiv & (\ghost_{\textsf{t1}} = 1) \land ((\ghost_{\textsf{main}} = 0 \land g \geq 1) \lor (\ghost_{\textsf{main}} = 1 \land g \geq 2))
\end{array}
\]
Instead of an analysis flow-insensitively accumulating all possible values of a given global variable $g$ at some
unknown $[g]$~\cite{Vojdani2010,DBLP:journals/jss/StievenartNMR19}, we consider a slightly improved analysis that
tracks values flow-sensitively while in an atomic block, and only contributes the values of globals to
respective unknowns at the end of such atomic blocks.

When applying this analysis to the program from \cref{fig:running} using the interval
domain without exploiting that the ghost variables are phase ghosts, the analysis will (still) compute the interval $[1,\infty]$
for the global variable $g$, meaning that the assertion at the end of main cannot be shown to succeed
and none of the invariants $\phi_A$ through $\phi_Y$ provided in a ghost witness can be validated.

Consider now an improved analysis that is aware of the phases the different threads are in, and separately accumulates
the values of the global variable $g$ that can be read depending on the phases each of the threads are in,
changing the domain associated with $[g]$ to maps from pairs of phases to intervals (see right half of \cref{fig:phase-worked-example}).
Similarly, for local states, the domain is a map from pairs of phases to abstract states (left half of \cref{fig:phase-worked-example}).
The dotted arrows in the figure indicate which values are contributed to global unknowns at the end of atomic blocks.

\begin{figure}[t]
\centering
\newcommand{\phaseA}{\text{\Circled{\textnormal{A}}}}
\newcommand{\phaseB}{\text{\Circled{\textnormal{B}}}}
\newcommand{\phaseX}{\text{\Circled{\textnormal{X}}}}
\newcommand{\phaseY}{\text{\Circled{\textnormal{Y}}}}
\newcommand{\pAX}{(\phaseA,\phaseX)}
\newcommand{\pAY}{(\phaseA,\phaseY)}
\newcommand{\pBX}{(\phaseB,\phaseX)}
\newcommand{\pBY}{(\phaseB,\phaseY)}
\newcommand{\absTmp}[1]{\{\,\mathtt{tmp} \mapsto #1\,, \dots\}}
\newcommand{\absG}[1]{\{\,\mathtt{g} \mapsto #1, \dots\}}
\newcommand{\stateann}[1]{{\scriptstyle #1}}
\begin{minipage}[t]{0.4\textwidth}
\begin{minted}[escapeinside=@@]{C}
main:
  assume(g > 0);
  // $\stateann{\pAX \mapsto \top}$
  create(t1);
  // $\stateann{\pAX \mapsto \top \qquad \pAY \mapsto \top}$
  // $\textcolor{red}{\phi_A}$
  atomic {
    // $\stateann{\pAX \mapsto \top \qquad \pAY \mapsto \top}$
    tmp = g;
    // $\stateann{\pAX \mapsto \absTmp{[1,\infty]}}$
    // $\stateann{\pAY \mapsto \absTmp{[2,\infty]}}$
    tmp++;
    // $\stateann{\pAX \mapsto \absTmp{[2,\infty]}}$
    // $\stateann{\pAY \mapsto \absTmp{[3,\infty]}}$
    g = tmp;
    @$\ghost_{\textsf{main}}$@ = 1;
    // $\stateann{\pAX \mapsto \absG{[2,\infty]}}$ \tikzmark{Apublish1}
    // $\stateann{\pAY \mapsto \absG{[3,\infty]}}$ \tikzmark{Apublish2}
  }
  // $\stateann{\pBX \mapsto \top \qquad \pBY \mapsto \top}$
  // $\textcolor{red}{\phi_B}$
  join(t1);
  // $\stateann{\pBY \mapsto \top}$
  assert(g > 2);
\end{minted}
    \end{minipage}\begin{minipage}[t]{0.3\textwidth}
\begin{minted}[firstnumber=25,escapeinside=@@]{C}
t1:
  // $\stateann{\pAX \mapsto \top \qquad \pBX \mapsto \top}$
  a = 5;
  // $\stateann{\pAX \mapsto \top \qquad \pBX \mapsto \top}$
  // $\textcolor{red}{\phi_X}$
  atomic {
    // $\stateann{\pAX \mapsto \top \qquad \pBX \mapsto \top}$
    tmp = g;
    // $\stateann{\pAX \mapsto \absTmp{[1,\infty]}}$
    // $\stateann{\pBX \mapsto \absTmp{[2,\infty]}}$
    tmp++;
    // $\stateann{\pAY \mapsto \absTmp{[2,\infty]}}$
    // $\stateann{\pBY \mapsto \absTmp{[3,\infty]}}$
    g = tmp;
    @$\ghost_{\textsf{t1}}$@ = 1;
    // $\stateann{\pAY \mapsto \absG{[2,\infty]}}$\tikzmark{Ypublish1}
    // $\stateann{\pBY \mapsto \absG{[3,\infty]}}$\tikzmark{Ypublish2}
  }
  // $\stateann{\pAY \mapsto \top \qquad \pBY \mapsto \top}$
  // $\textcolor{red}{\phi_Y}$
  return;
\end{minted}
\end{minipage}\begin{minipage}[t]{0.2\textwidth}
    {\footnotesize
    \[
    \begin{array}{l}
        \relax[g]:
        \left\{
        \begin{array}{rcl}
            \pAX \;\mapsto\; [1,\infty], \\
            \tikzmark{AY}\pAY \;\mapsto\; [2,\infty], \\
            \tikzmark{BX}\pBX \;\mapsto\; [2,\infty], \\
            \tikzmark{BY}\pBY \;\mapsto\; [3,\infty]\phantom{,}
        \end{array}
        \right\}
    \end{array}
    \]
    }
\end{minipage}
\tikz[remember picture,overlay]{
\draw[->,dotted] (pic cs:Apublish1) -- (pic cs:BX);
\draw[->,dotted] (pic cs:Apublish2) -- (pic cs:BY);
\draw[->,dotted] (pic cs:Ypublish1) -- (pic cs:AY);
\draw[->,dotted] (pic cs:Ypublish2) -- (pic cs:BY);
}
\caption{Phase-sensitive abstract states and corresponding published values for the running example.
Here, the tuples of circled letters represent the corresponding phase-maps mapping threads to the
respective values of the phase ghosts. To make the example easier to follow, assignments to globals
are pulled apart here.}
\label{fig:phase-worked-example}
\end{figure}

To check whether the assert holds, the abstract interpreter accesses the value
of $g$ for the current phases, namely $(\Circled{\mathrm{B}},\Circled{\mathrm{Y}})$.
Thus, the value $[3,\infty]$ is read, and the assertion can be established.

Consider the validation of the invariant at
$\phi_Y \equiv (\ghost_{\textsf{t1}} = 1) \land ((\ghost_{\textsf{main}} = 0 \land g \geq 1) \lor (\ghost_{\textsf{main}} = 1 \land g \geq 2))$.
For this program point, we have two different reaching local states, one associated with the phases $(\Circled{\mathrm{A}},\Circled{\mathrm{Y}})$
and one associated with the phases $(\Circled{\mathrm{B}},\Circled{\mathrm{Y}})$.
\begin{itemize}
    \item For $(\Circled{\mathrm{A}},\Circled{\mathrm{Y}})$, we have $\ghost_{\textsf{main}} = 0$ and $\ghost_{\textsf{t1}} = 1$.
    The first conjunct thus holds and it remains to consider the disjunction. As $\ghost_{\textsf{main}} = 0$, the assertion holds for
    this state, iff $g \geq 1$ holds. The entry for $(\Circled{\mathrm{A}},\Circled{\mathrm{Y}})$ at $[g]$
    is $[2,\infty]$, and the invariant can be validated.
    \item For $(\Circled{\mathrm{B}},\Circled{\mathrm{Y}})$, we have $\ghost_{\textsf{main}} = 1$ and $\ghost_{\textsf{t1}} = 1$. 
    Once more the first conjunct holds, and it remains to establish which of the disjuncts holds.
    As only the entry for $(\Circled{\mathrm{B}},\Circled{\mathrm{Y}})$ needs to be read, the second clause is known to hold
    and the invariant can be validated.
\end{itemize}

\subsection{Checking Whether a Ghost Variable Is a Phase Ghost}
To establish whether an integer-typed ghost variable $\ghost_0$ from a ghost witness $\Witness$
is a phase ghost for some thread with (unique) abstract thread \emph{id} $i_0$, we
perform an abstract interpretation of $P_{\ghost}$
and check the following conditions for a solution $(\lambda,\mu)$ of the constraint system:
\begin{enumerate}
    \item We collect the set of abstract thread \emph{id}s of threads writing to $\ghost_0$ and check that it consists
        of a single unique abstract thread \emph{id} $i_0$, i.e.,
        \[
            \exists i^\sharp_{\ghost_0}:
                \unique\,i^\sharp_{\ghost_0} \land
                \mbox{$\bigcup_{(u,\stmt,v) \text{ edge in CFG, } \GhostUpdate\,(u,\stmt,v) \text{ updates } \ghost_0}$} (\projTIDs\,(\lambda\,[u])) = \{i^\sharp_{\ghost_0}\}.
        \]
        Then, the thread with the abstract thread \emph{id} $i^\sharp_{\ghost_0}$ is said to \emph{own} the ghost $\ghost_0$.
    \item We check that the initial value and all updates to $\ghost_0$ are bounded by constants $L$ and $U$.
        That is, the initial value lies in $[L,U]$, and for all $u \in \Nodes$ such that along an edge
        $(u,\stmt,v)$ in the CFG,
        a ghost update sets $\ghost_0$ to some expression $e$ after executing the statement $\stmt$, we have
        $\aSemEB{L <= e\,\&\&\,e <= U}\left( \aSem{\stmt} (\lambda\,[u])\,\mu \right)_1 = \aTrue$.
\end{enumerate}
\begin{proposition}
    Provided the base analysis is sound, a ghost variable $\ghost_0$ satisfying the above conditions is a phase ghost
     for the thread with the abstract thread \emph{id} $i^\sharp_{\ghost_0}$.
\end{proposition}
\begin{proof}
    By construction of $P_{\ghost}$ (e.g., no assignment to ghosts through pointers) and soundness of the
    base analysis, the claim follows.
\end{proof}
\begin{remark}
    The definition allows flexibility in how verifiers encode phase information, as assignments need not be constant,
    as long as the abstract interpreter can infer constant lower and upper bounds on their values.
    This flexibility is, e.g., crucial when phase ghosts encode whether the final iteration of a
    loop has happened or not.
    Phase changes may decrease the value or revisit a previous value. Thus, phase ghosts can encode
    arbitrary per-thread finite-state automata, including cyclic behavior.
\end{remark}

\subsection{Exploiting Phase Ghosts by Lifting the Analysis}\label{ss:exploiting-phase-ghosts}

The key idea in exploiting phase ghosts is tracking the possible phases that
threads that own phase ghosts can be in at each program point, and
maintaining disjunctions of abstract states.
In this way, the analysis is not only flow-sensitive w.r.t. the program point of
the current thread, but also w.r.t. the \emph{phases} of other threads,
as encoded in the values of the respective phase ghosts.
For global unknowns, this makes the analysis flow-sensitive in the phases of all threads that own
phase ghosts.

We first make a few useful definitions. Let $\phaseGhostTIDs$ denote the set of abstract thread \emph{id}s which own a
phase ghost for a given program $P$ and witness $\Witness$. W.l.o.g., we assume that each thread owns at most one
phase ghost.\footnote{Where a thread owns multiple phase ghosts, they can be combined into one,
by encoding the finite Cartesian product of their value sets as integers.}
We also assume that while checking whether a given ghost variable is a phase ghost,
some auxiliary information about the possible phase changes it performs is computed:
For each unique abstract thread \emph{id} $i$, a new dedicated unknown $[\Phases^\rightarrow_i]$ meant to record information
about what phase changes can possibly happen for the phase ghost it owns is introduced.
This unknown then ranges over $2^{\Phases_i \times \Phases_i}$, where a tuple
$(a,b)$ (alternatively denoted by $a \to_i b$) denotes that the thread can change its phase from $a$ to $b$.
Here, $\Phases_{i} \subset \mathbb{Z}$ denotes the (finite) set of possible values
of the phase ghost owned by the thread with the abstract thread \emph{id} $i$.
In our implementation, which intertwines both phases, this auxiliary information is computed on-the-fly.

\begin{definition}[Phase Map]
    A \emph{phase ghost mapping} (\emph{phase map} for short) $\bar P \in \PhaseMaps = \phaseGhostTIDs \to_{i} \Phases_i$
    then maps each thread
    \emph{id} $i$ that owns a phase ghost to one of the phases $\Phases_i$ associated with that thread.
\end{definition}
Thus, a phase map encodes, for each thread in a program that owns a phase ghost,
in which phase that thread currently is.
The initial phase map $\bar P_0$ maps each thread \emph{id} that owns a phase ghost
to its corresponding initial value $v_i^0$:
$\bar P_0 = \{ i \mapsto v_i^0 \mid i \in \phaseGhostTIDs \}$.
The \emph{phase-lifted domain} then maps phase maps to abstract states from the base domain $\Domain$.
\begin{definition}[Phase-Lifted Domain]
    Given some base domain $\Domain$, its phase-lifted domain is given by $\Domain_{\PhaseMaps}: \PhaseMaps \to \Domain$ and maps each phase map to
    an abstract state from the base domain. All operations of the base domain are lifted to the phase-lifted domain pointwise.
\end{definition}
\begin{example}
    Consider again the example program in \cref{fig:phase-worked-example},
    assuming the main thread received the abstract thread \emph{id} $i^\sharp_{\textsf{main}}$,
    and the thread $t_1$ received the abstract thread \emph{id} $i^\sharp_{t_1}$.
    The state from the phase-lifted domain associated with line 2, when written out, corresponds to
    $\{ (\{ i^\sharp_{\textsf{main}} \mapsto 0, i^\sharp_{t_1} \mapsto 0  \}) \mapsto \top \}$
    where we (as we will also do later) omit all bindings that map to $\bot$.
\end{example}

$\DomainPhaseMap$ tracks disjunctions of abstract values for each possible phase map.
To obtain a lifted analysis exploiting phase information, we lift the transfer functions
of the base analysis: 
\begin{itemize}[leftmargin=2em]
    \item After every step, if any other thread is active, that thread can transition to any reachable phase.
        By contrast, the current thread only changes its phase when its phase ghost is assigned to.
    \item Upon hitting an update of a phase ghost, the thread publishes its current view of the world
        associated with the new phase map to the global unknowns.
    \item Because of this publishing, values at globals associated with phase maps
        that disagree with the current phase map never need to be read, as at phase change
        all previous values that are still relevant are published to the new phase map.
\end{itemize}

The resulting abstract interpretation is path-sensitive in the phases and can be precise enough to
validate a witness for a property it cannot establish without guidance from the witness.
\medskip

To lift the transfer functions, we first
define a helper function $\psuccs$. For a given phase map $\bar P$ and abstract state $\sigma^\sharp$,
$\psuccs\,\bar P\,\sigma^\sharp$ returns the set of possible phase maps that can be reached from $\bar P$
by other threads changing their phases, given the abstract state $\sigma^\sharp$ at the current program
point:
%
\begin{equation}\label{eq:succs}
    \begin{array}{lll}
        \psuccs\,\bar P\,\sigma^\sharp = \{
            \bar P' &\mid& \bar P'\in \PhaseMaps \land
            (\exists i \in \phaseGhostTIDs: \projTIDs\,\sigma^\sharp = \{ i \}  \implies \bar P'\,i = \bar P\,i) \\
            && \land (\forall j \in \phaseGhostTIDs: \bar P'\,j \in \psuccs_j\,(\bar P\,j)\,\sigma^\sharp)
         \}
    \end{array}
\end{equation}
where the first condition fixes the phase of the current thread and for other threads,
we have
\begin{equation}\label{eq:psuccs}
    \begin{array}{lll}
        \psuccs_j\,a_0\,\sigma^\sharp = \{a_0\} \cup \{ a_n \mid
            a_0 \to_{j,\sigma^\sharp} \dots \to_{j,\sigma^\sharp} a_n
        \}
    \end{array}
\end{equation}
where $a_0 \to_{j,\sigma^\sharp} a_1$ denotes that the thread with the abstract thread \emph{id} $j$ can change its phase
from $a_0$ to $a_1$
given the abstract value $\sigma^\sharp$ at the current program point. In a first approximation, we can define
$a \to_{j,\sigma^\sharp} a'$ to hold if $a \to_{j} a'$ is in the set of possible phase changes for thread $j$
accumulated at the $[\Phases^\rightarrow_j]$. This corresponds to assuming that all the other threads may advance
their phase at any time.
The transition relation need not respect the integer order and may contain cycles.
We will replace this with a more precise definition in \cref{ss:mhp}, where we also take into account the \emph{concurrency contexts}
of the respective program points.

\paragraph{Precision with Cyclic Phases.}
The phase-lifted domain keeps different phase maps separate,
and the base domain can preserve properties common to the states associated with each phase map.
Thus, recurring protocol states can remain distinguishable even when their data changes between iterations.
However, the successor closure may allow another thread to reach every phase in a cycle,
weakening the information conveyed by its previously observed phase.
Concurrency contexts (\cref{ss:mhp}) can restrict these transitions,
for example by ruling out phase changes while a shared mutex is held.
\cref{sec:cyclic-phase-example} illustrates this with a string-buffer protocol.

\medskip
\input{6-phases-rhs.tex}

The analysis is given as a generic lifting of the base analysis to the phase-lifted domain.
For the initial program point, we replace the constraint
$(\lambda\,[u_0], \mu) \sqsupseteq (\sigma^\sharp_0, \mu_0)$ by the lifted constraint
$(\lambda\,[u_0], \mu) \sqsupseteq \lift_{\bar P_0} (\sigma^\sharp_0, \mu_0)$
wrapping the right-hand side in a helper function $\lift_{\bar P_0}$ (see \cref{fig:lift-unlift})
to associate all values with the initial phase map $\bar P_0$.
For constraints $(\lambda\,\relax[v], \mu) \sqsupseteq \aSem{\stmt}(\lambda\,\relax[u], \mu)$ arising from edges $(u,\stmt,v)$,
we lift the
original transfer function
$\aSem{\stmt}: \Domain \to (\GlobalUnknowns\to\Domain) \to \Domain \times (\GlobalUnknowns\to\Domain)$
to a new transfer function
$\aSemP{\stmt}: \DomainPhaseMap \to (\GlobalUnknowns\to\DomainPhaseMap) \to \DomainPhaseMap \times (\GlobalUnknowns\to\DomainPhaseMap)$
obtaining new constraints
$(\lambda\,\relax[v], \mu) \sqsupseteq \aSemP{\stmt}\,(\lambda\,\relax[u])\, \mu$.

\medskip\noindent
The definition of $\aSemP{\stmt}$ depends on the type of statement $\stmt$:
\begingroup
\setlength{\leftmargini}{1.5em}
\begin{itemize}
    \item \textbf{For statements other than assigning to a phase ghost}, we lift the
    original transfer function by applying a function $\succs$ to all bindings (see \cref{fig:lift-ordinary}).
    Given a phase map $\bar P$, a corresponding abstract state $\sigma^\sharp$,
    and a valuation $\mu$ of the values of global unknowns associated with the current phase map $\bar P$
    by the function $\unlift$\footnote{
        This function is parametric in a predicate $\phaseBound$ supplied by the analysis to determine whether,
        for the unknown given as an argument,
        it is sound to access only the binding associated with the current phase map.
        For an example where this may not be the case, consider an unknown associated with returning from a thread.
        A call to \texttt{thread\_join} may need to access values associated with an earlier phase, as the return
        from the thread and the join may happen in different phases.
    },
    $\succs$ computes the map from possible successor phase maps to the new abstract states as well as the new contributions to global
    unknowns.

    In the call to $\psuccs$, it is important to use the respective successor states; while for the first definition
    given in \cref{eq:psuccs}, the current abstract state $\sigma^\sharp$ is only used to obtain thread \emph{id}s, later
    this argument is used to determine whether a phase change is possible given the concurrency context, which can
    change after a statement is executed, e.g., when a mutex is released.

    In our setting (see \cref{sec:ai}), a statement refers to at most one global variable:
    Thus, the case of the phase changing in the middle of a statement need not be considered.


\item \textbf{For a statement assigning to a phase ghost} that changes the current thread's phase from $a$ to $b$,
    the analysis needs to ensure that the values of all global variables are published to the unknowns associated
    with the new phase map.
    We first provide a generic lifting before remarking that analyses may provide bespoke definitions
    of this transfer function to increase precision.
    W.l.o.g., let the non-ghost global variables be given by $\Globals = \{ g_0, \dots, g_{n-1} \}$,
    and let $\bar{g}_0, \dots, \bar{g}_{n-1}$ be fresh local variables.
    Then, the corresponding lifted transfer function is given in \cref{fig:lift-assign-phase-ghost}.
    Reading in all global variables and then writing these values again ensures that it suffices
    for other transfer functions
    to only read the values associated with the current phase map.
\end{itemize}
\endgroup
\input{6-phases-ghost-assign.tex}
\noindent We want to prove this lifted analysis sound. To this end, we first transform the witness-instrumented program
$P_{\ghost}$
to obtain a modified program $P'_{\ghost}$ where, for each update of a phase ghost variable, we add statements reading
all global variables before the update and writing them back after.
\noindent\begin{minipage}[t]{0.25\linewidth}
\vspace{-2pt}
\begin{minted}[
  linenos=false,
  fontsize=\scriptsize,
  highlightlines={3,5},
  highlightcolor=gray!20
]{C}
atomic {
  stmt;
  @$\bar{g}_0$@ = @$g_0$@; ... @$\bar{g}_{n-1}$@ = @$g_{n-1}$@;
  @$\ghost_{\textsf{main}}$@ = exp;
  @$g_0$@ = @$\bar{g}_0$@; ... @$g_{n-1}$@ = @$\bar{g}_{n-1}$@;
}
\end{minted}
\end{minipage}\hfill
\begin{minipage}[t]{0.74\linewidth}
\vspace{0pt}
The transformed fragment is shown on the left, with the added statements highlighted in gray.
Here, $\bar{g}_0, \dots, \bar{g}_{n-1}$ are fresh local variables introduced to temporarily store
the values of
the global variables.
In contrast to $P_{\ghost}$, in the modified program $P'_{\ghost}$ all reads are now phase-local in the sense
\end{minipage}
that the value read for any global variable has been written within the same phase.
We remark that performing this program transformation will yield a constraint system that corresponds to the compositions
of transfer functions in the lifted analysis.
We first need to establish that the modified program is equivalent to the original program w.r.t.\ the invariants
we are interested in.
\begin{proposition}\label{prop:ghost-ghostp}
    Any invariant from a witness expressed as a C expression holds at its program point in the original
    witness-instrumented program $P_{\ghost}$ if and only if it
    holds at the corresponding program point in the modified program $P'_{\ghost}$.
    Furthermore, $P_{\ghost}$ satisfies $\varphi$ if and only if $P'_{\ghost}$ satisfies $\varphi$.
\end{proposition}
\begin{proof}
    As the variables $\bar{g}_0, \dots, \bar{g}_{n-1}$ are \emph{fresh}, they cannot appear in any invariant of $P_{\ghost}$.
    Reading global variables into locals and writing them back \emph{in an atomic block} does not change
    the values of global variables visible at any program point.
    The second statement is by robustness of $\varphi$.
\end{proof}
Thus, to validate the correctness of witnesses, one can consider $P'_{\ghost}$ instead of the
original witness-instrumented program $P_{\ghost}$.
Recall that in $P'_{\ghost}$, all reads are phase-local. With this insight, one can give a refinement of the
base analysis with a digest~\cite{schwarz2023clustered,schwarzthesis,hammerandnail} that tracks the
(definite) thread \emph{id} of the executing thread (if there is one) and the current phase map and restricts
all incorporation of non-local information to those values originating from the same phase.
\begin{theorem}\label{thm:phase-lifted-sound}
    Consider a sound base analysis with monotonic transfer functions that admits refinement by digests.
    Assume sound collected phase-transition relations and that the least solution
    $(\lambda_{\PhaseMaps},\mu_{\PhaseMaps})$ of the lifted constraint system satisfies
    the following \emph{entry-closure condition}:
    \begin{equation}\label{eq:entry-closure}
        \bar Y\in\psuccs\,\bar X\,((\lambda_{\PhaseMaps}\,[u])\,\bar X)
        \implies
        (\lambda_{\PhaseMaps}\,[u])\,\bar X
        \sqsubseteq (\lambda_{\PhaseMaps}\,[u])\,\bar Y
    \end{equation}
    for all phase maps $\bar X,\bar Y$ and every non-initial program point $u$ outside
    atomic blocks, including their entry points. Thus, each phase-map entry is subsumed
    by every entry reachable through $\psuccs$ at these program points.
    Then, the lifted analysis with phase-lifted domain and transfer functions is sound.
\end{theorem}
\begin{proof}
    The proof is by construction of an appropriate digest and relating solutions of the digest-refined
    analysis and the lifted analysis. The key challenge is that the lifted analysis eagerly advances the
    phases of other threads, whereas the digest-refined analysis insists on computing all phases of all threads
    precisely, also transitively communicating phase changes via threads that do not own phase ghosts
    and necessitating precisely tracking the values of globals and locals.
    The proof-only digest tracks a list of visited phases per thread.
    This is then abstracted to the last phase visited by each thread in the phase-lifted analysis.
    We defer this proof to \cref{t:lifting-sound} in \cref{app:sound}, whose proof also
    establishes the entry-closure condition \eqref{eq:entry-closure} for the coarse
    phase-successor relation.
\end{proof}

Observe that from a solution of the constraint system for $P'_{\ghost}$
a solution of the constraint system for the analysis of $P_{\ghost}$ using the combined transfer function for phase change
(\cref{fig:lift-assign-phase-ghost}) can be obtained by projecting away intermediate program points.
With this generic lifting defined, and its soundness established, we can now turn to the validation of witnesses with
phase ghosts.
\paragraph{Validating Invariants.}
Let $(\lambda_{\PhaseMaps},\mu_{\PhaseMaps})$ be a solution of the constraint system arising from the ghost-instrumented program $P_{\ghost}$ and
let $\sigma^\sharp_{\PhaseMaps} = \lambda_{\PhaseMaps}\,[u]$ be the abstract state for program point $u$. 
Then, we can check whether the invariant $\GhostInvariant\,u := e_{\phi}$ is known to hold at program point $u$ by checking whether
\begin{equation}\label{eq:phase-invariant}
    \bigsqcup \left\{
        \aSemB{e_{\phi}}\,(\sigma^\sharp \sqcap (\alpha_{\PhaseMaps}\,\bar P))  \mid
        (\bar P \mapsto \sigma^\sharp) \in \sigma^\sharp_{\PhaseMaps}
    \right\} = \aTrue
\end{equation}
where $\alpha_{\PhaseMaps}: \PhaseMaps \to \Domain$ is a helper function that maps a phase map to an abstract state
in the base domain where the values of the phase ghost variables are set to the values corresponding to the phase map.
Thus, the invariant is known to hold at program point $u$ if the expression evaluates to $\aTrue$ for all possible
pairs of phase maps and abstract states at that program point.
The additional meet with $\alpha_{\PhaseMaps}\,\bar P$ slightly improves precision by inserting precise information about
the values of the phase ghost variables into the abstract state before evaluating the invariant. As phase ghost variables
are globals, their precise values may sometimes not be known in the abstract state, especially for phase ghosts associated
with other threads, but they can be recovered from the phase map $\bar P$.
\begin{remark}\label{rem:phase-meet}
    Such a meet with $\alpha_{\PhaseMaps}\,\bar P$ can also be performed after each application of
    $\succs$ to improve precision, which we do in our implementation.
    This relies on every reachable concrete state having a
    representing entry $\bar P\mapsto\sigma^\sharp$ whose phase map agrees with its ghost
    values. Both $\sigma^\sharp$ and $\alpha_{\PhaseMaps}\,\bar P$ then represent that state,
    so a meet soundly overapproximating intersection preserves its representation.
\end{remark}
\begin{theorem}
    Under the hypotheses of \cref{thm:phase-lifted-sound},
    if for a program point $u$,
    a solution $(\lambda_{\PhaseMaps},\mu_{\PhaseMaps})$ and
    the invariant $\GhostInvariant\,u \equiv \texttt{e$_\phi$}$,
    \cref{eq:phase-invariant} evaluates to $\aTrue$, then
    $\GhostInvariant\,u$ holds at program point $u$ in $P_{\ghost}$.
\end{theorem}
\begin{proof}
    Directly from \cref{prop:ghost-ghostp}, the soundness of the base analysis, and \cref{thm:phase-lifted-sound}.
\end{proof}

\paragraph{Analysis Cost and Termination.}
The analysis exploiting phase ghosts fundamentally changes the
shape of the abstract interpretation.
Since this makes the analysis considerably more precise, when considering the cost, the comparison
should not be drawn to the analysis
that does not distinguish any interleavings at all,
but to the product construction.
Let us first turn to the product construction (where that is possible). Assuming there is a maximum of
$P$ program points per thread, and there are at most $t_d$ runtime threads, abstract values are computed
for $\mathcal{O}(P^{t_d})$ program points of the product program, where each abstract value is
from $\Domain$.
The phase-lifted analysis on then other hand computes abstract values for
$\mathcal{O}(P \cdot t)$ program points, where $t$ corresponds to the number
of thread \emph{prototypes}. Each of these program points
then has an associated value from the phase-lifted domain. Let us consider the
maximum number of keys that can have a value from $\Domain$ associated with them.
This is bounded by $\Pi_{i \in \phaseGhostTIDs}(|\Phases_i|)$,
i.e., is in $\mathcal{O}(N^{\left|\phaseGhostTIDs\right|})$ where $N$ is the maximum number of phases for any thread.
Overall, $\mathcal{O}(P \cdot t \cdot N^{\left|\phaseGhostTIDs\right|})$ abstract values from $\Domain$
are to be computed.
We usually have $t \leq t_d$, $N \ll P$, and $|\phaseGhostTIDs| \ll t$
rendering our approach practical, where performing a product construction is not.
This is in particular the case for programs that spawn an unbounded number of threads
that, however, do not have phases associated.
To overcome potential scalability issues, one angle of attack would be to compactly represent maps $\PhaseMaps \to \Domain$,
e.g., by employing a coarser relational domain than the map domain.
Where the full precision is required one may, taking inspiration from
BDDs~\cite{DBLP:journals/scp/DimovskiAL22,jeannetbddapron},
use Multi-Value Decision Diagrams (MDDs)~\cite{DBLP:conf/iccad/SrinivasanKMB90}
with elements from $\Domain$ as terminals,
exploiting that many abstract states for different phase maps are expected to be largely similar if not identical.
Our construction inherits the termination requirements of the base analysis and its
solver: each phase
ghost has a finite range, and the program contains only finitely many phase ghosts.
%
\Goblint uses the TD3 solver~\cite{SeidlV21}, for which termination
has been established for finite sets of unknowns~\cite{TilscherGSS26}.
For recursive programs, the analysis must therefore also ensure that only finitely many calling
contexts are distinguished.

%% file: 6-phases-rhs.tex
\begin{figure}[t]
    \[
        \begin{array}{lll}
            \lift_{\bar P}\,(\sigma^\sharp, \mu) &=&
            (
                \{ \bar P \mapsto \sigma^\sharp \},
                \{ \relax[v] \mapsto \{ \bar P \mapsto v^\sharp \}  \mid (\relax[v] \mapsto v^\sharp) \in \mu \}

            )\\[1ex]
            \unlift_{\bar P}\,\mu_{\PhaseMaps} & = & \left\{ [g] \mapsto
            \begin{cases}
                v^\sharp\,\bar{P} & \text{if } \phaseBound\,[g]\\
                \bigsqcup_{\bar{P'} \in \PhaseMaps} v^\sharp\,\bar{P}' & \text{otherwise}
            \end{cases}
                \;\middle|\;
                    ([g] \mapsto v^\sharp) \in \mu_{\PhaseMaps}
                \right\}
        \end{array}
    \]

    \caption{Helper functions to (a) \emph{lift} pairs of abstract states and contributions to the phase-lifted domain and
        (b) to \emph{unlift} a valuation of global unknowns $\mu_{\PhaseMaps}$ in the phase-lifted domain
         to a valuation of global unknowns $\mu$ in the base domain that can be used in the original
         transfer function $\aSem{\stmt}$.}
    \label{fig:lift-unlift}
\end{figure}

\begin{figure}[b]
        \begin{mdframed}
    \[
    \begin{array}{lll}
        \aSemP{\stmt}\,\sigma^\sharp_{\PhaseMaps}\,\mu_{\PhaseMaps} &=&
        \bigsqcup \{
            \succs\,\bar P\,\sigma^\sharp\,(\unlift_{\bar P}\,\mu_{\PhaseMaps})

            \mid

            ({\bar P} \mapsto \sigma^\sharp) \in \sigma^\sharp_{\PhaseMaps}
        \} \qquad\qquad \text{where}\\[1ex]
            \succs\,\bar P\,\sigma^\sharp\,\mu &=& \\
            \quad \begin{cases}
                \lift_{\bar P}\,(\aSem{\stmt}\,\sigma^\sharp\,\mu) & \stmt \text{ is in an atomic block}\\
                \bigsqcup \left\{
                \begin{array}{c@{\;}|@{\;}l}
                    \lift_{{\bar P}'}\,(\sigma^{\sharp'},\mu') &
                    \begin{array}{l}
                        (\sigma^{\sharp'},\mu') = \aSem{\stmt}\,\sigma^\sharp\,\mu, \\
                        {\bar P}' \in \psuccs\,\bar P\,\sigma^{\sharp'}
                    \end{array}
                \end{array}
                \right\} & \text{otherwise}
            \end{cases}\span\span
        \end{array}
    \]
    \end{mdframed}
    \caption{Lifted transfer function for ordinary statements $\stmt$ other than assignment to phase ghosts.
    }
    \label{fig:lift-ordinary}
\end{figure}

%% file: 6-phases-ghost-assign.tex
\begin{figure}
    \begin{mdframed}
    \[
        \begin{array}{lll}
            \aSemP{\texttt{$\ghost_j$ = exp;}} \sigma^\sharp_{\PhaseMaps}\,\mu_{\PhaseMaps} &=&
                    \aSemP{\texttt{$g_0$ = $\bar{g}_0$;}} \circ \cdots
        \aSemP{\texttt{$g_{n-1}$ = $\bar{g}_{n-1}$;}} \circ \aSemP{\overline{\texttt{$\ghost_j$ = exp;}}} \circ\\
        && \aSemP{\texttt{$\bar{g}_0$ = $g_0$;}} \circ \cdots
        \aSemP{\texttt{$\bar{g}_{n-1}$ = $g_{n-1}$;}}\qquad\qquad\qquad\qquad\\[2ex]

        \text{where}\quad
        (f \circ g)\,\sigma^\sharp\,\mu = \Let\;(\sigma^{\sharp'},\mu_0) = g\,\sigma^\sharp\,\mu\;\In\;
                    \Let\;(\sigma^{\sharp''},\mu_1) = f\,\sigma^{\sharp'}\,\mu\;\In\;
                    (\sigma^{\sharp''}, \mu_0 \sqcup \mu_1)      \quad\text{and} \span\span\\[2ex]
        \aSemP{\overline{\texttt{$\ghost_j$ = exp;}}} \sigma^\sharp_{\PhaseMaps}\,\mu_{\PhaseMaps} &=&\\
            \qquad \bigsqcup \left\{

                \left\{
                    \lift_{\left(\bar P \oplus \{i^\sharp_{\ghost_j} \mapsto v \}\right)}
                    \aSem{\texttt{$\ghost_j$ = } v} \sigma^\sharp\,(\unlift_{\bar P}\,\mu_{\PhaseMaps})
                    \mid

                v \in \aSemInt{exp}\,\sigma^\sharp \right\}
              \mid
                    ({\bar P} \mapsto  \sigma^\sharp) \in \sigma^\sharp_{\PhaseMaps}
            \right\} \span\span
        \end{array}
    \]
    \end{mdframed}
    \caption{Lifted transfer function for assignment to a phase ghost
    where
    $i^\sharp_{\protect\ghost_j}$
    is the abstract thread \emph{id} owning
    $\protect\ghost_j$,
    $\oplus$ denotes updating a map, and, by abuse of notation, in
    $\texttt{$\protect\ghost_j$ = } v$,
    $v$ denotes the constant integer \emph{expression} corresponding to the value $v$.
    Here, $\aSemInt{exp}$ is finite as the analysis establishes phase ghosts to be bounded.
    }
    \label{fig:lift-assign-phase-ghost}
\end{figure}

%% file: 6a-further-refine.tex
\subsection{Further Refinements of the Phase-Lifted Analysis}\label{ss:refinements}
The lifting of the base analysis to the phase-lifted analysis as described in \cref{sec:phase-ghosts}
still has limitations, most notably the very coarse overapproximation of when other
threads can change their phase. Here, we improve precision by leveraging an
off-the-shelf may-happen-in-parallel (MHP) analysis to determine when other threads can change their phase.
Lastly, we discuss how analyses can be made more precise by providing bespoke definitions of the
transfer function for phase changes and by propagating additional information with phase changes.

\paragraph{Taking MHP Information into Account.}\label{ss:mhp}
In the first definition of the function $\psuccs$ (\cref{eq:psuccs}), we assume that all other threads may change their
phases at any time, which is a coarse overapproximation: often, threads cannot change their phase because they, for instance,
have not yet been created, have definitely already been joined, or can change phases only while holding a specific mutex.
We provide a generic way to take such \emph{may-happen-in-parallel} information into account by
introducing the notion of \emph{concurrency contexts}
and a predicate to determine whether two program points with given concurrency contexts can be active concurrently.
Let us assume that the concurrency context from some set $\ConcCtx$ can be extracted from an abstract state of the base analysis via some
function $\projConcCtx: \Domain \to \ConcCtx$.
Given two concurrency contexts, there should be a predicate
$\textsf{may\_jointly\_active}: \ConcCtx \to \ConcCtx \to \textbf{bool}$
which determines whether two threads at program points with the respective concurrency contexts can be jointly active.
\begin{example}
    If $\ConcCtx$ corresponds to sets of held mutexes,
    $\textsf{may\_jointly\_active}\, S_0\, S_1$ can be set to $S_0 \cap S_1 = \emptyset$
    to encode that program points with overlapping locksets cannot be active concurrently.
\end{example}

We then modify the information maintained at unknowns $[\Phases^\rightarrow_i]$ to not only track possible phase changes,
but also the concurrency contexts in which these phase changes can happen.\footnote{
    We assume for simplicity that the concurrency context does not change upon a statement corresponding to a phase change.
}
This unknown then ranges over $2^{(\Phases_i \times \ConcCtx \times \Phases_i)}$, where a tuple
$(a,c,b)$ (alternatively written $a \to_{i,c} b$) denotes that the thread can change its phase from $a$ to $b$ in its local concurrency context $c$.
This then allows replacing the definition of $\to_{j,\sigma^\sharp}$ in $\psuccs_j$ (\cref{eq:psuccs}) with a more precise version:
\[
a \to_{j,\sigma^\sharp} b  \Leftrightarrow \exists c \in \ConcCtx: (a,c,b) \in [\Phases^\rightarrow_j] \land
    \textsf{may\_jointly\_active}\,c\,(\projConcCtx\,\sigma^\sharp)
\]
This definition leverages MHP information to rule out phase changes of other threads that cannot happen at a given program point.
By tracking program points in $\ConcCtx$, any existing MHP analysis
(see, e.g., \cite{Naumovich1999,Barik2006,Agarwal07,Di2015,Zhou2018,Albert12,Albert15,Albert17})
can be plugged into the phase-lifted analysis to increase precision.

Our implementation maintains path sensitivity in concurrency contexts inside the base domain.
The lifting applies the transfer to each path component and queries $\psuccs$ separately for each
resulting context component. Components with different concurrency contexts remain separate at joins;
the partition also retains the ego-thread information used by $\psuccs$.
Thus joining values within a component does not change its phase-successor relation.
The soundness argument for MHP-sensitive phase successors is given in \cref{app:mhp-soundness}.

\paragraph{Bespoke Definitions.}\label{sss:bespoke}
For a given base analysis, precision can be improved with a bespoke transfer function for phase changes.
Such a transfer function can exploit the fact that dedicated global unknowns abstract only values of global variables
that satisfy additional properties, e.g., values written with a given thread \emph{id}.
The generic definition above treats all values of global variables that the thread making the phase change can read as written by this
thread at the phase change, which may decrease precision for other threads:
For instance, an analysis may, for another unique thread, maintain a local map that tracks its
last written values and only consult these instead of the global unknown.
A bespoke definition may copy values to appropriate global unknowns associated with the new phase without conflating
the origins of values,
making analyses more precise---at the cost of also requiring bespoke soundness proofs.
In our implementation, we use bespoke definitions for analyses that
maintain different abstract values for the contributions of different threads.

\paragraph{Propagating Additional Information with Phase Changes.}
Some analyses, e.g., the mutex-meet analysis~\cite{schwarz2023clustered,schwarzthesis}, can benefit from additional
information being propagated along phase changes.
For that analysis, it is whether the initial values of global variables have definitely
been overwritten.
Such information from some auxiliary domain $\DomainAux$ (extracted from an abstract value of the base domain by
$\projDomainAux: \Domain \to \DomainAux$)
can be propagated with a phase change by extending the unknowns $[\Phases^\rightarrow_i]$ to range over
$2^{(\Phases_i \times \ConcCtx \times \DomainAux \times \Phases_i)}$ and recording this information during the initial analysis.
Assuming that the auxiliary information of multiple changes can be combined with $\sqcap$ and that there is an operation
$\odot: \Domain \to \DomainAux \to \Domain$ to combine the auxiliary information with the abstract value of the base analysis,
we can then modify the definition of $\psuccs$
and $\psuccs_j$ to also propagate the auxiliary information with a phase change (see \cref{fig:phase-aux-propagation}).
\begin{figure}[t]
    \begin{mdframed}
    \[
        \begin{array}{lll}
            \succs\,\bar P\,\sigma^\sharp\,\mu = \\
            \quad \begin{cases}
                \lift_{\bar P}\,(\aSem{\stmt}\,\sigma^\sharp\,\mu) & \stmt \text{ is in an atomic block}\\
                \bigsqcup \left\{
                \begin{array}{c}
                    \lift_{{\bar P}'}\,(\sigma^{\sharp'} \odot a^\sharp,\mu')
                \end{array}
                \;\middle|\;
                \begin{array}{l}
                    (\sigma^{\sharp'},\mu') = \aSem{\stmt}\,\sigma^\sharp\,\mu, \\
                    ({\bar P}',a^\sharp) \in \psuccs\,\bar P\,\sigma^{\sharp'}
                \end{array}
                \right\} & \text{otherwise}
            \end{cases}
        \end{array}
    \]
    \[
        \begin{array}{lll}
            \psuccs\,\bar P\,\sigma^\sharp = \{
                (\bar P', a^\sharp) &\mid& \bar P'\in \PhaseMaps \land
                (\exists i \in \phaseGhostTIDs: \projTIDs\,\sigma^\sharp = \{ i \}  \implies \bar P'\,i = \bar P\,i) \\
                && \land (\forall j \in \phaseGhostTIDs: (\bar P'\,j, a^\sharp_j) \in \psuccs_j\,(\bar P\,j)\,\sigma^\sharp)
                \land a^\sharp = \bigsqcap_{j \in \phaseGhostTIDs} a^\sharp_j
             \}
        \end{array}
    \]
    \[
        \begin{array}{lll}
            \psuccs_j\,a_0\,\sigma^\sharp = \{(a_0,\top)\} \cup \{ (a_n, \bigsqcap_{k=0}^{n-1} a^\sharp_k) \mid
                a_0 \to_{j,\sigma^\sharp,a^\sharp_0} \dots \to_{j,\sigma^\sharp,a^\sharp_{n-1}} a_n
            \} \qquad\text{ where }
        \end{array}
    \]
    \[
    a \to_{j,\sigma^\sharp,a^\sharp} b  \Leftrightarrow \exists c \in \ConcCtx, a^\sharp \in \DomainAux:
    (a,c,a^\sharp,b) \in [\Phases^\rightarrow_j] \land
        \textsf{may\_jointly\_active}\,c\,(\projConcCtx\,\sigma^\sharp)
    \]
    \end{mdframed}
    \caption{Auxiliary-information propagation with phase changes.}
    \label{fig:phase-aux-propagation}
\end{figure}
While this construction allows increasing the precision of the analysis by propagating additional information with phase changes,
the soundness of an analysis exploiting this information needs to be established separately.

%% file: 7-unassume.tex
\section{Integrating Unassume Operators}\label{sec:unassume}

While the previous sections focused on exploiting information encoded in ghost instrumentation
when deriving a constraint system and when checking the invariants from the witness after computing
a solution, the question of how to integrate \emph{unassume} operators, as proposed
by Saan et al.~\cite{SaanSESTV24,SaanPhD}, to speed up fixpoint iteration and arrive at a more precise solution remains.

In fact, \emph{unassume} operators can be incorporated into our validation of ghost witnesses
in the same way as in the setting without ghosts: At program locations
where an invariant is provided, an \emph{unassume} statement is inserted into the CFG which relaxes
the abstract state computed during fixpoint iteration.
Thus, any off-the-shelf unassume operator 
can soundly be used in our setting. Precision, however, depends on how the operator interacts with the ghost instrumentation.

\paragraph{Unassume for Mutex Ghosts.}

In preliminary experiments, applying off-the-shelf unassume operators to invariants containing mutex ghosts
often 
made witness validation fail where it succeeded without unassume.
This is tied to their non-local nature: As they encode global invariants, they can be placed at any program location.
Unassuming such an invariant amounts to relaxing the current abstract state
(as if the corresponding global was written). In our examples, the resulting state was often too imprecise
to show correctness \emph{at a given program location}, whereas without unassume
the necessary precision is attained.
This echoes the problem described in~\cref{sss:bespoke} that motivated bespoke definitions of
transfer functions for phase changes.
While dedicated unassume operators for mutex ghosts are conceivable, 
this issue should
be tackled by increasing the quality of witnesses by ensuring that invariants are only placed
at program locations where treating the information they contain as written by the current thread
will not result in a precision loss.
When this is not the case, it is preferable to use unassume only for invariants
that do not contain mutex ghosts.

\begin{example}
  Using unassume for invariants not involving mutex ghosts remains crucial also when validating
  witnesses with mutex ghosts. Consider again~\cref{fig:mutex-ghost-example}: Here, if
  \code{0 <= x && x <= 42} is not unassumed, the abstract interpretation may overshoot the fixpoint
  (cf.\ \cref{ex:unassume}). Overshooting this fixpoint means that the resource invariant
  \code{g == h} cannot be validated.
\end{example}

\paragraph{Unassume for Phase Ghosts.}
As invariants using phase ghosts are tied to program locations, the issue
outlined above does not arise. However, here, the unassume operator needs to
be lifted to the new domain $\Domain_{\PhaseMaps}$.
The most intuitive way to lift an operation $\aSem{\unassume{e}}$ on $\Domain$
to $\Domain_{\PhaseMaps}$ is by 
\begin{equation}\label{eq:unassume-phase-map-naive}
    \aSemP{\unassume{e}}\,\sigma^\sharp_{\PhaseMaps}
    =
    \left\{
        \bar{P} \mapsto
        \aSem{\unassume{e}}\,\sigma^\sharp
        \mid
        (\bar P \mapsto \sigma^\sharp) \in \sigma^\sharp_{\PhaseMaps}
    \right\}.
\end{equation}
However, this risks undoing the benefits of tracking different abstract values per phase map:
\begin{example}\label{ex:unassume-phase-map-naive}
    Consider an abstract state
    \[
        \sigma^\sharp_{\PhaseMaps} = \left\{
            \begin{aligned}
                \left(\{i_0 \mapsto 0, i_1 \mapsto 0\}\right) &\mapsto (\{\ghost_{i_0} \mapsto [0,0], \ghost_{i_1} \mapsto [0,0], x \mapsto [0,0]\}), \\
                \left(\{i_0 \mapsto 0, i_1 \mapsto 1\}\right) &\mapsto (\{\ghost_{i_0} \mapsto [0,0], \ghost_{i_1} \mapsto [1,1], x \mapsto [1,1]\})
            \end{aligned}
        \right\}
    \]
    for some program point, and assume that the witness provides the invariant
    \mintinline[escapeinside=@@]{C}!(@$\ghost_{i_1}$@ == 0 && x == 0) ||!
    \mintinline[escapeinside=@@]{C}!(@$\ghost_{i_1}$@ == 1 && x >= 1)!
    Applying the \emph{unassume} operator to each element of the abstract state $\sigma^\sharp_{\PhaseMaps}$
    yields
    \[
        \left\{
            \begin{aligned}
                \left(\{i_0 \mapsto 0, i_1 \mapsto 0\}\right) &\mapsto (\{\ghost_{i_0} \mapsto [0,0], \ghost_{i_1} \mapsto [0,1], x \mapsto [0,\infty]\}), \\
                \left(\{i_0 \mapsto 0, i_1 \mapsto 1\}\right) &\mapsto (\{\ghost_{i_0} \mapsto [0,0], \ghost_{i_1} \mapsto [0,1], x \mapsto [0,\infty]\})
            \end{aligned}
        \right\}
    \]
    Thus, the benefit of tracking $i_1$ is lost.
    This remains the case when, in line with \cref{rem:phase-meet}, information from the phase map is restored after
    applying \emph{unassume} as the values for $x$ remain too imprecise.
    Excluding phase ghosts from the variables havoced by \emph{unassume} has the same effect
    (see \cite{SaanSESTV24}).
\end{example}
One could define dedicated \emph{unassume} operators that, e.g., carefully
unassume only information about disjuncts that \emph{may} hold in a given phase. Here, we provide a simpler solution that strengthens
the invariant to be unassumed for each phase by conjoining it with the information from the phase map
and remains fully parametric in the unassume operator, i.e., we define
\begin{equation}\label{eq:unassume-phase-map}
    \begin{aligned}
        \aSemP{\unassume{e}}\,\sigma^\sharp_{\PhaseMaps}
        =
        \bigl\{
            \bar{P} \mapsto
            \aSem{\unassume{e && e_p}}\,\sigma^\sharp
            \;\bigm|\;&
            (\bar P \mapsto \sigma^\sharp) \in \sigma^\sharp_{\PhaseMaps}, \\
            \qquad \qquad\qquad\qquad\qquad\qquad
            \code{e_p} = \&\&_{i \in \phaseGhostTIDs} (\ghost_i == \bar P(i))
        \bigr\},
    \end{aligned}
\end{equation}
where $\ghost_i$ denotes the phase ghost associated with thread \emph{id} $i$
and $\&\&$ concatenates expressions.
%
\begin{proposition}
    If the base unassume operator $\aSem{\unassume{e}}$ is sound on $\Domain$,
    then the lifted operator defined in \cref{eq:unassume-phase-map} is sound on $\Domain_{\PhaseMaps}$.
\end{proposition}
\begin{proof}
    By the soundness of unassume for $\Domain$,
    $\gamma(\sigma^\sharp) \subseteq \gamma(\aSem{\unassume{e && e_p}}\,\sigma^\sharp)$.
    Thus,
    \[
                \gamma(\sigma^\sharp_{\PhaseMaps})
        \subseteq
        \textstyle\bigcup_{(\bar P \mapsto \sigma^\sharp) \in \sigma^\sharp_{\PhaseMaps}} \gamma(\aSem{\unassume{e && e_p}}\,\sigma^\sharp)
        =
        \gamma(\aSemP{\unassume{e}}\,\sigma^\sharp_{\PhaseMaps})\qedhere
    \]
\end{proof}
\begin{example}
    Consider again the abstract state from \cref{ex:unassume-phase-map-naive}, and assume that the unassume operator from \cref{eq:unassume-phase-map}
    is used and that the inner unassume operator performs some boolean simplification on the invariant to be unassumed.
    Then, the resulting abstract state, as desired, is:
    \[
        \left\{
            \begin{aligned}
                \left(\{i_0 \mapsto 0, i_1 \mapsto 0\}\right) &\mapsto (\{\ghost_{i_0} \mapsto [0,0], \ghost_{i_1} \mapsto [0,0], x \mapsto [0,0]\}), \\
                \left(\{i_0 \mapsto 0, i_1 \mapsto 1\}\right) &\mapsto (\{\ghost_{i_0} \mapsto [0,0], \ghost_{i_1} \mapsto [1,1], x \mapsto [1,\infty]\})
            \end{aligned}
        \right\}\qedhere
    \]
\end{example}

%% file: 8-experiments.tex
\section{Implementation and Experimental Evaluation}\label{sec:experiments}
We have implemented our approach in \Goblint,
a thread-modular abstract interpreter. It already supports the validation of
sequential correctness witnesses using the \emph{unassume} approach~\cite{SaanSESTV24}.

\subsection{Implementation Considerations}
The analysis is context-sensitive (\emph{functional approach}~\cite{pnueli1981two} with partial tabulation)
and partially flow-sensitive. It handles most {C} features including dynamic memory allocation, recursion, structures, and indirect calls.
For concurrency, \Goblint implements various analyses in the mutex-meet style~\cite{schwarz2023clustered,schwarzthesis},
as well as flavors of protection-based analysis~\cite{Vojdani2010}.
We instantiate these analyses with various relational and non-relational domains
to track both numerical and pointer information.

Our implementation combines the \textbf{Check} and \textbf{Exploit} phases into a single pass, as
much of the information established during \textbf{Check}, such as the reachability of program
locations or points-to sets, is also needed during \textbf{Exploit} and would otherwise have to be recomputed.
To ensure soundness, we introduce a dedicated unknown ranging over $\{\aTrue,\top\}$ for each hypothesis about how a ghost variable is used.
The constraints corresponding to the \textbf{Check} phase ensure that this unknown can take the value $\aTrue$ only if the hypothesis is valid,
and $\top$ otherwise.
Any exploitation of the hypothesis is guarded by the corresponding unknown having the value $\aTrue$.
If a hypothesis can no longer be established during analysis, this ensures that all transitively dependent unknowns are recomputed.
Thus, \emph{any} solution to the constraint system is sound. The solver computes \emph{one} such solution.
Where the analyzer is instantiated with non-Noetherian domains, widening and narrowing are used to ensure termination,
meaning that the computed solution will generally not coincide with a solution computed by a strict two-phase approach.
%
%
%
The implementation maintains \emph{concurrency contexts} consisting of dynamically computed thread \emph{id}s
and sets of held mutexes and must-joined threads. For the latter, we have implemented an optimization
that makes the phase of a thread that has been must-joined more precise: At a separate unknown, we collect the
phases a thread may be in when it returns. Then, in the join, we can set the phase of the joined thread to this phase.
We have, in fact, silently made use of this when discussing the example from~\cref{fig:phase-worked-example}.

\subsection{Experimental Evaluation}

We conducted two experiments, targeting witnesses with mutex ghosts and targeting phase ghosts.

\paragraph{Experiment 1: Validating Witnesses with Mutex Ghosts.}

We build on previous work~\cite{ErhardBHKSSSTV2025} extending \Goblint
with the generation of ghost witnesses. It uses ghost variables in two ways: Most are \emph{mutex ghosts}
and are used to encode resource invariants. In addition, it introduces a ghost variable \code{multithreaded}
initialized to \code{false} and set to \code{true} when the first thread is created.
As \Goblint is path-sensitive in the threading mode,
this ghost variable does not carry information beyond what is already tracked by the analysis, and no special handling is
required to validate invariants involving this ghost variable.\footnote{We observe that, conceptually, such a ghost variable
is a phase ghost.}
We compare the validation of such witnesses by \Goblint with \Ultimate,
which is the only other validator for such witnesses.
The experiment was conducted on a machine with two \textsc{Intel Xeon Platinum 8260 @
2.40GHz} processors with 24 cores each and 512~GB of RAM running Ubuntu 24.04.3 LTS, where
we used \textsc{BenchExec}~\cite{BeyerLW19} to get reliable numbers.

We re-use the set of witnesses produced for programs from the \textsc{ConcurrencyMain} category of \SVComp 2024~\cite{svcomp24}.
These witnesses are categorized along two dimensions:
the underlying analysis technique (\emph{mutex-meet} or \emph{protection-based}) and whether \Goblint can verify
the reachability specifications from \SVComp (\emph{spec-(un)confirmed}).
For the spec-confirmed cases, we validate the generated witness and the specification.
For the spec-unconfirmed cases, where \Goblint was not able to verify the property or the expected verdict is false,
\Goblint still generates a witness (for the trivial property \textsf{true}).
In this case, we validate the generated witnesses only.

For \Goblint, we use the mutex-meet analysis to validate.
Our implementation follows the construction in \cref{sec:mutex-ghosts} directly.
Its soundness is established by \cref{thm:mutex-ghost-check}.
Moreover, as the way the invariants with ghosts are placed caused the issues described in \cref{sec:unassume},
we configured \Goblint to not unassume invariants that contain ghost variables.
Each tool had a timeout of 15 minutes and a memory limit of 16~GB per benchmark.
We answer the following research questions:
\begin{itemize}
     \item (\RQ1) Is \Goblint able to validate the generated witness (in both spec-confirmed and -unconfirmed cases), as well as the original specification (in the spec-confirmed cases)?
    \item (\RQ2) How does its performance compare to \Ultimate in terms of the number of validated witnesses and the time taken for validation?
\end{itemize}

Table~\ref{tab:ghost-mutex-validation} shows the validation outcomes.
Regarding (\RQ1), in the spec-confirmed cases \Goblint validated all but one witness together with the original specification.
In the spec-unconfirmed cases,
\Goblint validated 402 out of 474 witnesses in the \texttt{mutex-meet-spec-unconfirmed} benchmark group and 380 out of 448 witnesses in \texttt{protection-spec-unconfirmed}.
This confirms that our approach can validate the witnesses and re-establish the original specification in most of the cases.
Regarding (\RQ2), \Ultimate validated all witnesses in the spec-confirmed cases, and validated 316 out of 474 witnesses in the \texttt{mutex-meet-spec-unconfirmed}
benchmark group and 343 out of 448 witnesses in \texttt{protection-spec-unconfirmed},
which is close, but still less than what \Goblint validated.
More importantly, \Goblint is significantly faster than \Ultimate in all cases.
On witnesses validated by both tools, \Goblint is on average around $90\times$, $8.3\times$, $97.7\times$, and $9.6\times$
 faster than \Ultimate in the four groups, respectively.
We further analyze the agreement between \Goblint and \Ultimate on validation, as shown in Table~\ref{tab:2x2-validation}.
\Goblint and \Ultimate uniquely validate 126 and 40 witnesses in the
\texttt{mutex-meet-spec-unconfirmed} benchmark group, respectively, and uniquely validate 89 and 52 witnesses in the
\texttt{protection-spec-unconfirmed} benchmark group, respectively.
This matches the intuition that tools based on different techniques may be able to validate different witnesses,
and underscores the need for a diverse set of validators.
\input{table-mutex-ghosts}
\input{table-mutex-ghosts-comparison}

\paragraph{Experiment 2: Validating Witnesses with Phase Ghosts.}
We start from \Goblint's standard
configuration for validating witnesses for sequential programs in \SVComp.
The underlying privatization is \texttt{protection-atomic-ghost}, which extends the
framework's \texttt{protection-atomic} analysis with ghost-witness validation.
Phase changes use the generic lifting from \cref{ss:exploiting-phase-ghosts}, whose soundness
is established by \cref{thm:phase-lifted-sound}; no bespoke transfer function is used.
Of the refinements from \cref{ss:refinements}, we use MHP-sensitive phase successors and
must-join reasoning, but do not propagate additional information with phase changes.
The MHP information is obtained from the thread \emph{id} and must-join
information supplied by \Goblint~\cite{schwarz2023clustered,hammerandnail,DBLP:conf/vmcai/SchwarzE26};
for its use in the lifted analysis see \cref{lem:history-phase-reachability-mhp}.
Must-join reasoning fixes the phase of a joined thread to one of its possible return phases;
we justify this refinement in \cref{app:must-join-soundness}.
Finally, we use the exact phase-value restoration described in \cref{rem:phase-meet},
which also sketches its soundness argument.

While phase ghosts are what, e.g., \textsc{Ultimate} aspires for,
no automated tools for the generation of such witnesses exist yet.
We thus manually craft witnesses with phase ghosts for 5 tasks from different
subsets of the \texttt{unreach-call.Concurrency} category of \SVComp~2026~\cite{svcomp26}
that are out of reach for \Goblint without witness-guidance,
and attempt to validate them with \Goblint and \Ultimate.
In each of five subfolders of this category, we examined tasks unsolved by unguided \Goblint
in alphabetical order, selecting the first for which we could readily construct a suitable phase witness.
For each task, we consider two witnesses: one with ghost instrumentation only ($\ghost$),
and one that also contains invariants about ghost and program variables ($\ghost \land \textbf{I}$).
\cref{tab:crafted-phase-ghost-overview} provides an overview of the tasks,
and how long verification and validation took.
For $\ghost \land \textbf{I}$, we also report on validation with \Goblint when additionally using
\emph{unassume}.
With witness guidance, \Goblint can validate witnesses where verifying the property without a witness is out of reach.
For the task marked $\star$, the phase-aware definition of unassume
(\cref{sec:unassume}~\eqref{eq:unassume-phase-map}) was crucial when using unassume;
the na\"ive definition (\cref{sec:unassume}~\eqref{eq:unassume-phase-map-naive}) caused validation to fail.
While (as expected), considering some aspects of interleavings as directed by the phase ghosts is slower than
not considering them at all, the scalability benefits of the thread-modular approach are retained:
Where \Ultimate terminates, \Goblint is over an order of magnitude faster.
For a detailed discussion see \cref{sec:crafted-appendix}.

\input{table-crafted.tex}

To complement these hand-crafted examples and spur the further development of validation techniques,
we also provide a set of 50 programs generated with AI support with invariants
based on real-world patterns whose correctness is out of reach for \Goblint without witness-guidance,
but which can be verified with the help of phase ghosts.\footnote{
    \url{https://github.com/michael-schwarz/phase-witnesses}. We inspected and confirmed each program
    manually.
} See \cref{sec:crafted-ai} for details.
We will contribute them, as well as the hand-crafted ones, to next year's \SVComp
validation track.

\paragraph{Threats to Validity.}
Our evaluation is limited by the availability of ghost witnesses, which is due to
the lack of validators that our work addresses.
We used witnesses generated by \Goblint for previous work~\cite{ErhardBHKSSSTV2025}.
They were generated for validation by a model checker,
mitigating concerns that they are tailored to abstract interpreters.
For phase ghosts, no automated generation technique exists, and we therefore hand-crafted witnesses.
We selected tasks from different subsets of the widely-used \SVComp suite.
We hope that our work will spur further development in their generation. 

%% file: table-mutex-ghosts.tex
\begin{table}[t]
	\footnotesize
	\centering
	\begin{tabular*}{\linewidth}{@{\extracolsep{\fill}}llrrr@{}}
		\toprule
		Benchmark Group & Technique & \multicolumn{1}{c}{\shortstack{Validation\\(validated / total)}} & \multicolumn{1}{c}{\shortstack{Avg time\\(both validated)}} & \multicolumn{1}{c}{\shortstack{Avg time\\(total, excl.\ timeout)}} \\
		\midrule
		\texttt{mutex-meet-spec-confirmed} & \GoblintAnon & \textcolor{gray}{\scriptsize (8\,/\,29)} 28\,/\,29 & 0.3\,s & 0.3\,s \\
		 & \Ultimate & 29\,/\,29 & 27.0\,s & 27.9\,s \\
		\midrule
		\texttt{mutex-meet-spec-unconfirmed} & \GoblintAnon & \textcolor{gray}{\scriptsize (152\,/\,474)} 402\,/\,474 & 18.1\,s & 66.0\,s \\
		 & \Ultimate & 316\,/\,474 & 150.1\,s & 175.2\,s \\
		\midrule
		\texttt{protection-spec-confirmed} & \GoblintAnon & \textcolor{gray}{\scriptsize (8\,/\,15)} 15\,/\,15 & 0.3\,s & 0.3\,s \\
		 & \Ultimate & 15\,/\,15 & 29.3\,s & 29.3\,s \\
		\midrule
		\texttt{protection-spec-unconfirmed} & \GoblintAnon & \textcolor{gray}{\scriptsize (190\,/\,448)} 380\,/\,448 & 10.4\,s & 24.1\,s \\
		 & \Ultimate & 343\,/\,448 & 100.2\,s & 136.6\,s \\
		\bottomrule
	\end{tabular*}
	\caption{Witness validation using \GoblintAnon and \Ultimate.
		Gray parenthesized counts show the ablation results for \GoblintAnon with
		(\cref{eq:mutex-ghost-check}) disabled, i.e., evaluating the
		invariant without first locking the corresponding mutex.
		The average time (column 4)
		is computed over the witnesses \emph{both} tools validated.
		The average time (column 5) is computed over all validated witnesses,
		excluding the benchmarks that timed out.}
	\label{tab:ghost-mutex-validation}
\end{table}

%% file: table-mutex-ghosts-comparison.tex
\begin{table}
	\footnotesize
	\centering
	\begin{subtable}{0.48\textwidth}
		\centering
		\begin{tabular}{lrrr}
			\toprule
			& \GoblintAnon: \textcolor{green!50!black}{\faCheckCircle} & \GoblintAnon: \textcolor{red}{\faTimesCircle} & Total \\
			\midrule
			\Ultimate: \textcolor{green!50!black}{\faCheckCircle} & 276 & 40 & 316 \\
			\Ultimate: \textcolor{red}{\faTimesCircle} & 126 & 32 & 158 \\
			\midrule
			Total & 402 & 72 & 474 \\
			\bottomrule
		\end{tabular}
		\caption{\texttt{mutex-meet-spec-unconfirmed} ($n=474$)}
		\label{tab:2x2-mutex-ghost-unconfirmed}
	\end{subtable}
	\hfill
	\begin{subtable}{0.48\textwidth}
		\centering
		\begin{tabular}{lrrr}
			\toprule
			& \GoblintAnon: \textcolor{green!50!black}{\faCheckCircle} & \GoblintAnon: \textcolor{red}{\faTimesCircle} & Total \\
			\midrule
			\Ultimate: \textcolor{green!50!black}{\faCheckCircle} & 291 & 52 & 343 \\
			\Ultimate: \textcolor{red}{\faTimesCircle} & 89 & 16 & 105 \\
			\midrule
			Total & 380 & 68 & 448 \\
			\bottomrule
		\end{tabular}
		\caption{\texttt{protection-spec-unconfirmed} ($n=448$)}
		\label{tab:2x2-protection-unconfirmed}
	\end{subtable}
	\caption{Validation agreement between \GoblintAnon and \Ultimate on the two spec-unconfirmed (validate-only)
		groups (witness-present benchmarks only)
		where \textcolor{green!50!black}{\faCheckCircle}
		(\textcolor{red}{\faTimesCircle}) indicate (un-)successful validation, respectively.}
	\label{tab:2x2-validation}
\end{table}

%% file: table-crafted.tex
\begin{table}
	\footnotesize
	\centering
	\resizebox{\textwidth}{!}{%
	\begin{tabular}{llrccccccc}
		\toprule
		& & & \multicolumn{4}{c}{\GoblintAnon} & \multicolumn{3}{c}{\textsc{Ultimate Gemcutter}} \\
		\cmidrule(lr){4-7}\cmidrule(lr){8-10}
		Suite & Task & Loc & \shortstack{Verification\\w/o Witness} &
			\shortstack{Validation\\ ($\ghost$)} &
			\shortstack{Validation\\ ($\ghost \land \textbf{I}$)} &
			\shortstack{Validation ($\ghost \land \textbf{I}$)\\ + \textbf{Unassume}} &
			\shortstack{Verification\\w/o Witness} &
			\shortstack{Validation\\ ($\ghost$)} &
			\shortstack{Validation\\ ($\ghost \land \textbf{I}$)} \\
		\midrule
		\texttt{ldv-races}~\cite{DBLP:journals/pcs/ZakharovMMNPK15} &  \texttt{race-1\_1-join.yml} & 1729
			& \textcolor{red}{\faTimesCircle} (0.1s)
			& \textcolor{green!50!black}{\faCheckCircle} (0.2s)
			& \textcolor{green!50!black}{\faCheckCircle} (0.2s)
			& \textcolor{green!50!black}{\faCheckCircle} (0.3s) \phantom{$\star$}
			& \textcolor{green!50!black}{\faCheckCircle} (20.7s)
			& \textcolor{green!50!black}{\faCheckCircle} (22.7s)
			& \textcolor{green!50!black}{\faCheckCircle} (26.2s) \\
		\texttt{pthread} &  \texttt{stateful01-2.yml} & 731
			& \textcolor{red}{\faTimesCircle} (0.1s)
			& \textcolor{green!50!black}{\faCheckCircle} (0.3s)
			& \textcolor{green!50!black}{\faCheckCircle} (0.3s)
			& \textcolor{green!50!black}{\faCheckCircle} (0.3s) \phantom{$\star$}
			& \textcolor{green!50!black}{\faCheckCircle} (20.4s)
			& \textcolor{green!50!black}{\faCheckCircle} (21.8s)
			& \textcolor{green!50!black}{\faCheckCircle} (31.8s) \\
		\texttt{pthread-deagle} &  \texttt{arith...-ok.yml} & 768
			& \textcolor{red}{\faTimesCircle} (0.1s)
			& \textcolor{green!50!black}{\faCheckCircle} (0.4s)
			& \textcolor{green!50!black}{\faCheckCircle} (0.4s)
			& \textcolor{green!50!black}{\faCheckCircle} (0.4s) $\star$
			& \textcolor{green!50!black}{\faCheckCircle} (56.9s)
			& \textcolor{green!50!black}{\faCheckCircle} (57.8s)
			& \textcolor{green!50!black}{\faCheckCircle} (71.3s) \\
		\texttt{pthread-C-DAC} &  \texttt{...-datarace-1.yml} & 1359
			& \textcolor{red}{\faTimesCircle} (0.6s)
			& \textcolor{green!50!black}{\faCheckCircle} ($\approx$ 11min)
			& \textcolor{green!50!black}{\faCheckCircle} ($\approx$ 9min)
			& \textcolor{green!50!black}{\faCheckCircle} ($\approx$ 9min)\phantom{$\star$}
			& \textcolor{green!50!black}{\faCheckCircle} (72.0s)
			& {\faHourglassEnd} (30min)
			& {\faHourglassEnd} (30min) \\
		\rowcolor{gray!15}\cellcolor{white}\texttt{} &  (smaller constants) & 1359
			& \textcolor{red}{\faTimesCircle} (0.1s)
			& \textcolor{green!50!black}{\faCheckCircle} (0.8s)
			& \textcolor{green!50!black}{\faCheckCircle} (0.8s)
			& \textcolor{green!50!black}{\faCheckCircle} (0.9s) \phantom{$\star$}
			& \textcolor{green!50!black}{\faCheckCircle} (28.5s)
			& \textcolor{green!50!black}{\faCheckCircle} (72.1s)
			& \textcolor{green!50!black}{\faCheckCircle} (77.4s)\\
		\texttt{weaver}~\cite{DBLP:conf/cav/FarzanV19} &  \texttt{test-easy8.wvr.yml} & 78
			& \textcolor{red}{\faTimesCircle}  (0.1s)
			& \textcolor{green!50!black}{\faCheckCircle} (0.1s)
			& \textcolor{green!50!black}{\faCheckCircle} (0.1s)
			& \textcolor{green!50!black}{\faCheckCircle} (0.1s) \phantom{$\star$}
			& \textcolor{green!50!black}{\faCheckCircle} (16.6s)
			& \textcolor{green!50!black}{\faCheckCircle} (16.6s)
			& \textcolor{green!50!black}{\faCheckCircle} (17.7s) \\
		\bottomrule
	\end{tabular}%
	}
	\caption{Validation of crafted phase-ghost witnesses with \GoblintAnon and \Ultimate.}
	\label{tab:crafted-phase-ghost-overview}
\end{table}


%% file: 9-related.tex
\section{Related Work}\label{sec:related}

We first discuss abstract interpretation,
before turning to witness generation and validation for the sequential case.
Lastly, we discuss existing work on concurrent correctness witnesses with ghosts. 

\paragraph{Thread-Modular Abstract Interpretation}
To the best of our knowledge, no existing approaches are concerned with the validation of correctness witnesses with ghosts.
However, there is a wide variety of
approaches~\cite{DBLP:journals/jss/StievenartNMR19,Vojdani2010,DBLP:conf/sas/SchwarzSSAEV21,CarreH09,
Sharma21,Farzan12,Kusano16,Kusano17,Mukherjee2017,valentinShape}
to thread-modular abstract interpretation.
One class of approaches pioneered by Miné and his co-authors~\cite{Mine2011,Mine2012,Mine2014,Mine2016,Mine2017,Mine2018}
proposes the notion of \emph{interferences} between threads that are emitted and applied during the
analysis of individual threads. The analysis then terminates when a fixpoint
is reached for all threads and the interferences.
These approaches are generic and allow for both relational and non-relational abstractions of
values and interferences.
Recently, a new interference domain has been proposed that tracks that some globals are only written by a thread when some other global
has a specific value~\cite{DBLP:conf/fm/ToblerS26}:
For a witness with mutex ghosts, this domain can infer protection in the sense of~\citet{Vojdani2010}.
When it comes to the phase-lifting construction, the work by \citet{Mine2017} is the most closely
related: There, interferences are sensitive to (abstractions of) program locations of other threads,
tracked in auxiliary variables as part of the numerical abstract state, allowing, e.g., for the verification
of mutual exclusion algorithms. We remark that while values of phase ghosts sometimes cleanly correspond to sets
of program points that a thread is at, this is not always the case, in particular where procedure calls are involved
or when phases encode more semantic information.
While several of these approaches compute invariants in the form of
resource invariants~\cite{Mine2014,schwarz2023clustered,Vojdani2010,VojdaniARSVV16}, none of them support
outputting these as witnesses or tackle the problem of witness validation.

Our approach for exploiting phase ghosts
(after having confirmed that a ghost variable is a phase ghost) 
is by lifting mixed-flow sensitive~\cite{Stemmler25,DBLP:conf/fm/SeidlVES26}
thread-modular abstract interpretations~\cite{hammerandnail,schwarz2023clustered,schwarzthesis,DBLP:conf/sas/SchwarzSSAEV21}
which admit refinement by history abstractions (\emph{digests}).
We prove soundness of our analysis by relating it to a theoretical digest-refined analysis.
There, digests precisely propagate phase information with every action of the program
(which necessitates also tracking values of variables \emph{precisely}), making it
effectively equivalent to the concrete semantics.
Phase maps on their own do not constitute a digest and our analysis is not an instance
of the digest framework.
Instead, the \emph{structure} of our domain more closely resembles
cardinal power domain constructions~\cite{Cortesi13,Giacobazzi99,Cousot79}
as encountered in the \emph{view framework}~\cite{DBLP:journals/toplas/KimRR18}.
Neither considers concurrent programs.

We lastly remark that non-thread-modular approaches that bound the number of context switches
(e.g.,~\citet{DBLP:conf/cav/LalR08}), are not applicable, 
even if there is one phase
ghost for each thread: While the number of \emph{phases} is bounded, threads can be interrupted while in a given
phase. Thus, the number of context switches within a phase is not bounded, and neither is their overall number.

\paragraph{Software Verification Witnesses.}
We focus on witnesses enabling information exchange across techniques rather than those specific to,
e.g., abstract interpretation~\cite{DBLP:conf/iclp/AlbertAPH06,DBLP:journals/tcs/BessonJP06,DBLP:conf/lpar/AlbertPH04,DBLP:conf/esop/BessonJT07}.
Witnesses were introduced to \SVComp
by~\citet{DBLP:conf/sigsoft/0001DDHS15,DBLP:conf/sigsoft/0001DDH16,DBLP:journals/tosem/BeyerDDHLT22},
with \citet{DBLP:conf/spin/AyaziovaBRSS24} proposing the invariant-based format 
which has since been extended to
(non-)termination~\cite{DBLP:conf/kbse/AdamABBJRS25,CAV26a},
function contracts, and claims of inductiveness~\cite{DBLP:journals/corr/abs-2501-12313}.
%
In the 2026 edition of \SVComp, there were 12 tools that generated (non-trivial) correctness
witnesses for single-threaded programs, 
and 9 validators~\cite{svcomp26}.
We build on the \emph{unassume} approach of Saan et al.~\cite{SaanSESTV24,SaanPhD}
to validating witnesses for single-threaded programs.
Like all other competitors, this approach does not support witnesses with ghost variables.

\paragraph{Ghost Witnesses}
\citet{ErhardBHKSSSTV2025} showed that ghost variables allow information exchange between
thread-modular abstract interpreters 
and model checkers. 
They demonstrated generating witnesses with mutex ghosts with
an abstract interpreter and validating them with a model checker. The opposite
direction was not studied.
When it comes to the creation of concise ghost instrumentation from the proofs of an interleaving-based tool,
there has been some recent progress by~\citet{SchusseleZLK26}. The pattern of phase ghosts has been
partly inspired by this line of work, though further steps are necessary to obtain portable witnesses
from their concise (internal) proofs.
%
The \textsc{Ultimate} family of model
checkers~\cite{DBLP:conf/cav/HeizmannHP13,DBLP:conf/tacas/BenteleBEKDHKSP26,DBLP:conf/pldi/FarzanKP22,DBLP:conf/tacas/KlumppDHSEFP22,DBLP:conf/tacas/DietschHKSP23,DBLP:conf/sas/GreitschusDP17}
can, in principle, validate ghost witnesses.
They do so by inserting the ghost instrumentation and then employing their off-the-shelf verification approaches,
treating witnesses as additional proof burdens.
Our approach is the first to exploit the semantic information encoded by the ghost variables to aid witness validation.

%% file: 10-conclusion.tex
\section{Conclusion and Future Work}\label{sec:conclusion}

Previous approaches treat ghost instrumentation and invariants as additional proof obligations.
We instead presented a thread-modular abstract interpretation that checks and exploits semantic information
encoded in ghost variables.
Instantiated for resource invariants and phase information, it validates properties beyond the reach of
unguided analysis while preserving scalability.

Beyond producing and validating ghost witnesses with other verification techniques,
further directions include:
Ghost variables to guide analysis of sequential programs by inducing path splitting, which is
especially relevant for validating results from tools that search for effective
trace partitionings~\cite{DBLP:conf/sas/BabuLBM24,Rival07}.
Our phase-based approach already subsumes this pattern as well as encodings of context sensitivity.
As ghost witnesses become more widely used, it will be important to assess whether our current classes of ghost variables
will continue to cover practice and,
 if not, to develop techniques for richer forms such as ghosts that encode thread-specific information. 

%% file: appendix/5b-proof.tex
\section{Soundness Proof for the Phase-Lifted Analysis}\label{app:sound}
Consider an analysis that is based on the local trace semantics and admits refinement by digests.
To provide this proof, we rely on some results from~\cite{hammerandnail,schwarzthesis,schwarz2023clustered}
where the local trace semantics, the notion of \emph{digests}, and their admissibility were introduced.
While our general construction outlined in the main part is generic with respect to the flavor of thread-modular abstract
interpretation used, providing a worked-out soundness proof necessitates fixing some formalism and some assumptions.

We thus briefly recap some results from \cite{hammerandnail,schwarzthesis,schwarz2023clustered}
insofar as we rely on them in the following.

\subsection{Digest-Driven Abstract Interpretation}
The local trace semantics takes the perspective of a single thread and considers only the part
of the execution that the given thread can know about. Each thread knows about its own history
and learns further facts about the global execution by communicating with other threads.
Such communication occurs by executing pairs of observable and observing statements, where observing
statements may, e.g., be locking some mutex with the corresponding observable action being unlocking the
same mutex or program initialization. In the local trace semantics, the values of globals are not modeled
directly: Instead \emph{atomicity mutexes} are introduced that protect the values of globals and are acquired
before reading or writing a global variable and are released immediately after this action.
The atomicity mutex for a global $g$ is denoted $m_g$.
Inside a generated atomic block, all atomicity mutexes are acquired before copying global values
into the fresh locals. They remain held throughout the phase assignment and are released only
after all values have been written back.

We here introduce additional operations $\atomicBegin$ (observing) and $\atomicEnd$ (observable)
that mark the beginning and end of an atomic block introduced by the ghost instrumentation, respectively.
These operations act like mutex acquisitions and releases, respectively.

Let $\Traces$ be the set of local traces. The local traces of a given program can then be defined as the
least fixpoint of a constraint system
which has unknowns corresponding to program points and observable statements.
\[
    \begin{array}{lllr}
        (\lambda\,[u_0],\mu) &\sqsupseteq& (\textsf{init}, \emptyset)  & \\
        (\lambda\,[u'],\mu) &\sqsupseteq& (\sem{\texttt{new}} (\lambda\,[u])) \span \stmt \textsf{ creates a new thread starting at u'}  \\
        (\lambda\,[v],\mu) &\sqsupseteq& (\sem{\stmt} (\lambda\,[u]),\emptyset) & \stmt \textsf{ not observing or observable}  \\
        (\lambda\,[v],\mu) &\sqsupseteq& (\sem{\stmt} (\lambda\,[u], \mu\,[\stmt']),\emptyset) & \stmt \textsf{ observes } \stmt'  \\
        (\lambda\,[v],\mu) &\sqsupseteq& (\sem{\stmt} (\lambda\,[u]), \{[\stmt] \mapsto \sem{\stmt} (\lambda\,[u])  \}) & \stmt \textsf{ observable }  \\
    \end{array}
\]
where $u_0$ is the initial program point and $(u,\stmt,v)$ are edges from the control-flow graph.
Here, $\textsf{init} \subset \Traces$ is some set of initial local traces that consists of configurations
of the initial thread only.
Transfer functions then describe how to prolong traces.
For example, the function $\sem{\lock{a}}: 2^\Traces \to 2^\Traces \to 2^\Traces$ describes how local
traces ending at some program point $u$ are combined with other local traces ending in the observable
action $\unlock{a}$ to yield new local traces ending at program point $v$.
Here, one trace ending at $u$ is combined with one trace ending at $\unlock{a}$, where this is possible.
Instances where it may not be possible include, e.g., the case where the $\unlock{a}$ happens in a thread
that has definitely not been created yet in the execution of the considered trace ending at $u$.
The local trace semantics has been shown to be equivalent to a, perhaps more conventional,
interleaving semantics~\cite{ErhardBHKSSSTV2025}.

To obtain an abstract interpretation, unknowns are then modified to range over abstract
values and concrete transfer functions are replaced by suitable abstract counterparts.
As long as these abstract transfer functions overapproximate their concrete counterparts,
the resulting abstract interpretation is sound.

Digests can then be used to refine an abstract interpretation.
Let $\Digests$ be a set that can be extracted from a local trace by an abstraction
function $\alpha_\Digests: \Traces \to \Digests$.
There are digest counterparts $\aSemDigest{\stmt}: \Digests \to \Digests \to 2^{\Digests}$
for functions $\sem{\stmt}$ as well as
a set $\textsf{init}^\sharp_{\Digests}$ of initial digests.
Functions $\aSemDigest{\stmt}$ here are \emph{deterministic}, i.e., they return either a singleton
set or the empty set.
\begin{definition}\label{def:adm}
  A digest $\Digests$ is \emph{admissible} if the following hold:
  \begin{itemize}
    \item $\forall t \in \textsf{init}: \alpha_\Digests\,t \in \textsf{init}^\sharp_{\Digests}$
    \item $\forall t_0: \forall t' \in \sem{\textsf{new}}(\{t_0\}):
      \aSemDigest{\textsf{new}} (\alpha_{\Digests}\,t_0)  = \{ \alpha_{\Digests} t' \}$
    \item $\forall t_0: \forall t' \in \sem{\stmt}(\{t_0\}):
      \aSemDigest{\stmt} (\alpha_{\Digests}\,t_0)  = \{ \alpha_{\Digests} t' \}  $, for
      $\sem{\stmt}$ non-observing.
    \item $\forall t_0, t_1 \in \Traces: \forall t' \in \sem{\stmt}(\{t_0\},\{t_1\}):
      \aSemDigest{\stmt} (\alpha_{\Digests}\,t_0, \alpha_{\Digests}\,t_1)  = \{ \alpha_{\Digests} t' \}  $, for
      $\sem{\stmt}$ observing.
  \end{itemize}
\end{definition}
The refined constraint system then replaces unknowns $[u]$ for program points with
refined unknowns $[u,A]$, and unknowns for observable statements
$[\stmt]$ with refined unknowns $[\stmt,A]$ for $A\in\Digests$.
The constraints then take the following shape:
\[
    \begin{array}{lllr}
        (\lambda\,[u_0,A_0],\mu) &\sqsupseteq& (\textsf{init}^\sharp, \emptyset)  & \\[2ex]
        (\lambda\,[u',A'],\mu) &\sqsupseteq& (\aSem{\textsf{new}} (\lambda\,[u,A_0]),\emptyset) & \\
        & & \span (\stmt \textsf{ creates a new thread starting at $u'$},  A' \in \aSemDigest{\textsf{new}}\,A_0) \\[2ex]
        (\lambda\,[v,A'],\mu) &\sqsupseteq& (\aSem{\stmt} (\lambda\,[u,A_0]),\emptyset) & \\
        & & \span (\stmt \textsf{ not observing or observable},  A' \in \aSemDigest{\stmt}\,A_0) \\[2ex]
        (\lambda\,[v,A'],\mu) &\sqsupseteq& (\aSem{\stmt} (\lambda\,[u,A_0], \mu\,[\stmt',A_1]),\emptyset) & \\[0.5ex]
        & & \span (\stmt \textsf{ observes } \stmt',  A' \in \aSemDigest{\stmt}\,(A_0,A_1)) \\[2ex]
        (\lambda\,[v,A'],\mu) &\sqsupseteq& (\aSem{\stmt} (\lambda\,[u,A_0]), \{[\stmt,A'] \mapsto \aSem{\stmt,A'} (\lambda\,[u])  \}) & \\[0.5ex]
        & & \span (\stmt \textsf{ observable }, A' \in \aSemDigest{\stmt}\,A_0)
    \end{array}
\]
\begin{theorem}\label{t:refinesound}
  Refining a sound analysis with an admissible digest yields a sound analysis.
\end{theorem}
\begin{proofsketch}
  The proof proceeds by constructing a digest-refined concrete semantics
  and showing it equivalent to the original concrete semantics. Then, we show
  that soundness of the original analysis w.r.t. the non-refined semantics implies
  soundness of the refined analysis w.r.t.\ the refined semantics.
  We do not detail this proof here; see~\cite{schwarzthesis} for the details.
\end{proofsketch}

\subsection{Soundness Proof for the Phase-Lifted Analysis}

The soundness proof for the lifted analysis proceeds according to the following outline:
\begin{enumerate}
    \item We define a digest to track the current phases of the program a given thread has \emph{observed}.
    \item We show that this digest is admissible for programs to which the program transformation inserting
        reads and writes for globals on assignments to phase ghosts has been applied
        (\cref{prop:phase-ghost-digest-admissible}). From \cref{prop:phase-ghost-digest-admissible} and \cref{t:refinesound},
        we conclude that the digest-refined analysis is sound (\cref{thm:phase-ghost-digest-sound}).
    \item We relate this digest-refined analysis to the phase-lifted analysis from the main text,
        and show that from a solution of the phase-lifted analysis,
         one can construct a solution of the digest-refined analysis, implying that it is sound (\cref{t:lifting-sound}).
\end{enumerate}

We first define the phase-ghost digest and show that it is admissible.
\subsubsection{Phase-Ghost Digest and Its Admissibility}

To define the phase-ghost digest, we record the complete sequence of phase assignments observed for each thread.
For a local trace $t$ and a thread \emph{id} $i$ that owns a ghost,
i.e.,  $i \in \phaseGhostTIDs$, let $\mathsf{hist}_i(t)$ be the nonempty sequence consisting of the initial value of
$\ghost_i$, followed by the values assigned to $\ghost_i$ in $t$, in the order of the owning thread.
Every assignment contributes an element, including assignments that repeat the current value.
If no such assignment occurs in $t$, the sequence contains only the initial value.
\begin{definition}[Extended Phase Map]
    For each $i \in \phaseGhostTIDs$, let
    \[
        \mathcal H_i = \{ \mathsf{hist}_i(t) \mid t \in \Traces \}
    \]
    be the set of valid phase histories of thread $i$ in the transformed program $P'_{\ghost}$.
    An \emph{extended phase map}
    $\hat P \in \ExtendedPhaseMaps = \phaseGhostTIDs \to_i \mathcal H_i$
    maps each thread \emph{id} owning a phase ghost to a valid phase history.
\end{definition}
\begin{remark}
    The histories in an extended phase map are individually valid;
    for an arbitrary extended phase map, no trace containing all of these histories may exist.
\end{remark}
Write $h \preceq h'$ when $h$ is a prefix of $h'$, and further let
 $h \mathbin{\cdot}\langle v\rangle$ denote appending $v$,
and let $\mathsf{last}(h)$ denote taking the last element of $h$.

We define the phase-ghost digest $\Digests = \TIDs \times \ExtendedPhaseMaps$ that tracks the
(definite) thread \emph{id} of the executing thread (if there is one) and the extended phase map.
These histories describe phase assignments already \emph{observed} by the thread, rather than necessarily
all assignments that other threads have performed.
\begin{remark}
    Here, we assume that the thread \emph{id} digest does not change between program points that belong to the same
    thread. The proofs generalize to a setting where the thread \emph{id} digest contains some helper
    information that may evolve between program points of the same thread, though some of the arguments in the proof
    relating this refined analysis to the lifted analysis become more involved.
\end{remark}
For now, let us also assume that the assignments to phase ghosts are given as constants in the program.
We lift this restriction using concrete local traces in \cref{app:trace-digest-refinement}.

The abstraction function $\alpha_\Digests: \Traces \to \Digests$ is given by
\[
    \alpha_\Digests\,t =
    (\alpha_{\Digests_\TIDs}\,t,\{i \mapsto \mathsf{hist}_i(t) \mid i \in \phaseGhostTIDs\}),
\]
where $\alpha_{\Digests_\TIDs}\,t$ extracts the thread \emph{id} from $t$.
The initial digests are the images of the initial local traces; all their histories consist only of the initial phase.
For non-observing statements other than assignments to phase ghosts, we propagate the digest unchanged:
\begin{equation}
    \begin{array}{lll}
        \aSemDigest{\stmt} (i_0, \hat{P_0}) &=&
        \{(i_0, \hat{P_0})\}
    \end{array}
\end{equation}
For $\aSemDigest{\textsf{new}}$, the function returns the new thread \emph{id} as computed by the thread \emph{id}
component of the digest and copies the current extended phase map.
When an assignment to a phase ghost $\ghost_{i_0}$ occurs
(which, by definition of phase ghosts, only occurs for the current thread),
we update the phase map accordingly.
\begin{equation}
    \begin{array}{lll}
        \aSemDigest{\ghost_{i_0} = v}(i_0,\hat P_0) &=&
        \begin{cases}
        \{(i_0,\hat P_0 \oplus \{i_0 \mapsto (\hat P_0\,i_0)\mathbin{\cdot}\langle v\rangle\})\}
        & \text{if } (\hat P_0\,i_0)\mathbin{\cdot}\langle v\rangle \in \mathcal H_{i_0} \\
        \emptyset & \text{otherwise}
        \end{cases}
    \end{array}
\end{equation}
Thus, an assignment returns the empty set if its proposed history extension is not valid.
This cannot exclude a concrete successor, whose history belongs to $\mathcal H_{i_0}$ by definition.
The situation is more interesting for observing statements:
For observing statements $\stmt$ that do not correspond to locking atomicity mutexes or entering an atomic block,
we define:
\begin{equation}
    \begin{array}{lll}
        \aSemDigest{\stmt}(i_0,\hat P_0)\,(\_,\hat P_1) &=&
        \begin{cases}
            \emptyset & \text{if } i_0 \in \phaseGhostTIDs \land
                \hat P_1\,i_0 \not\preceq \hat P_0\,i_0,\\
            \emptyset & \text{if } \exists i \in \phaseGhostTIDs:
                \hat P_0\,i \not\preceq \hat P_1\,i \land \hat P_1\,i \not\preceq \hat P_0\,i,\\
            \{(i_0,\mathsf{merge}(\hat P_0,\hat P_1))\} & \text{otherwise},
        \end{cases}
    \end{array}
\end{equation}
where, for componentwise prefix-comparable histories,
\[
    \mathsf{merge}(\hat P_0,\hat P_1)(i) =
    \begin{cases}
        \hat P_1\,i & \text{if } \hat P_0\,i \preceq \hat P_1\,i,\\
        \hat P_0\,i & \text{otherwise}.
    \end{cases}
\]
Each resulting history is one of the input histories, so the result remains in $\ExtendedPhaseMaps$.
This definition accounts for the fact that one can, e.g., acquire mutexes whose last unlock
was in an earlier phase,
but respects that a thread can never learn about \emph{itself} having entered a new phase from making observations (first
$\emptyset$ case) and that two different phases cannot simultaneously be the $k$-th phase a given thread is currently in
(second $\emptyset$ case).
By construction, we know that reads of global variables can only happen from the same phase for the types of
programs we consider.
For locking an atomicity mutex outside an atomic block we define:
\begin{equation}\label{eq:phase-ghost-digest}
    \begin{array}{lll}
        \aSemDigest{\textsf{lock}\,(m_g)} (i_0, \hat{P_0})\,(\_, \hat{P_1}) &=&
        \begin{cases}
            \emptyset & \text{if } i_0 \in \phaseGhostTIDs \land \hat{P_0}\,i_0 \neq \hat{P_1}\,i_0 \\
            \emptyset & \text{if } \exists i \in \phaseGhostTIDs: \hat{P_0}\,i \not\preceq \hat{P_1}\,i,\\
            \{(i_0,\hat P_1)\} & \text{otherwise}.
        \end{cases}
    \end{array}
\end{equation}

The first case covers situations where the ego thread would learn from another thread that it had itself
made a step, which is not possible. The second case with an empty set of resulting digests covers the case where the
value to be read would be from a phase that is in the past or incompatible. This is the point where
we exploit the property that all values are read from the same phase.

For entering an atomic block, we likewise define:
\begin{equation}\label{eq:phase-ghost-digest-atomic}
    \begin{array}{lll}
        \aSemDigest{\atomicBegin} (i_0, \hat{P_0})\,(\_, \hat{P_1}) &=&
        \begin{cases}
            \emptyset & \text{if } i_0 \in \phaseGhostTIDs \land \hat{P_0}\,i_0 \neq \hat{P_1}\,i_0 \\
            \emptyset & \text{if } \exists i \in \phaseGhostTIDs: \hat{P_0}\,i \not\preceq \hat{P_1}\,i,\\
            \{(i_0,\hat P_1)\} & \text{otherwise}.
        \end{cases}
    \end{array}
\end{equation}

For an acquisition $\stmt=\lock{m_g}$ \emph{inside} a generated atomic block for a phase change,
we instead require equality with the current history map:
\begin{equation}\label{eq:phase-ghost-digest-inside}
    \aSemDigest{\stmt}(i_0,\hat P_0)\,(\_,\hat P_1)=
    \begin{cases}
        \{(i_0,\hat P_0)\} & \text{if }\hat P_0=\hat P_1,\\
        \emptyset & \text{otherwise}.
    \end{cases}
\end{equation}

\begin{proposition}\label{prop:phase-ghost-digest-admissible}
    The phase-encoding ghost digest is admissible for programs $P_{\ghost}'$ to which the program transformation inserting
    the reads and writes for globals at phase changes is applied.
\end{proposition}
\begin{proof}
    The proof is by verifying the admissibility conditions (\cref{def:adm}) for the phase-encoding ghost digest.
    For the initial digests, non-observing statements, and thread creation, this is straightforward.
    Let us consider the more interesting case of observing statements:
    \medskip

    First, for observing statements other
    than locking atomicity mutexes or entering an atomic block. Consider the case where two traces $t_0$ and $t_1$ are compatible, i.e.,
    there is some $t' \in \sem{\stmt}(\{t_0\},\{t_1\})$, and let
    $(i_0,\hat{P_0}) = \alpha_\Digests\,t_0$ and $(i_1,\hat{P_1}) = \alpha_\Digests\,t_1$.
    We distinguish three cases:
    \begin{itemize}
        \item Suppose $i_0\in\phaseGhostTIDs$ and $\hat P_1\,i_0\not\preceq\hat P_0\,i_0$.
            Since $t_0,t_1$ are compatible, their histories for the ego thread are prefixes of the same
            sequence of assignments. Hence $\hat P_0\,i_0$ must be a strict prefix of $\hat P_1\,i_0$.
            The trace $t_1$ would then contain an assignment performed by the ego thread after its current
            endpoint in $t_0$. Observing that assignment would reveal the ego thread's own future.
            Thus $\sem{\stmt}(\{t_0\},\{t_1\})=\emptyset$ (contradiction).
        \item Suppose that, for some $i\in\phaseGhostTIDs$, neither $\hat P_0\,i\preceq\hat P_1\,i$
            nor $\hat P_1\,i\preceq\hat P_0\,i$ holds.
            These histories disagree at some assignment position that occurs in both histories.
            They therefore cannot be prefixes of the assignment sequence of thread $i$ in one execution.
            Consequently, $t_0,t_1$ cannot be compatible, and $\sem{\stmt}(\{t_0\},\{t_1\})=\emptyset$ (contradiction).
        \item Otherwise, the histories are component-wise prefix-comparable.
            For each $i$, the assignments to $\ghost_i$ in $t'$ are the union of those in $t_0$ and $t_1$,
            with common assignments counted only once. Therefore,
            \[
                \mathsf{hist}_i(t')=
                \begin{cases}
                    \hat P_1\,i & \text{if }\hat P_0\,i\preceq\hat P_1\,i,\\
                    \hat P_0\,i & \text{otherwise}.
                \end{cases}
            \]
            Since the observing statement preserves the ego-thread \emph{id}, we obtain
            \[
                \begin{array}{lll}
                    \alpha_\Digests\,t'
                    &=& (\alpha_{\Digests_\TIDs}\,t',\{i\mapsto\mathsf{hist}_i(t')\mid i\in\phaseGhostTIDs\})\\
                    &=& (i_0,\{i\mapsto\mathsf{merge}(\hat P_0,\hat P_1)(i)\mid i\in\phaseGhostTIDs\})\\
                    &=& (i_0,\mathsf{merge}(\hat P_0,\hat P_1)).
                \end{array}
            \]
    \end{itemize}
    \medskip

    \noindent Next, consider the case of entering an atomic block.
    Let $t_0,t_1$ be compatible traces with $t'\in\sem{\atomicBegin}(\{t_0\},\{t_1\})$, and let
    $(i_0,\hat P_0)=\alpha_\Digests\,t_0$ and $(i_1,\hat P_1)=\alpha_\Digests\,t_1$.
    We examine the history-prefix and ego-thread rejection cases of \cref{eq:phase-ghost-digest-atomic},
    followed by the additional equality requirement inside generated blocks from \cref{eq:phase-ghost-digest-inside}:
    \begin{itemize}
        \item Suppose $\hat P_0\,i\not\preceq\hat P_1\,i$ for some $i\in\phaseGhostTIDs$.
            Incomparable histories are excluded by compatibility, as above.
            Thus $\hat P_1\,i$ must be a strict prefix of $\hat P_0\,i$.
            Consider the first assignment to $\ghost_i$ recorded in $\hat P_0\,i$ but missing from $\hat P_1\,i$.
            Observe that by construction of $P'_{\ghost}$, all assignments to phase ghosts are totally ordered by the sequence
            of $\atomicBegin$ and $\atomicEnd$ statements in the program.
            Thus, $\hat P_0$ records a \emph{later} assignment to a phase ghost than $\hat P_1$.
            Therefore, the trace $t_1$ cannot end in the \emph{last} action $\atomicEnd$, and the traces are thus
            not compatible, i.e.,
            $\sem{\atomicBegin}(\{t_0\},\{t_1\})=\emptyset$ (contradiction).
        \item Suppose $i_0\in\phaseGhostTIDs$ and $\hat P_0\,i_0\neq\hat P_1\,i_0$.
            The preceding case establishes $\hat P_0\,i_0\preceq\hat P_1\,i_0$ for compatible inputs.
            If these histories differ, the prefix is strict, so $t_1$ contains an assignment from the
            ego thread's own future. Once more, $\sem{\atomicBegin}(\{t_0\},\{t_1\})=\emptyset$ (contradiction).
        \item Otherwise, $\hat P_0\,i\preceq\hat P_1\,i$ for every $i\in\phaseGhostTIDs$.
            The longer history is always the one from $t_1$, so
            $\mathsf{merge}(\hat P_0,\hat P_1)=\hat P_1$ and
            \[
                \begin{array}{lll}
                    \alpha_\Digests\,t'
                    &=& (\alpha_{\Digests_\TIDs}\,t',\{i\mapsto\mathsf{hist}_i(t')\mid i\in\phaseGhostTIDs\})
                    = (i_0,\{i\mapsto\mathsf{merge}(\hat P_0,\hat P_1)(i)\mid i\in\phaseGhostTIDs\})\\
                    &=& (i_0,\hat P_1)
                    \in \aSemDigest{\atomicBegin}(i_0,\hat P_0)\,(\_,\hat P_1).
                \end{array}
            \]
    \end{itemize}

    \medskip
    Next, consider locking an atomicity mutex $m_g$ for a global $g$.
    Let $t_0,t_1$ be compatible traces with $t'\in\sem{\lock{m_g}}(\{t_0\},\{t_1\})$, and let
    $(i_0,\hat P_0)=\alpha_\Digests\,t_0$ and $(i_1,\hat P_1)=\alpha_\Digests\,t_1$.
    We examine the history-prefix and ego-thread rejection cases of \cref{eq:phase-ghost-digest},
    followed by the additional equality requirement inside generated blocks from \cref{eq:phase-ghost-digest-inside}:
    \begin{itemize}
        \item Suppose $\hat P_0\,i\not\preceq\hat P_1\,i$ for some $i\in\phaseGhostTIDs$.
            Incomparable histories are excluded by compatibility, as above.
            Thus $\hat P_1\,i$ must be a strict prefix of $\hat P_0\,i$.
            Consider the first assignment to $\ghost_i$ recorded in $\hat P_0\,i$ but missing from $\hat P_1\,i$.
            By construction of $P'_{\ghost}$, this assignment is instrumented with a read and write of $g$,
            acquiring $m_g$ before the phase assignment and releasing it afterwards.
            This instrumented access publishes a history containing the assignment.
            A trace ending in an unlock of $m_g$ whose history omits the assignment therefore cannot supply
            the last unlock preceding the current acquisition: it has been superseded by the instrumented access.
            Hence $\sem{\lock{m_g}}(\{t_0\},\{t_1\})=\emptyset$, a contradiction.
        \item Suppose $i_0\in\phaseGhostTIDs$ and $\hat P_0\,i_0\neq\hat P_1\,i_0$.
            The preceding case establishes $\hat P_0\,i_0\preceq\hat P_1\,i_0$ for compatible inputs.
            If these histories differ, the prefix is strict, so $t_1$ contains an assignment from the
            ego thread's own future. As above, this contradicts compatibility.
            Thus, the ego-thread histories must agree.
        \item Suppose $\stmt$ is an acquisition inside a generated atomic block,
            and $\hat P_0\neq\hat P_1$.
            From the previous cases, we have $\hat P_0\,i \preceq\hat P_1\,i$ for every $i\in\phaseGhostTIDs$.
            Consider thus one $i$ for which $\hat P_0\,i \neq \hat P_1\,i$.
            In this case, there is a phase assignment that is part of $t_1$ that is not part of $t_0$. Phase assignments
            only happen when inside an atomic section and when holding all atomicity mutexes.
            The ego thread in $t_0$ is currently inside an atomic block. Thus an additional assignment must either
            have been recorded in $t_0$ (contradiction), occur after $t_0$ releases the atomicity mutex again
            (contradiction), or belong to a different execution.
            Thus, $\sem{\lock{m_g}}(\{t_0\},\{t_1\})=\emptyset$ (contradiction).
        \item Otherwise, $\hat P_0\,i\preceq\hat P_1\,i$ for every $i\in\phaseGhostTIDs$.
            The longer history is always the one from $t_1$, so
            $\mathsf{merge}(\hat P_0,\hat P_1)=\hat P_1$ and
            \[
                \begin{array}{lll}
                    \alpha_\Digests\,t'
                    &=& (\alpha_{\Digests_\TIDs}\,t',\{i\mapsto\mathsf{hist}_i(t')\mid i\in\phaseGhostTIDs\})\\
                    &=& (i_0,\{i\mapsto\mathsf{merge}(\hat P_0,\hat P_1)(i)\mid i\in\phaseGhostTIDs\})\\
                    &=& (i_0,\hat P_1) \in \aSemDigest{\lock{m_g}}(i_0,\hat P_0)\,(\_,\hat P_1).
                \end{array}
            \]
    \end{itemize}
    The proof proceeds analogously for the remaining cases.
\end{proof}
\begin{theorem}\label{thm:phase-ghost-digest-sound}
    The solution $(\lambda^\sharp, \mu)$ of the digest-refined analysis with unknowns $[u,(i,\hat{P})]$ and $[act,(i,\hat{P})]$
    is sound w.r.t.\ the local trace semantics of the program $P_{\ghost}'$.
\end{theorem}
\begin{proof}
    By \cref{prop:phase-ghost-digest-admissible}, the soundness of the original analysis, and the soundness of
    refining a sound analysis with an admissible digest~(\cref{t:refinesound}).
\end{proof}
\subsubsection{Relating the phase-lifted analysis to the digest-refined analysis}
It thus remains to relate solutions of the phase-lifted analysis to the digest-refined analysis. To this end,
we observe that the lifted analysis is \emph{eager} in the phase changes, i.e., it overapproximates the phase changes
by eagerly advancing to the next phase, whereas the digest-refined analysis only advances to the next phase when
a local trace from some other thread is incorporated where that phase change has already been observed.

For the following reachability argument, we use the coarse phase-successor relation of
\cref{eq:succs,eq:psuccs}, before its refinement by concurrency contexts.
We require the collected relation $[\Phases^\rightarrow_j]$ to contain every concrete change from one
phase value to another at an assignment to $\ghost_j$.

\begin{lemma}\label{lem:history-phase-reachability}
    Suppose the collected phase-transition relations satisfy this soundness requirement.
    If $h,h'\in\mathcal H_j$ and $h\preceq h'$, then $\mathsf{last}(h')$ is reachable from
    $\mathsf{last}(h)$ in $[\Phases^\rightarrow_j]$ by zero or more transitions.
\end{lemma}
\begin{proof}
    By definition of $\mathcal H_j$, some concrete local trace has history $h'$ for thread $j$.
    The suffix extending $h$ consists of assignments occurring in that trace.
    Each pair of consecutive distinct phase values is therefore included in $[\Phases^\rightarrow_j]$.
    Repeated values require no transition for reachability, and an empty suffix gives reflexive reachability.
\end{proof}

Putting things together, we obtain:

\begin{theorem}\label{t:lifting-sound}
    Consider a sound base analysis with monotonic transfer functions that admits refinement by digests.
    Assume sound collected phase-transition relations and the coarse phase-successor relation described above.
    Then, the lifted analysis with phase-lifted domain and transfer functions is also sound.
\end{theorem}
\begin{proof}
We present the argument for Scott-continuous transfer functions,
so that the least solution is obtained as the limit of Kleene iterates.
For merely monotone transfer functions, the same construction applies using the transfinite Kleene sequence.
As the transfer functions of the lifted analysis are assembled from
the transfer functions of the base analysis, they are also Scott-continuous.
A \emph{least} solution of the constraint system $C_{\PhaseMaps}$ of the phase-lifted analysis
exists and can be computed by Kleene iteration.
Let $(\lambda_{\PhaseMaps}, \mu_{\PhaseMaps})$ be this \emph{least} solution. 
We construct from it a solution $(\lambda, \mu)$ of the digest-refined analysis (with constraint system $C$) by setting:
\[
    \begin{array}{lll}
        \lambda [u,(i,\hat{P})] &=& (\lambda_{\PhaseMaps}\,[u])\,(\projToPhaseMap\,\hat{P})\\
        \mu\,[\stmt,(i,\hat{P})] &=& (\mu_{\PhaseMaps}\,[\stmt])\,(\projToPhaseMap\,\hat{P})
    \end{array}
\]
where $\projToPhaseMap\,\hat{P} = \{ i \mapsto \mathsf{last}(\hat{P}\,i) \mid i \in \phaseGhostTIDs \}$,
i.e., it projects each history to its last phase. The construction only indexes valid histories from $\mathcal H_i$;
it does not assign values to arbitrary sequences ending in the same phase.

We then show that $(\lambda, \mu)$ as constructed above is a solution of the digest-refined analysis
by inductively checking that all constraints of the digest-refined analysis are satisfied by $(\lambda, \mu)$.
    We first consider the constraints corresponding to initialization:
    the digest-refined analysis starts only in the initial digest $(i_0,\hat{P}_0)$.
    Thus, the only initialization constraint to check in $C$ is
    \[
        (\lambda\,[u_0,(i_0,\hat{P}_0)],\mu)
        \sqsupseteq
        (\sigma_0^\sharp,\mu_0).
    \]
    The corresponding constraint in $C_{\PhaseMaps}$ is
    \[
        (\lambda_{\PhaseMaps}\,[u_0],\mu_{\PhaseMaps})
        \sqsupseteq
        \lift_{\bar{P}_0}(\sigma_0^\sharp,\mu_0).
    \]
    By unfolding $\lift$, we obtain
    \[
        (\lambda_{\PhaseMaps}\,[u_0])\,\bar{P}_0
        \sqsupseteq
        \sigma_0^\sharp
    \]
    and, for every initialized global unknown $\relax[x]$,
    \[
        (\mu_{\PhaseMaps}\,[x])\,\bar{P}_0
        \sqsupseteq
        \mu_0\,[x].
    \]
    As $\projToPhaseMap\,\hat{P}_0 = \bar{P}_0$ and by construction
    of $(\lambda,\mu)$, this is precisely
    \[
        \lambda\,[u_0,(i_0,\hat{P}_0)] \sqsupseteq \sigma_0^\sharp
        \qquad\text{and}\qquad
        \mu\,[x,(i_0,\hat{P}_0)] \sqsupseteq \mu_0\,[x],
    \]
    so the initialization constraints are satisfied.

    Consider first a local action $\stmt$ along an edge $(u,\stmt,v)$ which does not assign to the phase ghost.
    In the concrete and in
    the digest-refined analysis, the phase does not change here, whereas it potentially does in the lifted analysis.
    The constraint of $C$ is then given by
    \[
        (\lambda\,[v,(i,\hat{P})],\mu) \sqsupseteq \aSem{\stmt} (\lambda\,[u,(i,\hat{P})])\,\mu
    \]
    Consider the corresponding constraint in $C_{\PhaseMaps}$, which is given by
    \[
        (\lambda_{\PhaseMaps}\,[v],\mu_{\PhaseMaps}) \sqsupseteq \aSemP{\stmt} (\lambda_{\PhaseMaps}\,[u])\,\mu_{\PhaseMaps}
    \]
    with
    \[
    \begin{array}{lll}
        \aSemP{\stmt}\,\sigma^\sharp_{\PhaseMaps}\,\mu_{\PhaseMaps} &=&
        \bigsqcup \{
            \succs\,\bar P\,\sigma^\sharp\,(\unlift_{\bar P}\,\mu_{\PhaseMaps})

            \mid

            ({\bar P} \mapsto \sigma^\sharp) \in \sigma^\sharp_{\PhaseMaps}
        \}
    \end{array}
    \]
    Let $\bar P = \projToPhaseMap\,\hat{P}$.
    By construction of $(\lambda,\mu)$, we have
    $(\lambda_{\PhaseMaps}\,[u])\,\bar{P} = \lambda\,[u,(i,\hat{P})]$.
    Consider thus one element of the join over
    \[
        \succs\,\bar P\,(\lambda\,[u,(i,\hat{P})])\,(\unlift_{\bar P}\,\mu_{\PhaseMaps}),
    \] namely the one where the phase map stays the same, i.e., $\bar P$.
    If the statement is inside an atomic block, this is the only returned element (\cref{fig:lift-ordinary}); otherwise
    it appears in the join in \cref{fig:lift-ordinary} because
    $\forall \sigma^\sharp \in \Domain: \bar P \in \psuccs\,\bar P\,\sigma^\sharp$
    by \cref{eq:succs,eq:psuccs}.
    \[
        \lift_{{\bar P}}\,(\aSem{\stmt}\,(\lambda\,[u,(i,\hat{P})])\,(\unlift_{\bar P}\,\mu_{\PhaseMaps})),
    \]
    which, because executing a local statement neither accesses any global unknowns nor contributes to them,
    equals
    \[
        \lift_{{\bar P}}(\aSem{\stmt} (\lambda\,[u,(i,\hat{P})])\,\mu).
    \]
    We thus have
    \[
        \begin{array}{lll}
        \lambda\,[v,(i,\hat{P})] &=&
        (\lambda_{\PhaseMaps}\,[v])\,\bar{P}\\
        &\sqsupseteq& (\aSemP{\stmt} (\lambda_{\PhaseMaps}\,[u])\,\mu_{\PhaseMaps})_1\,\bar{P}\\
        &\sqsupseteq& (\succs\,\bar P\,(\lambda\,[u,(i,\hat{P})])\,(\unlift_{\bar P}\,\mu_{\PhaseMaps}))_1\,\bar{P}\\
        &\sqsupseteq& (\lift_{{\bar P}}(\aSem{\stmt} (\lambda\,[u,(i,\hat{P})])\,\mu))_1\,\bar{P}\\
        &=& (\aSem{\stmt} (\lambda\,[u,(i,\hat{P})])\,\mu)_1
        \end{array}
    \]
    Recall that $(\cdot)_k$ denotes the $k$-th component of the pair.
    The constraint holds for the first element of the pair, and as no contributions to global unknowns are made, it also holds overall.

\medskip
\noindent
    Next, consider an \emph{observing} action $\stmt$ along an edge $(u,\stmt,v)$ outside an atomic block
    introduced by the ghost instrumentation. This action observes a statement $\stmt'$.
    Consider a constraint of $C$ of the shape
    \[
        (\lambda\,[v,(i,\hat{P}'')],\mu) \sqsupseteq \aSem{\stmt} (\lambda\,[u,(i,\hat{P})])\,\mu
    \]
    where a specific $[\stmt',(i',\hat{P}')]$ from $\mu$ is accessed and
    $\aSemDigest{\stmt}((i,\hat{P}),(i',\hat{P}')) = \{ (i,\hat{P}'')\}$.
    We rewrite the right-hand side of this constraint as
    \[
        \aSem{\stmt} (\lambda\,[u,(i,\hat{P})])\,\mu
        = \aSem{\stmt}_{I} (\lambda\,[u,(i,\hat{P})])\,(\mu\,[\stmt',(i',\hat{P}')])
    \]
    The corresponding constraint in $C_{\PhaseMaps}$ is given by
    \[
        (\lambda_{\PhaseMaps}\,[v],\mu_{\PhaseMaps}) \sqsupseteq \aSemP{\stmt} (\lambda_{\PhaseMaps}\,[u])\,\mu_{\PhaseMaps}
    \]
    with
    \[
    \begin{array}{lll}
        \aSemP{\stmt}\,\sigma^\sharp_{\PhaseMaps}\,\mu_{\PhaseMaps} &=&
        \bigsqcup \{
            \succs\,\bar P\,\sigma^\sharp\,(\unlift_{\bar P}\,\mu_{\PhaseMaps})

            \mid

            ({\bar P} \mapsto \sigma^\sharp) \in \sigma^\sharp_{\PhaseMaps}
        \}
    \end{array}
    \]
    For $\bar{P} = \projToPhaseMap\,\hat{P}$, $\bar{P}' = \projToPhaseMap\,\hat{P}'$,
    and $\bar{P}'' = \projToPhaseMap\,\hat{P}''$,
    by construction of $(\lambda,\mu)$, we have
    $(\lambda_{\PhaseMaps}\,[u])\,\bar{P} = \lambda\,[u,(i,\hat{P})]$ and
    $\mu_{\PhaseMaps}\,[\stmt']\,\bar{P}' = \mu\,[\stmt',(i',\hat{P}')]$.
    Consider the case where the unknown $[\stmt']$ satisfies $\phaseBound$, which, for our analysis,
    is the case for statements $\stmt'$ corresponding to unlocking atomicity mutexes which are observed
    by statements $\stmt$ corresponding to locking the respective atomicity mutex.

    Let us consider the more challenging case where $\hat P \neq \hat{P}'$ and
    $\bar P \neq \bar{P}'$.
    The property to establish is that
    \[
        \lambda\,[v,(i,\hat{P}'')] \sqsupseteq
        (\aSem{\stmt}_{I} (\lambda\,[u,(i,\hat{P})])\,(\mu\,[\stmt',(i',\hat{P}')]))_1
    \]
    As the global satisfies $\phaseBound$, the only global unknown considered by the phase-lifted transfer at phase map
    $\bar{P}'$ is the one associated with $\bar{P}'$.
    By construction of $(\lambda,\mu)$, this is the same observable contribution as in the digest-refined constraint:
    $(\unlift_{{\bar P}'}\,\mu_{\PhaseMaps})\,[\stmt'] = \mu\,[\stmt',(i',\hat{P}')]$.

    We show that the phase-lifted analysis, as it eagerly advances phases, contains at
    $(\lambda_{\PhaseMaps}\,[u])\,\bar{P}'$ a value that accounts for the value
    at $(\lambda\,[u,(i,\hat{P})])$ in the digest-refined analysis.

    \medskip

    To show that $(\lambda_{\PhaseMaps}\,[u])\,\bar{P}'$ subsumes $(\lambda\,[u,(i,\hat{P})])$,
    we consider a closure property for Kleene-iterates of the phase-lifted analysis:

    Let $\sigma^\sharp_e=(\lambda_{\PhaseMaps}\,[u])\,\bar P$.
    From $\aSemDigest{\stmt}((i,\hat{P}),(i',\hat{P}')) \neq \emptyset$, by \cref{eq:phase-ghost-digest}, we have that
    $\hat P\,j \preceq \hat P'\,j$ for every $j\in\phaseGhostTIDs$, and the ego-thread histories agree
    whenever the ego thread owns a phase ghost  (and thus also in $\bar{P}$ and $\bar{P}'$).
    By \cref{lem:history-phase-reachability}, for every $j$ there is a path in the collected phase-transition
    relation from $\bar P\,j$ to $\bar P'\,j$.
    Whenever the abstract state identifies a definite phase-ghost owner as the ego thread, that identity
    agrees with the thread-id component $i$ of the digest. The ego-thread condition in $\psuccs$ therefore
    fixes precisely a component on which $\bar P$ and $\bar P'$ agree.
    If no definite phase-ghost owner is identified, that condition imposes no restriction.
    By the coarse definition in \cref{eq:succs,eq:psuccs}, we therefore have
    $\bar{P}' \in \psuccs\,\bar{P}\,\sigma^\sharp_e$.
    Thus, the lifted analysis has already performed all eager phase advances. More precisely, for all Kleene iterates
    $(\lambda^i_{\PhaseMaps},\mu^i_{\PhaseMaps})$ starting at $(\bot,\bot)$, the following entry-closure invariant
    holds:
    \[
        \bar{Y} \in \psuccs\,\bar X\,((\lambda^i_{\PhaseMaps}\,[u])\,\bar{X}) \implies
        (\lambda^i_{\PhaseMaps}\,[u])\,\bar{X} \sqsubseteq (\lambda^i_{\PhaseMaps}\,[u])\,\bar{Y}
    \]
    for every non-initial program point $u$ that is not inside an atomic block.
    This invariant is established by induction over the Kleene iterates: whenever a contribution reaches such an
    entry point through a non-atomic incoming edge, the non-atomic branch of $\succs$ emits the successor contribution
    at every phase map in $\psuccs$; transitivity of $\psuccs$ gives the displayed closure.

    Altogether, we then get
    \[
        \begin{array}{lll}
        \lambda\,[v,(i,\hat{P}'')] &=&
        (\lambda_{\PhaseMaps}\,[v])\,\bar{P}'' \\
        &\sqsupseteq& (\aSemP{\stmt} (\lambda_{\PhaseMaps}\,[u])\,\mu_{\PhaseMaps})_1\,\bar{P}'' \\
        &\sqsupseteq& ((\succs\,\bar P'\,((\lambda_{\PhaseMaps}\,[u])\,\bar{P}')\,(\unlift_{{\bar P}'}\,\mu_{\PhaseMaps}))_1)\,\bar{P}'' \\
        &\sqsupseteq& (\lift_{\bar{P}''} (\aSem{\stmt}((\lambda_{\PhaseMaps}\,[u])\,\bar{P}')\,(\unlift_{{\bar P}'}\,\mu_{\PhaseMaps})))_1\,\bar{P}'' \\
        &=& (\aSem{\stmt}((\lambda_{\PhaseMaps}\,[u])\,\bar{P}')\,(\unlift_{{\bar P}'}\,\mu_{\PhaseMaps}))_1 \\
        &=& (\aSem{\stmt}_I ((\lambda_{\PhaseMaps}\,[u])\,\bar{P}')\,((\unlift_{{\bar P}'}\,\mu_{\PhaseMaps})\,[\stmt']))_1 \\
        &\sqsupseteq& (\aSem{\stmt}_I (\lambda\,[u,(i,\hat{P})])\,(\mu\,[\stmt',(i',\hat{P}')]))_1 \\
        \end{array}
    \]
    Thus, the property holds for the first element of the pair, and as no contributions to global unknowns are made, it also holds overall.
    The proof for the case where $\bar P = \bar{P}'$ is obtained as a simplification of the above proof,
    as is the case where the global does not satisfy $\phaseBound$.

    \medskip
    For locking of atomicity mutexes inside an atomic block, we observe that the entry-closure invariant is established
    when the atomic block is entered. All phase-bound reads refer to the same fixed phase, and no further phase changes
    are possible inside the block. Thus, subsumption upon entry into the block is sufficient.

    \medskip
    The proof continues analogously for the remaining transfer functions.
\end{proof}

\begin{remark}
    The proof for analyses that do not admit refinement by digests is more involved.
    For a protection-based analysis, this proof can be obtained by showing that from the result of a protection-based
    analysis, one can construct a solution of a mutex-meet-style analysis:
    Protection-based analyses~\cite{Vojdani2010,schwarzthesis} associate each global variable $g$ with a set
    of protecting mutexes, every one of which is held whenever $g$ is written.
    They then accumulate two invariants for every global variable:
    The values that can be read when holding one of the
    protecting mutexes (\emph{protected invariant}) and those that can be read without holding any of the protecting mutexes
    (\emph{unprotected invariant}).
    To prove soundness of such an analysis, we first perform a program transformation:
    We introduce a dedicated auxiliary read-write mutex $p_{\mathrm{g}}$
    for each global variable $g$ with a nonempty protecting set.
    If the set contains more than one mutex, a thread acquires $p_{\mathrm{g}}$ in read mode
    immediately after acquiring its first protecting mutex for $g$.
    It upgrades $p_{\mathrm{g}}$ to write mode immediately after acquiring the last missing protecting mutex.
    Before releasing the first protecting mutex from the complete set, it downgrades $p_{\mathrm{g}}$
    to read mode. It releases $p_{\mathrm{g}}$ immediately before releasing its last held protecting mutex.
    For a singleton protecting set, the thread instead acquires $p_{\mathrm{g}}$ directly in write mode
    after acquiring the protecting mutex and releases it before releasing that mutex.
    For an empty protecting set, no auxiliary mutex is introduced; all accesses are unprotected
    and remain modeled through the atomicity mutex $m_g$.
    When a thread holds all protecting mutexes for $g$, no other thread can hold any of them.
    Since every holder of $p_{\mathrm{g}}$ also holds a protecting mutex, no other thread can then
    hold $p_{\mathrm{g}}$ in either mode. Thus the upgrade, or the direct write-mode acquisition
    in the singleton case, can proceed immediately. Shared acquisitions likewise cannot conflict
    with a writer, since such a writer would hold every protecting mutex.
    This transformation preserves the validity of assertions over the original program variables.

    We can then translate a solution of the phase-refined protection-based analysis into a solution of the phase-refined
    mutex-meet analysis (which can be shown to be sound for read/write mutexes) of the transformed program.
    To this end, for each nonempty protecting set, we associate the protected invariant with $p_{\mathrm{g}}$.
    For every global $g$, we associate the unprotected invariant with the atomicity mutex $m_g$.
    We set unknowns for all other mutexes to $\top$.
    Soundness of the phase-lifted protection-based analysis then follows from the theorem
    establishing soundness of phase-lifted mutex-meet analysis.
\end{remark}

\subsubsection{Refinement by concrete local traces}\label{app:trace-digest-refinement}
Non-constant assignments to phase ghosts and MHP-sensitive phase successors require information
beyond the phase histories. We use the product of the phase-history digest and the concrete local
trace digest for both extensions. Local traces are themselves an admissible digest, and products
of admissible digests are admissible~\cite{hammerandnail}.
The reachable indices have the form $((i,\hat P),t)$ with
$(i,\hat P)=\alpha_\Digests\,t$.
This digest serves as an intermediate step in the proof, so it is not an issue that it is typically
not computable.

We remark that, for an observing action, the trace component admits a successor only when
\[
    t''\in\sem{\stmt}(\{t\},\{t'\}).
\]
This supplies compatibility of the concrete traces, which does not follow from acceptance by the
phase-history digest alone. This is a key ingredient for the MHP reasoning in \cref{app:mhp-soundness}.

\paragraph{Non-constant assignments.}
For an assignment $\ghost_{i_0}=e$, the trace determines the values of all variables on which $e$
depends. We extend the history with the value obtained by evaluating $e$ in that trace:
\[
    \begin{array}{lll}
        \aSemDigest{\ghost_{i_0} = e} ((i_0,\hat P_0),t)
        &=& \Let\;\{t'\}=\sem{\ghost_{i_0} = e}\,t\;\In \\
        && \{((i_0,\hat P_0\oplus\{i_0\mapsto h\}),t')
            \mid h=(\hat P_0\,i_0)\mathbin{\cdot}\langle\semInt{e}(t)\rangle,
            \ h\in\mathcal H_{i_0}\}.
    \end{array}
\]
Here, $\semInt{e}(t)$ denotes concrete expression evaluation in the local trace $t$.
For a concrete successor $t'$, its history is exactly the old history with this value appended
and belongs to $\mathcal H_{i_0}$ by definition. Thus the assignment satisfies the admissibility
condition for the product digest. Other statements propagate the phase-history component as before
and update the trace component according to the concrete semantics.

The proof of \cref{t:lifting-sound} for this generalization proceeds in the same way by showing that
\[
    \begin{array}{lll}
        \lambda [u,((i,\hat{P}),t)] &=& (\lambda_{\PhaseMaps}\,[u])\,(\projToPhaseMap\,\hat{P})\\
        \mu\,[\stmt,((i,\hat{P}),t)] &=& (\mu_{\PhaseMaps}\,[\stmt])\,(\projToPhaseMap\,\hat{P})
    \end{array}
\]
is a solution of the digest-refined analysis also considering the local trace component.
Just as the thread \emph{id} component of the digest was not used in the proof,
the same is true in the extended proof for the local trace component of the digest,
except that the auxiliary property that for any $\bar P=\projToPhaseMap\,\hat P_0$ and
$\sigma^\sharp=(\lambda_{\PhaseMaps}\,[u])\,\bar P$, $\semInt{e}(t)\in\aSemInt{e}\,\sigma^\sharp$
is also established to ensure that the lifted assignment includes the phase-map update
selected by the trace component.

\subsubsection{MHP-sensitive phase successors}\label{app:mhp-soundness}
The argument of \cref{t:lifting-sound} uses the coarse phase-successor relation.
We treat the MHP refinement separately, using concrete local traces to relate history extensions
to the observer's concurrency context.
For the MHP refinement of \cref{ss:mhp}, the corresponding reachability statement additionally
requires that the transitions in the history suffix are possible in the concurrency context being queried.
\begin{lemma}\label{lem:history-phase-reachability-mhp}
    Assume that the collection of context-labelled transitions is sound in the following sense:
    For every concrete phase change from $a$ to $b$ in some concrete state $\sigma$ of thread $j$,
    there is an element $(a,c,b)\in[\Phases^\rightarrow_j]$ such that there is some $\sigma^\sharp$
    with $\sigma\in\gamma\,\sigma^\sharp$ and $\projConcCtx\,\sigma^\sharp = c$.
    Furthermore, assume the MHP predicate ($\textsf{may\_jointly\_active}$) to be sound, i.e.,
    it is true whenever threads with the given concurrency contexts can be jointly active in some concrete execution.
    Then, let $h,h'\in\mathcal H_j$ with $h\preceq h'$, and let $\sigma^\sharp$ be an abstract state
    of another thread. If the assignments extending $h$ to $h'$ occur in a concrete execution
    with the other thread staying at a program point with a concurrency context
    represented by $\projConcCtx\,\sigma^\sharp$, then
    \[
        \mathsf{last}(h')\in\psuccs_j\,(\mathsf{last}(h))\,\sigma^\sharp
    \]
    for the MHP-sensitive definition of $\psuccs_j$.
\end{lemma}
\begin{proof}
    Write $a_0=\mathsf{last}(h)$ and let $a_1,\ldots,a_n$ be the additional phase values
    in the suffix extending $h$ to $h'$, where $a_n=\mathsf{last}(h')$.
    If $n=0$, the conclusion follows by reflexivity.
    For every $0\leq k<n$ with $a_k\neq a_{k+1}$, there is some $c_k^\sharp$ such that
    $(a_k,c_k^\sharp,a_{k+1})\in[\Phases^\rightarrow_j]$.
    At that assignment, thread $j$ and the stationary thread are jointly active in contexts
    represented by $c_k^\sharp$ and $\projConcCtx\,\sigma^\sharp$, respectively. Soundness of the MHP predicate gives
    \[
        \textsf{may\_jointly\_active}\,c_k^\sharp\,(\projConcCtx\,\sigma^\sharp).
    \]
    Consequently, $a_k\to_{j,\sigma^\sharp}a_{k+1}$ by \cref{ss:mhp}.
    Repeated consecutive values require no transition.
    Altogether, we have $\mathsf{last}(h')\in\psuccs_j\,(\mathsf{last}(h))\,\sigma^\sharp$.
\end{proof}

Our implementation maintains path sensitivity in concurrency contexts for values at unknowns associated
with program points. Thus, instead of maintaining a single value from $\Domain$ for each phase map entry,
we maintain a set of domain elements for each entry. The partition retains all information used by $\psuccs$,
including the concurrency context and the ego-thread information. Two elements are joined only when their
partition keys coincide, and the join preserves this information. The function $\psuccs$ is applied to each
element separately, with results joined only within the same partition.
In the following argument, local abstract values and their comparisons are understood componentwise for a
fixed partition key; we suppress the component index. Successor queries use this component's context and
ego-thread information. Transfers that change the context or ego-thread information place their results
in the corresponding output component. This convention applies throughout the argument, including the choice
of $\sigma^\sharp_e$, the application of \cref{lem:history-phase-reachability-mhp}, the entry-closure invariant,
and the final transfer comparison.

The proof then proceeds analogously to the proof of \cref{t:lifting-sound}, using the MHP-sensitive
phase-successor relation in place of the coarse phase-successor relation. The case of \emph{observing}
actions is of particular interest here:

\medskip
\noindent
Consider once again an \emph{observing} action $\stmt$ along an edge $(u,\stmt,v)$ outside an atomic block
introduced by the ghost instrumentation. This action observes a statement $\stmt'$.
Consider a constraint of $C$ of the shape
\[
    (\lambda\,[v,((i,\hat{P}''),t'')],\mu) \sqsupseteq \aSem{\stmt} (\lambda\,[u,((i,\hat{P}),t)])\,\mu
\]
where a specific $[\stmt',((i',\hat{P}'),t')]$ from $\mu$ is accessed and
$\aSemDigest{\stmt}(((i,\hat{P}),t),((i',\hat{P}'),t')) = \{ ((i,\hat{P}''), t'')\}$.
We rewrite the right-hand side of this constraint as
\[
    \aSem{\stmt} (\lambda\,[u,((i,\hat{P}),t)])\,\mu
    = \aSem{\stmt}_{I} (\lambda\,[u,((i,\hat{P}),t)])\,(\mu\,[\stmt',((i',\hat{P}'),t')])
\]
The corresponding constraint in $C_{\PhaseMaps}$ is given by
\[
    (\lambda_{\PhaseMaps}\,[v],\mu_{\PhaseMaps}) \sqsupseteq \aSemP{\stmt} (\lambda_{\PhaseMaps}\,[u])\,\mu_{\PhaseMaps}
\]
with
\[
\begin{array}{lll}
    \aSemP{\stmt}\,\sigma^\sharp_{\PhaseMaps}\,\mu_{\PhaseMaps} &=&
    \bigsqcup \{
        \succs\,\bar P\,\sigma^\sharp\,(\unlift_{\bar P}\,\mu_{\PhaseMaps})

        \mid

        ({\bar P} \mapsto \sigma^\sharp) \in \sigma^\sharp_{\PhaseMaps}
    \}
\end{array}
\]
For $\bar{P} = \projToPhaseMap\,\hat{P}$, $\bar{P}' = \projToPhaseMap\,\hat{P}'$,
and $\bar{P}'' = \projToPhaseMap\,\hat{P}''$,
by construction of $(\lambda,\mu)$, we have
$(\lambda_{\PhaseMaps}\,[u])\,\bar{P} = \lambda\,[u,((i,\hat{P}),t)]$ and
$\mu_{\PhaseMaps}\,[\stmt']\,\bar{P}' = \mu\,[\stmt',((i',\hat{P}'),t')]$.
Consider the case where the unknown $[\stmt']$ satisfies $\phaseBound$, which, for our analysis,
is the case for statements $\stmt'$ corresponding to unlocking atomicity mutexes which are observed
by statements $\stmt$ corresponding to locking the respective atomicity mutex.

Let us consider the more challenging case where $\hat P \neq \hat{P}'$ and
$\bar P \neq \bar{P}'$.
The property to establish is that
\[
    \lambda\,[v,((i,\hat{P}''),t'')] \sqsupseteq
    (\aSem{\stmt}_{I} (\lambda\,[u,((i,\hat{P}),t)])\,(\mu\,[\stmt',((i',\hat{P}'),t')]))_1
\]
As the global satisfies $\phaseBound$, the only global unknown considered by the phase-lifted transfer at phase map
$\bar{P}'$ is the one associated with $\bar{P}'$.
By construction of $(\lambda,\mu)$, this is the same observable contribution as in the digest-refined constraint:
$(\unlift_{{\bar P}'}\,\mu_{\PhaseMaps})\,[\stmt'] = \mu\,[\stmt',((i',\hat{P}'),t')]$.

We show that the phase-lifted analysis, as it eagerly advances phases, contains at
$(\lambda_{\PhaseMaps}\,[u])\,\bar{P}'$ a value that accounts for the value
at $(\lambda\,[u,((i,\hat{P}),t)])$ in the digest-refined analysis.

\medskip

To show that $(\lambda_{\PhaseMaps}\,[u])\,\bar{P}'$ subsumes $(\lambda\,[u,((i,\hat{P}),t)])$,
we consider a closure property for Kleene-iterates of the phase-lifted analysis:

Let $\sigma^\sharp_e=(\lambda_{\PhaseMaps}\,[u])\,\bar P$.
From $\aSemDigest{\stmt}(((i,\hat{P}),t),((i',\hat{P}'),t')) \neq \emptyset$, by \cref{eq:phase-ghost-digest}, we have that
$\hat P\,j \preceq \hat P'\,j$ for every $j\in\phaseGhostTIDs$, and the ego-thread histories agree
whenever the ego thread owns a phase ghost  (and thus also in $\bar{P}$ and $\bar{P}'$).
It remains to show that $\bar{P}' \in \psuccs\,\bar{P}\,\sigma^\sharp_e$.
To this end, by \cref{lem:history-phase-reachability-mhp}, we need to show that for every
$j \in \phaseGhostTIDs$, where $\hat P\,j \neq \hat P'\,j$,
the assignments extending $\hat P\,j$ to $\hat P'\,j$ occur in a concrete execution with the
ego thread staying at the program point $u$ with a
concurrency context represented by $\projConcCtx\,\sigma^\sharp_e$.
Fix a component with $\hat P\,j\neq\hat P'\,j$, and consider an assignment event $a$
in the suffix extending $\hat P\,j$ to $\hat P'\,j$.
The following cases concern causal ordering in the compatible combined trace $t''$,
not the chronological order of one particular interleaving:
\begin{itemize}
    \item If $a$ causally precedes an event already in $t$, it belongs to the causal history
        recorded by $t$. Its assignment would therefore already occur in $\hat P\,j$ (contradiction).
    \item If $a$ causally follows the new observing action $\stmt$, then incorporating $t'$
        at that action would require observing an event from the observer's own future.
        Since $a$ is in the history being incorporated, this would create a causal cycle,
        contradicting compatibility of $t$ and $t'$.
    \item Otherwise, the missing assignment is neither required before an event already in $t$
        nor dependent on the new observing action. It may thus occur in a concrete execution
        with the ego thread staying at the program point $u$ with a concurrency context represented by
        $\projConcCtx\,\sigma^\sharp_e$.
\end{itemize}
We thus have $\bar{P}' \in \psuccs\,\bar{P}\,\sigma^\sharp_e$.
Thus, the lifted analysis has already performed the necessary phase advances. More precisely, for all Kleene iterates
$(\lambda^i_{\PhaseMaps},\mu^i_{\PhaseMaps})$ starting at $(\bot,\bot)$, the following entry-closure invariant
holds:
\[
    \bar{Y} \in \psuccs\,\bar X\,((\lambda^i_{\PhaseMaps}\,[u])\,\bar{X}) \implies
    (\lambda^i_{\PhaseMaps}\,[u])\,\bar{X} \sqsubseteq (\lambda^i_{\PhaseMaps}\,[u])\,\bar{Y}
\]
for every non-initial program point $u$ that is not inside an atomic block.
This invariant is established by induction over the Kleene iterates: whenever a contribution reaches such an
entry point through a non-atomic incoming edge, the non-atomic branch of $\succs$ emits the successor contribution
at every phase map in $\psuccs$.
Within a fixed partition, all incoming contributions use the same phase-successor relation, which is
unchanged by their join. By transitivity of this fixed relation, every contribution reaching $\bar X$
is therefore also propagated to every $\bar Y$ admitted by the successor query in the displayed invariant.
Taking the join within the partition preserves the resulting inequality.
This entry-closure property relies on the analysis being path-sensitive in the concurrency contexts.
For an analysis that is not path-sensitive in the concurrency contexts, this closure does not hold,
in general, making it necessary to argue about individual contributions.

Altogether, we then get
\[
    \begin{array}{lll}
    \lambda\,[v,((i,\hat{P}''),t'')] &=&
    (\lambda_{\PhaseMaps}\,[v])\,\bar{P}'' \\
    &\sqsupseteq& (\aSemP{\stmt} (\lambda_{\PhaseMaps}\,[u])\,\mu_{\PhaseMaps})_1\,\bar{P}'' \\
    &\sqsupseteq& ((\succs\,\bar P'\,((\lambda_{\PhaseMaps}\,[u])\,\bar{P}')\,(\unlift_{{\bar P}'}\,\mu_{\PhaseMaps}))_1)\,\bar{P}'' \\
    &\sqsupseteq& (\lift_{\bar{P}''} (\aSem{\stmt}((\lambda_{\PhaseMaps}\,[u])\,\bar{P}')\,(\unlift_{{\bar P}'}\,\mu_{\PhaseMaps})))_1\,\bar{P}'' \\
    &=& (\aSem{\stmt}((\lambda_{\PhaseMaps}\,[u])\,\bar{P}')\,(\unlift_{{\bar P}'}\,\mu_{\PhaseMaps}))_1 \\
    &=& (\aSem{\stmt}_I ((\lambda_{\PhaseMaps}\,[u])\,\bar{P}')\,((\unlift_{{\bar P}'}\,\mu_{\PhaseMaps})\,[\stmt']))_1 \\
    &\sqsupseteq& (\aSem{\stmt}_I (\lambda\,[u,((i,\hat{P}),t)])\,(\mu\,[\stmt',((i',\hat{P}'),t')]))_1 \\
    \end{array}
\]
Thus, the property holds for the first element of the pair, and as no contributions to global unknowns are made, it also holds overall.
The proof for the case where $\bar P = \bar{P}'$ is obtained as a simplification of the above proof,
as is the case where the global does not satisfy $\phaseBound$.

\subsubsection{Must-join reasoning}\label{app:must-join-soundness}
The implementation additionally uses must-join information to restrict a joined thread to
the phases in which it may return. For example, if thread~\texttt{t} can return only in
phase~$3$, its phase is fixed to~$3$ after a successful \texttt{join(t)}.
The MHP analysis uses the must-join information to rule out any further phase changes by
\texttt{t} after the join, since the joined thread is no longer active.

This refinement can be understood by instrumenting \texttt{t} with a fresh flag that is
initially~$0$. At the program point where \texttt{t} returns, we set
\texttt{t\_returned = 1}. In the joining thread, consider adding the check
\texttt{while (!t\_returned) \{ \}} immediately after \texttt{join(t)}. This loop is never entered
in the concrete, because a successful join implies that \texttt{t} has returned.
In the abstract, its effect is that the phase map entry for \texttt{t} is forced to $3$.
Consequently, fixing the phase of \texttt{t}
to a phase from which a return can happen does not exclude any reachable concrete state.

%% file: appendix/crafted.tex
\section{Details on Crafted Phase-Ghost Witnesses for \SVComp Tasks}\label{sec:crafted-appendix}
Out of the 374 examples in the \texttt{C.unreach-call.Concurrency} category of the \SVComp competition from 2026,
\Goblint was able to solve 52 tasks without any witness-guidance.

\cref{tab:crafted-phase-ghost-overview} provides an overview of the tasks for which we crafted witnesses.
We remark that despite some task names making references to data races, all tasks are race-free
and the property of interest is establishing that the function \code{reach_error} is not called.
For the lines of code, we report the number of lines in the file that is the official
input for \SVComp (\code{.i} file except for weaver, where it is an unprocessed \code{.wcr.c} file).


The case of \texttt{pthread-C-DAC/pthread-demo-datarace-1.yml} is particularly interesting here:
The witness here encodes how often each of the two threads has incremented a given counter, leading
to having to distinguish between 20 different phases for each thread. This led
to a total of 400 different combinations of phases. While this is likely to
cause some slowdown, we believe that much of the slowdown here is due to the way the \GoblintAnon
framework on which we build implements path-sensitivity, which causes an additional quadratic
blow-up in the number of computations that is unrelated to our approach.
Additionally, this task is a prime example where we suspect a better relational abstraction of the
map from phase maps to abstract states can avoid the observed blowup.

For this task, the version of \Ultimate (\texttt{0.2.4-?-e94c40d})
used in the original publication~\cite{ErhardBHKSSSTV2025} on concurrent correctness
witnesses did not terminate within 30 minutes at all,
whereas version \texttt{0.3.2-dev-43c3bf1a5c} terminates only for the configurations without
a witness.
To see the impact of the constants, we generated a smaller
version of this task, setting the constants to 3 instead of to 20. Here, \GoblintAnon
terminates in around one second, whereas \Ultimate takes about a minute.

\Ultimate has some limitations about the type of statements that a ghost update can be associated with.
For \texttt{stateful01-2.yml}, we had to change some increment statements from \texttt{g++;}
to the statement \texttt{g+=1;} which is semantically equivalent in the considered program
to be able to associate the ghost updates with it.

\section{AI-Generated Programs}\label{sec:crafted-ai}

The suite contains 50 programs generated with AI support, each inconclusive for \Goblint without
witness guidance but provable using the supplied phase information. All programs use pthreads,
with synchronization through mutexes and joins.
The tasks cover compact phase protocols inspired by queues, caches, replicas, and other systems
patterns; nondeterministic background-thread creation and worker loop bounds; mutex access through
pointers; and assertions combining phase-sensitive values with points-to information.
Larger tasks include helper functions, additional control flow, and larger shared state while
requiring only a few phase distinctions. Further protocols draw inspiration from \textsc{SCTBench}
and systems idioms such as work-stealing deques, seqlocks, token buckets, priority inheritance,
and RCU-style retirement.
Half of the witnesses also contain location invariants; the remainder provide only ghost
instrumentation. The suite thus covers both phase guidance alone and validation of invariants
alongside phase guidance.

\cref{tab:crafted-ai-runtimes} reports the runtimes and outcomes for all 50 tasks.
We find that \Ultimate is also able to verify (most of) the tasks with and without
witness, but at significantly higher runtimes than \GoblintAnon.

\begin{table}[p]
    \footnotesize
    \centering
    \renewcommand{\arraystretch}{0.90}
    \begin{tabular*}{\textwidth}{@{\extracolsep{\fill}}lcccc@{}}
        \toprule
        & \multicolumn{2}{c}{\GoblintAnon} & \multicolumn{2}{c}{\textsc{Ultimate Gemcutter}} \\
        \cmidrule(lr){2-3}\cmidrule(lr){4-5}
        Task & \shortstack{Verification\\w/o Witness} &
            \shortstack{Validation\\with Witness} &
            \shortstack{Verification\\w/o Witness} &
            \shortstack{Validation\\with Witness} \\
        \midrule
        \texttt{01-bank-ledger}~\textbf{I}
            & \textcolor{red}{\faTimesCircle} (0.3s)
            & \textcolor{green!50!black}{\faCheckCircle} (1.4s)
            & \textcolor{green!50!black}{\faCheckCircle} (37.0s)
            & \textcolor{green!50!black}{\faCheckCircle} (33.0s) \\
        \texttt{02-warehouse-stock}~\textbf{I}
            & \textcolor{red}{\faTimesCircle} (0.3s)
            & \textcolor{green!50!black}{\faCheckCircle} (4.0s)
            & \textcolor{green!50!black}{\faCheckCircle} (215.5s)
            & \textcolor{green!50!black}{\faCheckCircle} (260.3s) \\
        \texttt{03-telemetry-rollup}~\textbf{I}
            & \textcolor{red}{\faTimesCircle} (0.2s)
            & \textcolor{green!50!black}{\faCheckCircle} (1.1s)
            & \textcolor{green!50!black}{\faCheckCircle} (169.8s)
            & \textcolor{green!50!black}{\faCheckCircle} (177.8s) \\
        \texttt{04-cache-statistics}~\textbf{I}
            & \textcolor{red}{\faTimesCircle} (0.3s)
            & \textcolor{green!50!black}{\faCheckCircle} (1.2s)
            & \textcolor{green!50!black}{\faCheckCircle} (159.9s)
            & \textcolor{green!50!black}{\faCheckCircle} (205.5s) \\
        \texttt{05-reference-lifecycle}~\textbf{I}
            & \textcolor{red}{\faTimesCircle} (0.3s)
            & \textcolor{green!50!black}{\faCheckCircle} (7.2s)
            & {\faHourglassEnd} (5min)
            & {\faHourglassEnd} (5min) \\
        \texttt{06-payment-batch}~\textbf{I}
            & \textcolor{red}{\faTimesCircle} (0.3s)
            & \textcolor{green!50!black}{\faCheckCircle} (3.5s)
            & {\faHourglassEnd} (5min)
            & {\faHourglassEnd} (5min) \\
        \texttt{07-log-rotation}~\textbf{I}
            & \textcolor{red}{\faTimesCircle} (0.2s)
            & \textcolor{green!50!black}{\faCheckCircle} (0.7s)
            & \textcolor{green!50!black}{\faCheckCircle} (132.5s)
            & {\faHourglassEnd} (5min) \\
        \texttt{08-sensor-calibration}~\textbf{I}
            & \textcolor{red}{\faTimesCircle} (0.3s)
            & \textcolor{green!50!black}{\faCheckCircle} (0.9s)
            & \textcolor{green!50!black}{\faCheckCircle} (63.2s)
            & \textcolor{green!50!black}{\faCheckCircle} (66.2s) \\
        \texttt{09-resource-pool}~\textbf{I}
            & \textcolor{red}{\faTimesCircle} (0.4s)
            & \textcolor{green!50!black}{\faCheckCircle} (7.5s)
            & {\faHourglassEnd} (5min)
            & {\faHourglassEnd} (5min) \\
        \texttt{10-sharded-counter}~\textbf{I}
            & \textcolor{red}{\faTimesCircle} (0.3s)
            & \textcolor{green!50!black}{\faCheckCircle} (3.1s)
            & \textcolor{green!50!black}{\faCheckCircle} (217.2s)
            & \textcolor{green!50!black}{\faCheckCircle} (251.0s) \\
        \texttt{11-index-maintenance}~\textbf{I}
            & \textcolor{red}{\faTimesCircle} (0.3s)
            & \textcolor{green!50!black}{\faCheckCircle} (0.7s)
            & {\faHourglassEnd} (5min)
            & {\faHourglassEnd} (5min) \\
        \texttt{12-media-pipeline}~\textbf{I}
            & \textcolor{red}{\faTimesCircle} (0.3s)
            & \textcolor{green!50!black}{\faCheckCircle} (1.1s)
            & \textcolor{green!50!black}{\faCheckCircle} (109.3s)
            & {\faHourglassEnd} (5min) \\
        \texttt{13-session-accounting}~\textbf{I}
            & \textcolor{red}{\faTimesCircle} (0.3s)
            & \textcolor{green!50!black}{\faCheckCircle} (1.5s)
            & \textcolor{green!50!black}{\faCheckCircle} (17.1s)
            & \textcolor{green!50!black}{\faCheckCircle} (13.3s) \\
        \texttt{14-replica-progress}~\textbf{I}
            & \textcolor{red}{\faTimesCircle} (0.2s)
            & \textcolor{green!50!black}{\faCheckCircle} (2.0s)
            & \textcolor{green!50!black}{\faCheckCircle} (76.9s)
            & \textcolor{green!50!black}{\faCheckCircle} (107.1s) \\
        \texttt{15-build-queue}~\textbf{I}
            & \textcolor{red}{\faTimesCircle} (0.3s)
            & \textcolor{green!50!black}{\faCheckCircle} (3.0s)
            & \textcolor{green!50!black}{\faCheckCircle} (34.1s)
            & \textcolor{green!50!black}{\faCheckCircle} (32.0s) \\
        \texttt{16-order-fulfillment}~\textbf{I}
            & \textcolor{red}{\faTimesCircle} (0.3s)
            & \textcolor{green!50!black}{\faCheckCircle} (0.8s)
            & \textcolor{green!50!black}{\faCheckCircle} (34.8s)
            & \textcolor{green!50!black}{\faCheckCircle} (38.8s) \\
        \texttt{17-network-bookkeeping}~\textbf{I}
            & \textcolor{red}{\faTimesCircle} (0.2s)
            & \textcolor{green!50!black}{\faCheckCircle} (0.6s)
            & \textcolor{green!50!black}{\faCheckCircle} (60.6s)
            & \textcolor{green!50!black}{\faCheckCircle} (65.0s) \\
        \texttt{18-snapshot-service}~\textbf{I}
            & \textcolor{red}{\faTimesCircle} (0.2s)
            & \textcolor{green!50!black}{\faCheckCircle} (1.4s)
            & \textcolor{green!50!black}{\faCheckCircle} (69.7s)
            & \textcolor{green!50!black}{\faCheckCircle} (88.4s) \\
        \texttt{19-job-scheduler}~\textbf{I}
            & \textcolor{red}{\faTimesCircle} (0.3s)
            & \textcolor{green!50!black}{\faCheckCircle} (5.2s)
            & {\faHourglassEnd} (5min)
            & {\faHourglassEnd} (5min) \\
        \texttt{20-audit-aggregation}~\textbf{I}
            & \textcolor{red}{\faTimesCircle} (0.3s)
            & \textcolor{green!50!black}{\faCheckCircle} (4.2s)
            & {\faHourglassEnd} (5min)
            & {\faHourglassEnd} (5min) \\
        \texttt{21-dynamic-metrics-daemon}
            & \textcolor{red}{\faTimesCircle} (0.3s)
            & \textcolor{green!50!black}{\faCheckCircle} (1.6s)
            & \textcolor{green!50!black}{\faCheckCircle} (38.3s)
            & \textcolor{green!50!black}{\faCheckCircle} (35.0s) \\
        \texttt{22-dynamic-cache-sweeper}
            & \textcolor{red}{\faTimesCircle} (0.4s)
            & \textcolor{green!50!black}{\faCheckCircle} (2.8s)
            & \textcolor{green!50!black}{\faCheckCircle} (131.1s)
            & \textcolor{green!50!black}{\faCheckCircle} (158.6s) \\
        \texttt{23-dynamic-broker-heartbeat}
            & \textcolor{red}{\faTimesCircle} (0.4s)
            & \textcolor{green!50!black}{\faCheckCircle} (4.9s)
            & {\faHourglassEnd} (5min)
            & {\faHourglassEnd} (5min) \\
        \texttt{24-dynamic-index-watchers}
            & \textcolor{red}{\faTimesCircle} (0.3s)
            & \textcolor{green!50!black}{\faCheckCircle} (1.6s)
            & \textcolor{green!50!black}{\faCheckCircle} (37.7s)
            & \textcolor{green!50!black}{\faCheckCircle} (33.7s) \\
        \texttt{25-dynamic-gc-assistants}
            & \textcolor{red}{\faTimesCircle} (0.3s)
            & \textcolor{green!50!black}{\faCheckCircle} (2.2s)
            & \textcolor{green!50!black}{\faCheckCircle} (130.7s)
            & \textcolor{green!50!black}{\faCheckCircle} (131.0s) \\
        \texttt{26-pointer-lock-ledger}~\textbf{I}
            & \textcolor{red}{\faTimesCircle} (0.3s)
            & \textcolor{green!50!black}{\faCheckCircle} (0.8s)
            & \textcolor{green!50!black}{\faCheckCircle} (23.9s)
            & \textcolor{green!50!black}{\faCheckCircle} (22.5s) \\
        \texttt{27-pointer-lock-router}~\textbf{I}
            & \textcolor{red}{\faTimesCircle} (0.3s)
            & \textcolor{green!50!black}{\faCheckCircle} (0.9s)
            & \textcolor{green!50!black}{\faCheckCircle} (28.4s)
            & \textcolor{green!50!black}{\faCheckCircle} (25.7s) \\
        \texttt{28-pointer-lock-quota}~\textbf{I}
            & \textcolor{red}{\faTimesCircle} (0.3s)
            & \textcolor{green!50!black}{\faCheckCircle} (0.7s)
            & \textcolor{green!50!black}{\faCheckCircle} (26.3s)
            & \textcolor{green!50!black}{\faCheckCircle} (21.6s) \\
        \texttt{29-pointer-lock-journal}~\textbf{I}
            & \textcolor{red}{\faTimesCircle} (0.3s)
            & \textcolor{green!50!black}{\faCheckCircle} (0.9s)
            & \textcolor{green!50!black}{\faCheckCircle} (25.9s)
            & \textcolor{green!50!black}{\faCheckCircle} (25.0s) \\
        \texttt{30-pointer-lock-allocator}~\textbf{I}
            & \textcolor{red}{\faTimesCircle} (0.3s)
            & \textcolor{green!50!black}{\faCheckCircle} (0.8s)
            & \textcolor{green!50!black}{\faCheckCircle} (24.2s)
            & \textcolor{green!50!black}{\faCheckCircle} (23.2s) \\
        \texttt{31-points-to-session}
            & \textcolor{red}{\faTimesCircle} (0.4s)
            & \textcolor{green!50!black}{\faCheckCircle} (1.0s)
            & \textcolor{green!50!black}{\faCheckCircle} (97.8s)
            & \textcolor{green!50!black}{\faCheckCircle} (121.6s) \\
        \texttt{32-points-to-buffer}
            & \textcolor{red}{\faTimesCircle} (0.3s)
            & \textcolor{green!50!black}{\faCheckCircle} (0.8s)
            & \textcolor{green!50!black}{\faCheckCircle} (81.6s)
            & \textcolor{green!50!black}{\faCheckCircle} (100.3s) \\
        \texttt{33-points-to-owner}
            & \textcolor{red}{\faTimesCircle} (0.3s)
            & \textcolor{green!50!black}{\faCheckCircle} (0.9s)
            & \textcolor{green!50!black}{\faCheckCircle} (104.1s)
            & \textcolor{green!50!black}{\faCheckCircle} (115.4s) \\
        \texttt{34-points-to-routing}
            & \textcolor{red}{\faTimesCircle} (0.4s)
            & \textcolor{green!50!black}{\faCheckCircle} (1.1s)
            & \textcolor{green!50!black}{\faCheckCircle} (120.7s)
            & \textcolor{green!50!black}{\faCheckCircle} (129.1s) \\
        \texttt{35-points-to-snapshot}
            & \textcolor{red}{\faTimesCircle} (0.3s)
            & \textcolor{green!50!black}{\faCheckCircle} (1.0s)
            & \textcolor{green!50!black}{\faCheckCircle} (102.3s)
            & \textcolor{green!50!black}{\faCheckCircle} (114.5s) \\
        \texttt{36-large-mailbox-drain}
            & \textcolor{red}{\faTimesCircle} (0.2s)
            & \textcolor{green!50!black}{\faCheckCircle} (1.1s)
            & {\faHourglassEnd} (5min)
            & {\faHourglassEnd} (5min) \\
        \texttt{37-large-report-rollup}
            & \textcolor{red}{\faTimesCircle} (0.4s)
            & \textcolor{green!50!black}{\faCheckCircle} (1.6s)
            & \textcolor{black}{$\blacktriangle$} (96.0s)
            & \textcolor{black}{$\blacktriangle$} (85.6s) \\
        \texttt{38-large-package-index}
            & \textcolor{red}{\faTimesCircle} (0.3s)
            & \textcolor{green!50!black}{\faCheckCircle} (1.5s)
            & \textcolor{black}{$\blacktriangle$} (113.5s)
            & \textcolor{black}{$\blacktriangle$} (97.8s) \\
        \texttt{39-large-security-scan}
            & \textcolor{red}{\faTimesCircle} (0.3s)
            & \textcolor{green!50!black}{\faCheckCircle} (1.3s)
            & \textcolor{black}{$\blacktriangle$} (81.2s)
            & \textcolor{black}{$\blacktriangle$} (95.3s) \\
        \texttt{40-large-backup-catalog}
            & \textcolor{red}{\faTimesCircle} (0.2s)
            & \textcolor{green!50!black}{\faCheckCircle} (1.1s)
            & {\faHourglassEnd} (5min)
            & {\faHourglassEnd} (5min) \\
        \texttt{41-large-event-normalizer}
            & \textcolor{red}{\faTimesCircle} (0.3s)
            & \textcolor{green!50!black}{\faCheckCircle} (1.4s)
            & \textcolor{black}{$\blacktriangle$} (103.9s)
            & \textcolor{black}{$\blacktriangle$} (87.5s) \\
        \texttt{42-large-image-manifest}
            & \textcolor{red}{\faTimesCircle} (0.3s)
            & \textcolor{green!50!black}{\faCheckCircle} (1.3s)
            & \textcolor{black}{$\blacktriangle$} (77.7s)
            & \textcolor{black}{$\blacktriangle$} (103.0s) \\
        \texttt{43-large-metrics-fold}
            & \textcolor{red}{\faTimesCircle} (0.3s)
            & \textcolor{green!50!black}{\faCheckCircle} (1.2s)
            & \textcolor{black}{$\blacktriangle$} (76.7s)
            & \textcolor{black}{$\blacktriangle$} (73.1s) \\
        \texttt{44-large-retention-policy}
            & \textcolor{red}{\faTimesCircle} (0.3s)
            & \textcolor{green!50!black}{\faCheckCircle} (1.0s)
            & {\faHourglassEnd} (5min)
            & {\faHourglassEnd} (5min) \\
        \texttt{45-large-symbol-table}
            & \textcolor{red}{\faTimesCircle} (0.3s)
            & \textcolor{green!50!black}{\faCheckCircle} (1.4s)
            & \textcolor{black}{$\blacktriangle$} (78.4s)
            & \textcolor{black}{$\blacktriangle$} (79.2s) \\
        \texttt{46-mixed-range-table}
            & \textcolor{red}{\faTimesCircle} (0.3s)
            & \textcolor{green!50!black}{\faCheckCircle} (1.7s)
            & \textcolor{green!50!black}{\faCheckCircle} (59.1s)
            & \textcolor{green!50!black}{\faCheckCircle} (79.5s) \\
        \texttt{47-mixed-mask-table}
            & \textcolor{red}{\faTimesCircle} (0.3s)
            & \textcolor{green!50!black}{\faCheckCircle} (0.7s)
            & \textcolor{green!50!black}{\faCheckCircle} (19.7s)
            & \textcolor{green!50!black}{\faCheckCircle} (20.2s) \\
        \texttt{48-mixed-window-table}
            & \textcolor{red}{\faTimesCircle} (0.3s)
            & \textcolor{green!50!black}{\faCheckCircle} (1.8s)
            & \textcolor{green!50!black}{\faCheckCircle} (21.4s)
            & \textcolor{green!50!black}{\faCheckCircle} (25.5s) \\
        \texttt{49-mixed-priority-table}
            & \textcolor{red}{\faTimesCircle} (0.3s)
            & \textcolor{green!50!black}{\faCheckCircle} (1.4s)
            & \textcolor{green!50!black}{\faCheckCircle} (37.4s)
            & \textcolor{green!50!black}{\faCheckCircle} (47.0s) \\
        \texttt{50-mixed-state-table}
            & \textcolor{red}{\faTimesCircle} (0.3s)
            & \textcolor{green!50!black}{\faCheckCircle} (1.1s)
            & \textcolor{green!50!black}{\faCheckCircle} (18.7s)
            & \textcolor{green!50!black}{\faCheckCircle} (21.1s) \\
        \bottomrule
    \end{tabular*}
    \caption{Runtimes and outcomes for the AI-generated programs with \GoblintAnon and \Ultimate.
        \textcolor{green!50!black}{\faCheckCircle}: success;
        \textcolor{red}{\faTimesCircle}: inconclusive;
        \faHourglassEnd: timeout (5 minutes);
        \textcolor{black}{$\blacktriangle$}: out of RAM (exit code 7).
        \textbf{I} marks witnesses containing location invariants.
        Runtimes are rounded to one decimal place.}
    \label{tab:crafted-ai-runtimes}
\end{table}

%% file: appendix/cyclic-phases.tex
\section{A String-Buffer Example with Cyclic Phases}\label{sec:cyclic-phase-example}
\cref{fig:cyclic-phase-program} illustrates the cyclic protocol from \cref{ss:exploiting-phase-ghosts}.
Here, \code{?} denotes nondeterministic choice.
The producer cycles between phases $1$ (not-ready) and $2$ (ready), with distinct
phases $0$ (initialization) and $3$ (end) before and after the loop.
In phase $1$, it fills the entire buffer with arbitrary characters and releases the mutex,
so the buffer can be unterminated when the consumer acquires the mutex.
Before publishing \code{ready = 1}, it writes a terminator into the final element.
Thus, phase $2$ always satisfies $\code{buffer}[N-1]=0$, even though the other characters
may differ between iterations. This invariant implies null termination within the buffer.
A base analysis that tracks the final array element can represent this property.

The consumer's bounded scan asserts that a terminator exists whenever \code{ready} is set.
The check of \code{ready} and the scan occur in the same critical section, during which
MHP information can exclude the producer's phase changes.
The example illustrates how a property common to all ready states can survive cyclic phase changes.

\begin{figure}[htbp]
    \begin{minipage}[t]{0.54\textwidth}
\begin{minted}{C}
#include <assert.h>
#include <pthread.h>

#define N 8
pthread_mutex_t mutex = PTHREAD_MUTEX_INITIALIZER;
char buffer[N];
int ready = 0;
@\colorbox{gray!20}{\strut\texttt{int }$\ghost$\texttt{ = 0;}}@

void *producer(void *arg) {
    while (?) {
        pthread_mutex_lock(&mutex);
        @\colorbox{gray!20}{\strut\texttt{atomic \{}}@ ready = 0; @\colorbox{gray!20}{\strut$\ghost$ \texttt{= 1; \}}}@
        for (int i = 0; i < N; ++i)
            buffer[i] = ?;
        pthread_mutex_unlock(&mutex);

        pthread_mutex_lock(&mutex);
        buffer[N - 1] = '\0';
        @\colorbox{gray!20}{\strut\texttt{atomic \{}}@ ready = 1; @\colorbox{gray!20}{\strut$\ghost$ \texttt{= 2; \}}}@
        pthread_mutex_unlock(&mutex);
    }
    pthread_mutex_lock(&mutex);
    @\colorbox{gray!20}{\strut\texttt{atomic \{}}@ ready = 0; @\colorbox{gray!20}{\strut$\ghost$ \texttt{= 3; \}}}@
    pthread_mutex_unlock(&mutex);
    return NULL;
}
\end{minted}
    \end{minipage}\hfill
    \begin{minipage}[t]{0.44\textwidth}
\begin{minted}{C}
void *consumer(void *arg) {
    pthread_mutex_lock(&mutex);
    if (ready) {
        int i = 0;
        while (i < N && buffer[i] != '\0')
            ++i;
        assert(i < N);
    }
    pthread_mutex_unlock(&mutex);
    return NULL;
}

int main(void) {
    pthread_t p, c;
    pthread_create(&p, NULL, producer, NULL);
    pthread_create(&c, NULL, consumer, NULL);
    pthread_join(p, NULL);
    pthread_join(c, NULL);
    return 0;
}
\end{minted}
    \scalebox{0.6}{
        \begin{tikzpicture}[
        phase/.style={draw,rounded corners=2pt,fill=gray!10,align=center,
            minimum width=1.7cm,minimum height=0.85cm,font=\small},
        every edge/.style={draw,-{Stealth}},
        node distance=1cm
    ]
        \node[phase] (init) {$\ghost==0$\\initialization};
        \node[phase,right=of init] (preparing) {$\ghost==1$\\not-ready};
        \node[phase,right=of preparing] (ready) {$\ghost==2$\\ready};
        \node[phase,right=of ready] (end) {$\ghost==3$\\end};
        \path (init) edge (preparing)
              (preparing) edge node[above,font=\scriptsize] {} (ready)
              (ready) edge node[above,font=\scriptsize] {} (end)
              (ready) edge[bend left=40] node[below,font=\scriptsize] {} (preparing);
        \draw[-{Stealth}] (init.north) to[out=20,in=155]
            node[above,font=\scriptsize] {} (end.north);
    \end{tikzpicture}}
    \end{minipage}

    \caption{A string-buffer protocol and its phase transitions.}
    \label{fig:cyclic-phase-program}
\end{figure}